\documentclass{aa}  

\usepackage{graphicx}
\usepackage{txfonts}
\usepackage{multirow}%
\usepackage{amsmath,amssymb,amsfonts}%
\usepackage{mathrsfs}%
\usepackage[title]{appendix}%
\usepackage{xcolor}%
\usepackage{textcomp}%
\usepackage{manyfoot}%
\usepackage{booktabs}%
\usepackage{algorithm}%
\usepackage{algorithmicx}%
\usepackage{algpseudocode}%
\usepackage{listings}%
\usepackage{ragged2e}
\usepackage{tikz}
\usepackage{natbib}

\usepackage{cancel}
\usepackage{placeins}

\definecolor{emerald}{rgb}{0.0,0.5,0.0}
\definecolor{smcolor}{rgb}{0.7,0.3,0.0}
\definecolor{blue-violet}{rgb}{0.54, 0.17, 0.89}
\newcommand{\rev}[1]{#1}

\newcommand{\LEt}[1]{}

\def\inumber{i}                                 
\def\define{\equiv}                             
\def\scale{\, \propto \,}                       

\newcommand{\notation}[4]{#1_{#2 ; #3}^{#4}}            
\newcommand{\ddroit}{{\rm d}}                           
\newcommand{\vect}[1]{\boldsymbol{#1}}          
\newcommand{\evect}[1]{\vect{{\rm e}}_{#1}}     
\newcommand{\normvar}[1]{\tilde{#1}}  
\newcommand{\dd}[2]{\partial_{#2} #1}           
\newcommand{\ddd}[3]{\partial_{#2 #3} #1}       
\newcommand{\DDn}[3]{\dfrac{\ddroit^{#3} #1}{\ddroit #2^{#3}}} 
\newcommand{\Dt}[1]{\dot{#1}}                           
\newcommand{\abs}[1]{\left| #1 \right|}         
\newcommand{\ceil}[1]{\lceil #1 \rceil}
\newcommand{\conj}[1]{\overline{#1}}
\newcommand{\matx}[1]{\boldsymbol{#1}}
\newcommand{\infvar}[1]{\ddroit #1}

\newcommand{\kron}[2]{\delta_{#1, #2}}
\newcommand{\mean}[1]{\overline{#1}}

\def\dotp{\cdot }
\def\crossp{\times}

\newcommand{\timeav}[1]{\left\langle #1 \right\rangle}

\def\ffunc{f}

\def\nab{\nabla}                                        
\def\grad{\nab}                                 
\def\lap{\nab^2}                                        
\def\div{\grad \dotp}                           
\def\XX{X}      

\newcommand{\cosi}[1]{C_{#1}}
\newcommand{\sini}[1]{S_{#1}}

\newcommand{\domainbd}[1]{\partial #1}
\def\surface{S}
\def\volume{V}
\def\surfacev{\vect{\surface}}

\def\domainstar{\domain_{*}}
\def\rdstar{\rr_{*}}
\def\domainstarbd{\domainbd{\domainstar}}

\def\angmomsymb{\mathcal{L}}
\def\amoperator{\boldsymbol{\angmomsymb}}
\newcommand{\amopi}[1]{\angmomsymb_{#1}}

\def\amopcoef{L}
\def\angmom{L}
\def\angmomv{\vect{\angmom}}

\def\angmomplav{\angmomv_{\ipla}}
\def\angmompert{\angmom_{\ipert}}
\def\angmompertv{\angmomv_{\ipert}}
\def\angmomtot{\angmom_{\itotal}}
\def\angmomtotv{\angmomv_{\itotal}}
\newcommand{\amopci}[3]{\amopcoef_{#1,#2}^{#3}}
\def\eplus{\evect{+}}
\def\eminus{\evect{-}}
\def\ezero{\evect{0}}
\def\compcoef{Z}

\def\llat{l}                                            
\def\lmax{N}

\def\mm{m}                                      
\def\kk{k}                                              

\def\slon{s}                                              
\def\jj{j}
\def\qq{q}
\def\pp{p}
\def\Legendre{P}                                
\def\SPH{Y}     

\newcommand{\Ylm}[2]{\SPH_{#1}^{#2}}

\newcommand{\wignerD}[3]{D_{#1,#2}^{#3}}
\newcommand{\wignerd}[3]{d_{#1,#2}^{#3}}
\def\wignermat{\matx{D}}
\newcommand{\wignerDmati}[3]{\wignermat_{#1,#2}^{#3}}

\def\cwrec{c}
\newcommand{\cwminusi}[3]{\cwrec_{#1,#2}^{#3,-}}
\newcommand{\cwplusi}[3]{\cwrec_{#1,#2}^{#3,+}}
\newcommand{\cwzeroi}[3]{\cwrec_{#1,#2}^{#3,0}}

\def\rotmat{\matx{R}}

\newcommand{\rotmati}[1]{\rotmat_{#1}}

\def\eulermat{\matx{R}}
\def\angalp{\alpha}
\def\angbet{\beta}
\def\anggam{\gamma}

\def\angapert{\angalp_{\ipert}}
\def\angbpert{\angbet_{\ipert}}
\def\anggpert{\anggam_{\ipert}}

\def\tidaltf{H}
\newcommand{\tidaltfi}[3]{\tidaltf_{#1,#2}^{#3}}

\def\nn{n}                                              
\def\spinpar{\nu}                        

\newcommand{\LegP}[1]{\Legendre_{#1}}                     
\newcommand{\LegF}[2]{\Legendre_{#1}^{#2}}                
\newcommand{\Hansen}[3]{X_{#1}^{#2,#3}}                 
\newcommand{\expo}[1]{{\rm e}^{#1}}                             
\newcommand{\integ}[4]{\int_{#3}^{#4} #1 \infvar{#2} }   
\newcommand{\ointeg}[4]{\oint_{#3}^{#4} #1 \infvar{#2} }
\newcommand{\Houghval}[3]{\Lambda_{#1}^{#2,#3}}         

\def\Ggrav{G}                                   

\def\ipla{{\rm p}}                              

\def\ipert{{\rm s}}

\def\iocean{{\rm oc}}                           
\def\isol{{\rm sol}}                            
\def\idrag{{\rm R}}                             

\def\ieq{{\rm eq}}
\def\iwater{{\rm w}}
\def\itotal{{\rm tot}}

\def\xx{x}
\def\yy{y}
\def\zz{z}
\def\XX{X}
\def\YY{Y}
\def\ZZ{Z}
\def\rr{r}                                           
\def\col{\theta}                                
\def\lon{\varphi}                             
\def\time{t}                                    

\def\angrot{\phi}
\def\colpla{\hat{\col}}
\def\lonpla{\hat{\lon}}
\def\rrrot{\hat{\rr}}
\def\colrot{\hat{\col}}
\def\lonrot{\hat{\lon}}
\def\colpert{\col_{\ipert}}
\def\lonpert{\lon_{\ipert}}
\def\rrpla{\hat{r}}

\def\er{\evect{\rr}}                            
\def\etheta{\evect{\col}}                       
\def\ephi{\evect{\lon}}                 

\def\rvect{\vect{r}}                            
\def\ex{\evect{\xx}}
\def\ey{\evect{\yy}}
\def\ez{\evect{\zz}}

\def\eyprim{\ey^\prime}
\def\eX{\evect{\XX}}
\def\eY{\evect{\YY}}
\def\eZ{\evect{\ZZ}}

\def\framesymb{\mathcal{R}}                     

\newcommand{\framei}[1]{\framesymb_{#1}}
\def\framestd{\framei{}}

\def\ifix{{\rm f}}
\def\framepla{\framestd}

\def\framefix{\framei{\ifix}}

\newcommand{\rframe}[5]{\framesymb_{#1} {:} \left( #2, #3, #4, #5 \right) }

\def\ecc{e}                                     
\def\inclination{i}                           
\def\argperi{\omega}                     
\def\meana{\mathcal{M}}                                   
\def\truea{v}                                   
\def\smaxis{a}                                  
\def\meanmotion{n}			     

\def\lonan{\vartheta}                     
\def\obli{\beta}                     
\def\spinrate{\Omega}                   
\def\spinvect{\vect{\spinrate}}         

\def\Mbody{M}                                   
\def\Rbody{R}                                   
\def\Mpla{\Mbody_\ipla}                 

\def\Mpert{\Mbody_{\ipert}}
\def\npert{\meanmotion_{\ipert}}

\def\rpert{\rr_{\ipert}}

\def\eccpert{\ecc_{\ipert}}
\def\incpert{\inclination_{\ipert}}
\def\meanapert{\meana_{\ipert}}
\def\meanapertc{\meana_{\ipert;0}}
\def\trueapert{\truea_{\ipert}}
\def\smaxispert{\smaxis_{\ipert}}
 
\def\lonanpert{\lonan_{\ipert}}
\def\norb{\meanmotion_\ipert}         
\def\argperipert{\argperi_{\ipert}}

\def\rpertvect{\vect{\rr}_{\ipert}}
\def\Rpla{\Rbody_\ipla}                 

\def\Hlayer{H}                                  
\def\density{\rho}                              
\def\chartime{\tau}                             
\def\freq{\sigma}                               
\def\ggravi{g}                                  
\def\Hoc{\Hlayer}                       
\def\fdrag{\freq_\idrag}                        

\def\rhowater{\density_{\iwater}}
\def\force{F}

\def\fcorio{\vect{f}}

\def\Cinertie{C}

\def\Cinertpla{\Cinertie}

\def\mupla{\mu}                                 
\def\tauA{\chartime_{\rm A}}            
\def\tauM{\chartime_{\rm M}}            
\def\alphaA{\alpha}                             
\def\rhocore{\mean{\density}}           

\def\zenithphi{\Psi}				

\def\ftide{\freq}                                       
\def\ftidepla{\hat{\ftide}}
\def\ftiderot{\hat{\ftide}}

\def\period{P}                                  
\def\Porb{\period_{\ipert}}                     
\def\Prot{\period_{\rm rot}}

\newcommand{\ftidefixi}[1]{\ftidefix_{#1}}
\newcommand{\ftideplai}[1]{\ftidepla_{#1}}

\newcommand{\ftideroti}[1]{\ftiderot_{#1}}

\def\gravpot{U}                                 
\def\gravpotn{\normvar{\gravpot}}
\def\vel{V}                                             
\def\iforcing{{\rm T}}                          
\def\iocdyn{{\rm D}}

\def\Utide{\gravpot_\iforcing}          
\def\zetaoc{\zeta}
\def\zetaeq{\zetaoc_{\ieq}}
\def\Vvect{\vect{\vel}}                 
\def\Vr{\vel_\rr}                                       
\def\Vtheta{\vel_\col}                          
\def\Vphi{\vel_\lon}                            
\def\Qtide{Q}
\newcommand{\vartide}[1]{\delta #1}
\def\rhotide{\vartide{\density}}

\newcommand{\Ulmrotsigj}[3]{\notation{\hat{\gravpot}}{\iforcing}{#1}{#2,#3}}
\newcommand{\Uplmrotsigj}[3]{\notation{\hat{\gravpot}}{\iresp}{#1}{#2,#3}}
\newcommand{\zetalmrotsigi}[3]{\hat{\zetaoc}_{#1}^{#2,#3}}
\newcommand{\Ulmsigj}[3]{\notation{\gravpotn}{\iforcing}{#1}{#2,#3}}
\newcommand{\Uplmsigj}[3]{\notation{\gravpotn}{\iresp}{#1}{#2,#3}}

\def\iresp{{\rm D}}
\def\Utiden{\gravpotn_{\iforcing}}
\def\Uresp{\gravpot_{\iresp}}
\def\Urespn{\gravpotn_{\iresp}}

\newcommand{\Utidensigi}[1]{\Utiden^{#1}}
\newcommand{\Urespnsigi}[3]{\Urespn^{#1,#2,#3}}
\newcommand{\Urespnk}[1]{\Urespn^{#1}}
\newcommand{\Urespnrotsigi}[1]{\hat{\gravpot}_{\iresp}^{#1}}
\newcommand{\Utidenrotsigi}[1]{\hat{\gravpot}_{\iforcing}^{#1}}
\newcommand{\zetaocrotsigi}[1]{\hat{\zetaoc}^{#1}}

\def\Utidensig{\Utidensigi{\kk}}
\def\Urespnsig{\Urespnsigi{\kk}{\qq}{\pp}}

\def\focn{\ftide_{\nn}}

\newcommand{\periodi}[1]{\period_{#1}}

\def\lovegrav{k}                                        
\def\torque{\mathcal{T}}                        

\def\iload{{\rm L}}

\newcommand{\klsigi}[2]{\lovegrav_{#1}^{#2}}
\newcommand{\kloadlsigi}[2]{\lovegrav_{\iload;#1}^{#2}}

\def\tiltoperator{\Gamma}

\def\tiltopd{\tiltoperator_{\iocdyn}}
\def\tiltopg{\tiltoperator_{\iforcing}}

\def\torquev{\boldsymbol{\torque}}

\newcommand{\torquei}[1]{\torque_{#1}}
\def\torquex{\torquei{\xx}}
\def\torquey{\torquei{\yy}}
\def\torquez{\torquei{\zz}}

\def\Ktorque{K}

\def\domain{\mathcal{V}}

\def\forcevect{\vect{\force}}

\def\Ivect{\vect{I}_{\ipert}}
\def\Jvect{\vect{J}_{\ipert}}
\def\Kvect{\vect{K}_{\ipert}}

\def\nnodev{\vect{n}}

\def\idiss{{\rm diss}}
\def\power{\mathcal{P}}
\def\powertide{\power_{\iforcing}}

\def\powerdiss{\power_{\idiss}}

\def\nuco{\nu}

 \newcommand{\hklmq}[4]{\mathcal{H}_{#1,#2}^{#3,#4}}
 
\def\energy{E}
\def\enertot{\energy_{\itotal}}

\def\enerpert{\energy_{\ipert}}

\newcommand{\majorised}[1]{#1^{\rm UB}}
\def\Urespub{\majorised{\abs{\Uresp}}}

\newcommand{\eq}[1]{Eq.~(\ref{#1})}
\newcommand{\eqs}[2]{Eqs.~(\ref{#1}) and~(\ref{#2})}
\newcommand{\eqsto}[2]{Eqs.~(\ref{#1}-\ref{#2})}
\newcommand{\eqsthree}[3]{Eqs.~(\ref{#1}), (\ref{#2}), and~(\ref{#3})}

\newcommand{\append}[1]{Appendix~\ref{#1}}

\newcommand{\fig}[1]{Fig.~\ref{#1}}
\newcommand{\figs}[2]{Figs.~\ref{#1} and~\ref{#2}}
\newcommand{\figsto}[2]{Figs.~\ref{#1}-\ref{#2}}
\newcommand{\sect}[1]{Sect.~\ref{#1}}

\newcommand{\comments}[1]{}

\usepackage{etoolbox}
\usepackage[refpage]{nomencl}

\providetoggle{nomsort}
\settoggle{nomsort}{true} 

\makeatletter
\iftoggle{nomsort}{%
    \let\old@@@nomenclature=\@@@nomenclature        
        \newcounter{@nomcount} \setcounter{@nomcount}{0}%
        \newcommand{\threedigits}[1]{\ifnum#1<100 0\two@digits{#1} \else \number#1\fi}
        \renewcommand\the@nomcount{\threedigits{\value{@nomcount}}}
        \def\@@@nomenclature[#1]#2#3{
          \addtocounter{@nomcount}{1}%
        \def\@tempa{#2}\def\@tempb{#3}%
          \protected@write\@nomenclaturefile{}%
          {\string\nomenclatureentry{\the@nomcount\nom@verb\@tempa @[{\nom@verb\@tempa}]%
          \begingroup\nom@verb\@tempb\protect\nomeqref{\theequation}%
          |nompageref}{\thepage}}%
          \endgroup
          \@esphack}%
      }{}
\makeatother

\newcommand{\mynomone}[3][section]{%
  \begingroup\edef\x{\endgroup
  \unexpanded{\nomenclature{#2}}%
    {\unexpanded{#3} \hspace*{\fill}  (\csname the#1\endcsname)}}\x}

\newcommand{\mynomtwo}[4][section]{%
  \begingroup\edef\x{\endgroup
  \unexpanded{\nomenclature[#2]{#3}}%
    {\unexpanded{#4} \hspace*{\fill}  (\csname the#1\endcsname)}}\x}

\renewcommand\nomgroup[1]{%
  \item[\bfseries
  \ifstrequal{#1}{A}{Acronyms}{%
  \ifstrequal{#1}{S}{Symbols}{%
  \ifstrequal{#1}{C}{Other Symbols}{}}}%
]}

\newcounter{logglabel}

\newcommand{\mynom}[3][S]{\nomenclature[#1]{#2}{#3~}}

\makenomenclature

\newcommand{\twolabelsat}[4]{%
\begin{tikzpicture}
\path (0,0) node {\vphantom{.}};
\path (#1,0) node[inner sep=0cm, outer sep=0cm] {#2};
\path (#3,0) node[inner sep=0cm, outer sep=0cm] {#4};
\end{tikzpicture}
}

\begin{document}

\title{Anisotropic tidal dissipation in misaligned planetary systems}


   \author{Pierre Auclair-Desrotour
          \and
          Gwenaël Boué 
          \and 
          Baptiste Loire
          }

   \institute{IMCCE, Observatoire de Paris, Université PSL, CNRS, Sorbonne Université, 77 Avenue Denfert-Rochereau, Paris, 75014, France. \\
              \email{pierre.auclair-desrotour@obspm.fr}
             }

   \date{Received; accepted}

 
  \abstract
   {Tides are the main driving force behind the long-term evolution of planetary systems. The associated energy dissipation and momentum exchanges are commonly described by Love numbers, which relate the exciting potential to the tidally perturbed potential. These transfer functions are generally assumed to depend solely on tidal frequency and body rheology, following the isotropic assumption, which presumes invariance of properties by rotation about the centre of mass. }
   {We examine the limitations of the isotropic assumption for fluid bodies, where Coriolis acceleration breaks spherical symmetry, resulting in \rev{rotational scattering and complex tidal responses}.}
   {Using angular momentum theory, we derive a new formalism to calculate the tidal rates of energy and momentum transfers in non-isotropic cases. We apply this formalism to the Earth-Moon system to assess the effects of anisotropy in planet-satellite systems with misaligned spin and orbital angular momenta.}
   {Our findings indicate that the isotropic assumption can introduce significant errors in planetary evolution models, particularly in the dynamical tide regime. These errors stem from forced wave resonances, with inaccuracies in energy dissipation scaling in proportion to resonance amplification factors.}
   {}


   \maketitle
%


\section{Introduction}


Tides are among the most fundamental mechanisms driving long-term celestial dynamics. They  originate from the differential gravitational forces exerted by orbiting bodies on one another, leading to mass redistribution and energy dissipation. This process induces the transfer of angular momentum and energy between the spin of celestial bodies and their orbits, resulting in internal heating and progressive evolution of planetary systems over extensive timescales \citep[e.g.,][]{Ogilvie2014}. In particular, tides shape system architecture by establishing stable equilibrium states. The main stabilising effects include (i) circularisation, which tends to make orbits circular; (ii) synchronisation, where orbital and rotational periods become equal to each other; and (iii) alignment, where equatorial and orbital planes converge \citep[e.g.,][]{Hut1980,Hut1981}. 

Tidal dissipation also impacts the climate and surface conditions of rocky planets by influencing their capacity to maintain liquid water on their surfaces \citep[see e.g.,][and references therein]{Bolmont2018}. The associated tidal heating can transform the thermal state of such bodies, potentially melting interiors and triggering volcanism, as seen on Io \citep[][]{Peale1979,Peale2003}. \rev{Similarly, \cite{Tyler2008} showed that tides can produce substantial heat fluxes on icy moons in the outer Solar System, and thus sustain long-lived subsurface oceans of liquid water \citep[see also][]{Tyler2009,Tyler2011,Tyler2014,Tyler2020}. The dissipative processes generating tidal heating and the interplay between the oceanic response and the upper icy crust were subsequently examined by several authors \citep[e.g.,][]{Matsuyama2014,Kamata2015,Beuthe2016,Matsuyama2018,HM2019,Rekier2019,Rovira-Navarro2023}.}

Finally, tidal interactions are crucial to understanding the Earth-Moon system's 4.5-billion-year climato-biologic evolution, especially through their effects on Earth's length of day (LOD)\mynom[A]{LOD}{Length of day} and obliquity \citep[e.g.,][]{Daher2021,Tyler2021,Farhat2022tides}. On Earth, ocean tides dominate, accounting for over 90\% of energy dissipation \citep[e.g.,][]{ER2000,ER2001}. 
As proposed by \cite{ZW1987}, atmospheric thermal tides -- namely tides caused by Solar radiative flux -- may have equally played an important role by partly counteracting the decelerating effect of gravitational tides on Earth's rotation during the Precambrian. However, despite sparse geological data suggesting that Earth's mid-Proterozoic LOD was stabilised \citep[][]{BS2016,MK2023,Wu2023}, tidal locking likely never occurred \citep[][]{Farhat2024,Laskar2024}, unlike Venus, which is maintained in asynchronous rotation by thermal tides \citep[][]{GS1969,ID1978,CL2001,CL2003,Leconte2015}.

Tides are usually considered as small perturbations around an equilibrium state, allowing for a linearised approach \citep[e.g.,][]{Mathis2013}. The linear response of a body to tidal forces is quantified by the potential Love number, named after A.E.H. Love, which relates the perturbed gravitational potential -- resulting from variations in self-attraction -- to the tide-raising potential \citep[e.g.,][]{Ogilvie2014}. This potential Love number is a complex-valued, dimensionless transfer function that varies with tidal frequency and accounts for both tidal deformation and energy dissipation in the frequency domain\footnote{\rev{The potential Love number should not be confused with the real-valued Love number accounting for the elastic elongation of the primary in the context of bodily tide theory, which is commonly used in literature \citep[e.g.,][]{MM1960book,Mathis2013}. By definition, this real-valued Love number is independent of dissipative processes. Therefore, it has to be complemented by the so-called tidal quality factor, usually denoted by $\Qtide$, to describe the action of tides on the long-term evolution of planetary systems \citep[e.g.,][]{MacDonald1964,GS1966,EW2009}. As the complex-valued Love number used here relies on a more general definition of the tidal response, this transfer function appears to be more suited to the study of fluid tides than the elastic Love number, and the term `Love number' thus systematically refers to it throughout the article.}}. Commonly denoted by~$\lovegrav_\llat^\mm$\mynom[S]{$\lovegrav_\llat^\mm$}{Potential Love number}, where $\llat$\mynom[S]{$\llat$}{Degree of the spherical harmonics} and $\mm$\mynom[S]{$\mm$}{Order of the spherical harmonics} represent the degree and order of the spherical harmonic respectively, the Love number is often simplified to a single parameter, $\lovegrav_2$\mynom[S]{$\lovegrav_2$}{Degree-2 potential Love number under the isotropic assumption}, or the degree-2 Love number, which depends solely on the tidal frequency and the internal physics of the tidally perturbed body \citep[e.g.,][]{Mathis2013,CV2022,VC2022}. Such an approximation implies rotational invariance around the body's centre of mass, meaning that the background fields and tidal dynamics equations are independent of the body's orientation relative to the perturber's orbit. Throughout this study, we refer to this simplification as the `isotropic assumption'\rev{, noting that isotropy designates uniformity in all orientations. Although alternative terminology can be used, we opt for this one because it evokes a meaningful analogy between tidal perturbations and light rays propagating through an optical medium.} 

In solid bodies, the isotropic assumption is generally reasonable as a first-order approximation, since their internal structure and visco-elastic properties are primarily determined by radial self-attraction and pressure forces \citep[e.g.,][]{Sotin2007}. The resulting bodily tide is essentially a quasi-static adjustment to tidal forces, without inertial effects\footnote{Tides are low-frequency disturbances relative to the vibrational (or seismic) modes of rocky bodies, which prevents these modes from being resonantly excited by tidal forces. For example, the typical periods of solid Earth's free oscillations are shorter than one hour \citep[e.g.,][]{Alterman1959}.}. Consequently, all tidal constituents are straightforwardly obtained from the equatorial degree-2 Love number describing the semidiurnal tidal bulge in the coplanar-circular configuration, which is just evaluated across several tidal frequencies -- one per constituent. As a corollary, the degree-2 Love number is a symmetric function of the tidal frequency, with an even real part and an odd imaginary part, as discussed by \cite{Efroimsky2012b}.

Unlike solid bodies, fluid layers exhibit anisotropic tidal responses -- even in spherically symmetric cases -- because fluid particle motion is influenced by Coriolis acceleration. By introducing a preferred direction (the spin axis), Coriolis acceleration distorts tidal flows, making the fluid response markedly different from that of solid materials. \rev{Notably, anisotropy induces scattering:  the spherical harmonics coefficients describing the fluid tidal response are energetically coupled together instead of being independent as in solid bodies.} This symmetry-breaking effect is intensified by the so-called dynamical tide -- the component of the response associated with wave propagation \citep[][]{Zahn1975} -- since each oscillatory fluid mode can resonate at a specific frequency \citep[e.g.,][]{LH1968,SW2002,OL2004,Ogilvie2013,Ogilvie2014,ADLML2018}. 

A wide variety of wave types may arise, mixing or remaining distinct depending on the fluid layer's properties \citep[e.g.,][]{Fuller2024}. For instance, ocean tides on rocky planets are predominantly governed by long-wavelength surface gravity modes \citep[][]{LH1968,Cartwright1977book}. These waves, restored by self-attraction, resemble the planetary-scale compressibility modes known as Lamb waves \citep[e.g.,][]{Bretherton1969,LB1972} that can be gravitationally or radiatively driven in planetary atmospheres, including Earth's \citep[e.g.,][]{Farhat2024}. In other cases, the dynamical tide may consist of inertial, gravity, Alfvén, or acoustic waves, each restored by different forces -- Coriolis acceleration, buoyancy (in stably stratified layers), magnetic tension, or pressure, respectively \citep[e.g.,][]{Mathis2013,Rieutord2015book} -- as observed in stars and the fluid envelopes of giant planets.

Tidal anisotropy is further amplified by complex geometries. On Earth, continental coastlines redirect tidal flows, shifting resonant frequencies and disrupting symmetries \citep[e.g.,][]{Green2017,Blackledge2020,Lyard2021,Auclair2023}, while shallow seas add dynamic coupling effects between shelf and ocean tides \citep[e.g.,][]{Arbic2009,AB2010,Wilmes2023}. Owing to these complexities, the role of anisotropic effects in planetary system evolution remains poorly understood. Yet, these effects may significantly alter the characteristic signatures of dynamical tide, the latter manifesting as staircase-like variations in orbital parameters during resonance episodes \rev{\citep[e.g.,][]{Auclair2014,Tyler2021,Motoyama2020,Farhat2022tides,Huang2024,Zhou2024,Wu2024}}. 

In the present study, we examine the limitations of the isotropic approximation to characterise their impact on the modelled orbital evolution of binary systems with misaligned spin and orbital angular momenta. Our approach extends previous research on the past history of the Earth-Moon system \citep[][]{Farhat2022tides,Farhat2022ellip,Farhat2024,Auclair2023}. In \sect{sec:torque_power}, building on insights from \cite{Ogilvie2013} and \cite{Boue2017}, we introduce a new formalism to calculate the time-averaged tidal rates of angular momentum and energy transfers in non-isotropic configurations. We further complement this formalism in \sect{sec:dynamics_system} with the equations governing the long-term tidal evolution of a planet-perturber system. Finally, in \sect{sec:application_earth_moon}, we apply these methods to an idealised Earth-Moon system, which allows us to probe the validity domain of the isotropic assumption and quantify related errors. Our findings reveal that the isotropic assumption results in notably inaccurate predictions for evolution models in the dynamical tide regime of fluid bodies, with errors scaling proportionally to the amplification factors associated with resonantly excited tidal waves. Notations and acronyms used throughout this paper are listed in the nomenclature in \append{app:nomenclature}.

\section{Tidal torque and tidally dissipated power}
\label{sec:torque_power}

\subsection{Tidal torque as a function of tidal potentials}
\label{ssec:torque_tidal_potentials}

In this work, we examine the tidal interaction of two bodies orbiting around their centre of mass. For simplicity, we assume the central body to be a planet of mass $\Mpla$\mynom[S]{$\Mpla$}{Planet mass} and radius $\Rpla$\mynom[S]{$\Rpla$}{Planet radius}, and the tidal perturber a satellite of mass $\Mpert$\mynom[S]{$\Mpert$}{Satellite's mass}. Nevertheless, the derivations presented in this section apply to any combination of central bodies and perturbers. The satellite induces a tidal potential, $\Utide$\mynom[S]{$\Utide$}{Tide-raising gravitational potential}, which disturbs the planet's mass distribution. This, in turn, generates a deformation potential, $\Uresp$\mynom[S]{$\Uresp$}{Perturbed gravitational potential}, which influences the system's orbital evolution. Our initial aim is to establish the relationship between these two gravitational potentials and the rates of angular momentum and energy exchanges between the planet and satellite. These derivations are framed within linear theory, where tides are treated as infinitesimal perturbations, and will later be used in \sect{sec:dynamics_system} for numerical calculations.

We adopt a non-rotating, geocentric frame of reference, $\rframe{\ifix}{O}{\ex}{\ey}{\ez}$\mynom[S]{$\framefix$}{Galilean frame of reference}, centred at the planet's centre of mass, O\mynom[S]{O}{Planet's centre of mass}. The corresponding Cartesian unit vectors, $\left(\ex , \ey , \ez \right)$\mynom[S]{$\ex,\ey,\ez$}{Cartesian unit vector associated with $\framefix$}, point towards distant stars. In $\framefix$, the position of any point M is described using spherical polar coordinates $\left( \rr, \col , \lon \right)$, where $\rr$\mynom[S]{$\rr$}{Radial coordinate in $\framefix$} is the radial distance, $\col$ the colatitude\mynom[S]{$\col$}{Colatitude in $\framefix$}, and $\lon$ the longitude\mynom[S]{$\lon$}{Longitude in $\framefix$}. The position vector of point M is denoted as $\rvect \define \rr \er$\mynom[S]{$\rvect$}{Position vector of the current point}, while the \rev{position vector of the satellite's centre of mass} is given by $\rpertvect = \rpert \er$\mynom[S]{$\rpertvect$}{Satellite's position vector}, where $\er$\mynom[S]{$\er$}{Unit radial vector} is the unit radial vector. Additionally, we define the rotating reference frame associated with the planet, $\rframe{}{O}{\eX}{\eY}{\eZ}$\mynom[S]{$\framepla$}{Frame of reference rotating with the planet}\mynom[S]{$\eX,\eY,\eZ$}{Cartesian unit vectors associated with $\framepla$}, where $\eZ$ aligns with the planet's spin axis and points towards the North Pole, while the other two unit vectors, $\eX$ and $\eY$, define orthogonal directions in the planet's equatorial plane. In this rotating frame, the position of a point is described by the spherical coordinates $\left( \rrpla, \colpla, \lonpla \right)$\mynom[S]{$\rrpla$}{Radial coordinate in $\framepla$}\mynom[S]{$\colpla$}{Colatitude in $\framepla$}\mynom[S]{$\lonpla$}{Longitude in $\framepla$}. 

\begin{figure}[t]
   \centering
   \includegraphics[width=0.48\textwidth,trim = 0.cm 0.cm 11.7cm 6.2cm,clip]{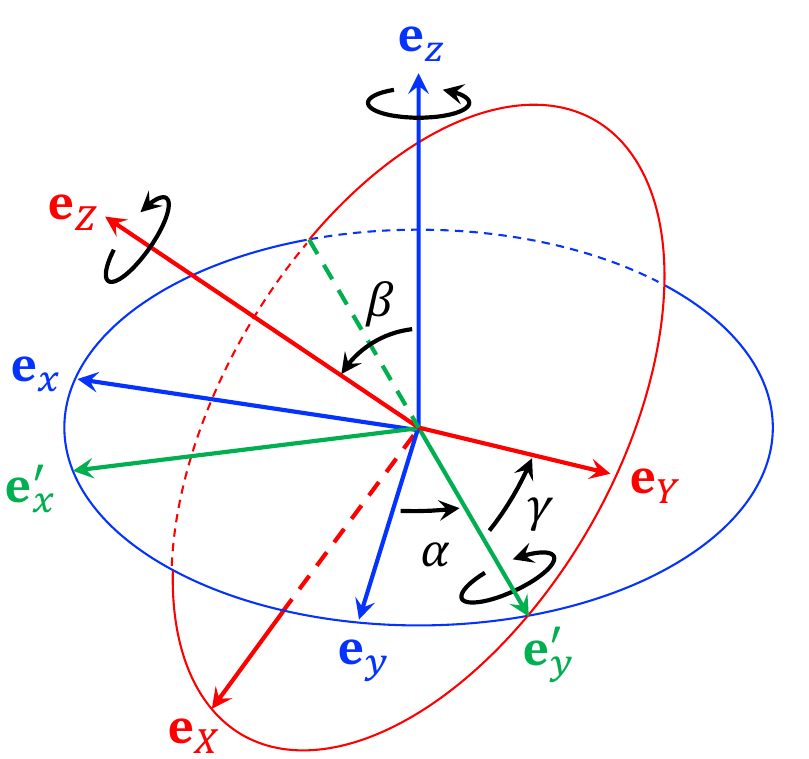}
      \caption{Euler angles $\left( \angalp, \angbet, \anggam \right)$ corresponding to the 3-2-3 Euler rotation matrix defined by \eq{eulermat}, which describes the change of coordinate systems $\rframe{\ifix}{O}{\ex}{\ey}{\ez} \rightarrow \rframe{}{O}{\eX}{\eY}{\eZ}$.}
       \label{fig:euler_angles}%
\end{figure}

As illustrated by \fig{fig:euler_angles}, the transformation between the two frames, $\rframe{\ifix}{O}{\ex}{\ey}{\ez} \rightarrow \rframe{}{O}{\eX}{\eY}{\eZ}$, is represented by the 3-2-3 Euler rotation matrix $\eulermat \left( \angalp, \angbet, \anggam \right)=  \rotmati{3} \left( \angalp \right) \rotmati{2} \left( \angbet \right) \rotmati{3} \left( \anggam \right) $\mynom[S]{$\eulermat$}{3-2-3 Eulerian rotation matrix}, where $\rotmati{2}$\mynom[S]{$\rotmati{\nn}$}{Rotation matrix around the $\nn$-th axis} and $\rotmati{3}$ are the rotation matrices around the second and the third axes, respectively. The matrix $\eulermat$ is parametrised by the Euler angles $\left( \angalp, \angbet, \anggam \right)$ as given by \citep[e.g.,][Eq.~(54) p.~30]{Varshalovich1988}
\begin{equation}
\eulermat \left( \angalp , \angbet , \anggam \right) \define 
\begin{bmatrix}
\cosi{\angalp} \cosi{\angbet} \cosi{\anggam} - \sini{\angalp} \sini{\anggam} & 
- \cosi{\angalp} \cosi{\angbet} \sini{\anggam} - \sini{\angalp} \cosi{\anggam} & 
\cosi{\angalp} \sini{\angbet} \\
\sini{\angalp} \cosi{\angbet} \cosi{\anggam} + \cosi{\angalp} \sini{\anggam} & 
- \sini{\angalp} \cosi{\angbet} \sini{\anggam} + \cosi{\angalp} \cosi{\anggam} & 
\sini{\angalp} \sini{\angbet} \\ 
- \sini{\angbet} \cosi{\anggam} & 
\sini{\angbet} \sini{\anggam} & \cosi{\angbet} 
\end{bmatrix},
\label{eulermat}
\end{equation}
with $\cosi{\angrot} = \cos \angrot$ and $\sini{\angrot} = \sin \angrot$ for $\angrot = \angalp, \angbet, $ or $\anggam$. Here, $\angalp$\mynom[S]{$\angalp$}{Planet's precession angle} and $\angbet$\mynom[S]{$\angbet$}{Planet's nutation angle} denote the planet's precession and nutation angles, respectively, while $\anggam$\mynom[S]{$\anggam$}{Planet's intrinsic rotation angle} represents the intrinsic rotation angle. In Kaula's theory, the angles $\angalp$ and $\angbet$ evolve over time scales that are much longer than typical tidal oscillation periods \citep[e.g.,][]{EW2009}. Therefore, they are treated as constants in the formulation of the tidal problem. This is not the case of the intrinsic rotation angle,~$\anggam$, which describes the planet's spin, and explicitly varies with time, $\time$\mynom[S]{$\time$}{Time}, as
\begin{equation}
\label{anggam}
\anggam \left( \time \right) = \spinrate \time, 
\end{equation}
where $\spinrate$\mynom[S]{$\spinrate$}{Planet's spin angular velocity} is the planet's spin angular velocity. 

The torque exerted by a force $\forcevect$\mynom[S]{$\forcevect$}{Gravitational force per unit mass} about the origin point O on a planet occupying a volume $\domain$\mynom[S]{$\domain$}{Domain occupied by the planet} is expressed in spherical coordinates $\left( \rr , \col , \lon \right)$ as\mynom[S]{$\torquev$}{3D torque vector}
\begin{equation}
    \torquev = \integ{\left( \rvect \crossp \forcevect \right) \density }{\volume}{\domain}{},
\end{equation}
where $\infvar{\volume}$\mynom[S]{$\infvar{\volume}$}{Infinitesimal volume element} is an infinitesimal volume element, $\density$\mynom[S]{$\density$}{Density} is the local density, and $\crossp$\mynom[S]{$\crossp$}{Cross product} denotes the cross product. Following the convention of \cite{Zahn1966a}, the tidal force is expressed as 
\begin{equation}
    \forcevect \left( \rvect, \time \right) = \grad \Utide,
\end{equation}
where $\grad$\mynom[S]{$\grad$}{Gradient operator} designates the gradient operator (see \append{app:vectorial_operators}). The tide-raising potential is given by \citep[e.g.,][]{Kaula1964,EW2009},
\begin{equation}
\Utide \define \Ggrav \Mpert \left( \frac{1}{\abs{\rvect- \rpertvect}} - \frac{\rvect \dotp \rpertvect}{\rpert^3}  - \frac{1}{\rpert} \right).
\label{tidalpot1}
\end{equation}
with $\Ggrav=6.67430\times10^{-11} \ {\rm m^3 \ kg^{-1} \ s^{-2} }$\mynom[S]{$\Ggrav$}{Universal gravitational constant} being the universal gravitational constant \citep[][]{tiesinga2021codata}. 

Since tidal oscillation periods are typically much shorter than the characteristic evolution time scales of the orbital system, we focus exclusively on the time-averaged tidal effects on the planet and satellite's motions. We denote by $\timeav{\ffunc}$\mynom[S]{$\timeav{\ffunc}$}{Average of the function $\ffunc$ over time} the average of any function of time, $\ffunc$\mynom[S]{$\ffunc$}{Any function of time}, defined as
\begin{equation}
\label{timeav}
\timeav{\ffunc} \define \lim_{\period \rightarrow + \infty} \frac{1}{\period} \integ{\ffunc \left( \time \right)}{\time}{0}{\period}.
\end{equation}
The time-averaged tidal torque exerted about the three directions of the Galilean frame ($\framefix$) is thus given by
\begin{equation}
\label{torquev_def}
\torquev = \timeav{  \integ{ \left( \amoperator \Utide \right) \rhotide  }{\volume}{\domain}{} },
\end{equation}
where $\amoperator \define \rvect \crossp \grad$\mynom[S]{$\amoperator$}{Angular momentum operator} is the angular momentum operator \citep[][Sect.~2.2]{Varshalovich1988}, and $\rhotide$\mynom[S]{$\rhotide$}{Tidal density variation} is the local tidal density variation. We can express $\rhotide$ in terms of the gravitational potential of the tidally distorted body, $\Uresp$, using Poisson's equation \citep[e.g.,][Sect.~1.14]{Arfken2005},
\begin{equation}
\label{Poisson_eq}
\lap \Uresp = - 4 \pi \Ggrav \rhotide ,
\end{equation}
with $\lap$\mynom[S]{$\lap$}{Laplacian operator} denoting the Laplacian operator, which leads to
\begin{equation}
\label{torquev_potentials}
\torquev = - \frac{1}{4 \pi \Ggrav} \timeav{  \integ{ \left( \amoperator \Utide \right) \left( \lap \Uresp \right)  }{\volume}{\domain}{} }.
\end{equation}
It is important to note that both $\Utide$ and $\Uresp$ are expressed in the system of coordinates associated with the non-rotating frame of reference ($\framefix$), specifically $\left( \rr , \col ,\lon \right)$. 

\begin{figure}[t]
   \centering
   \includegraphics[width=0.48\textwidth,trim = 0.cm 0.cm 24cm 18.5cm,clip]{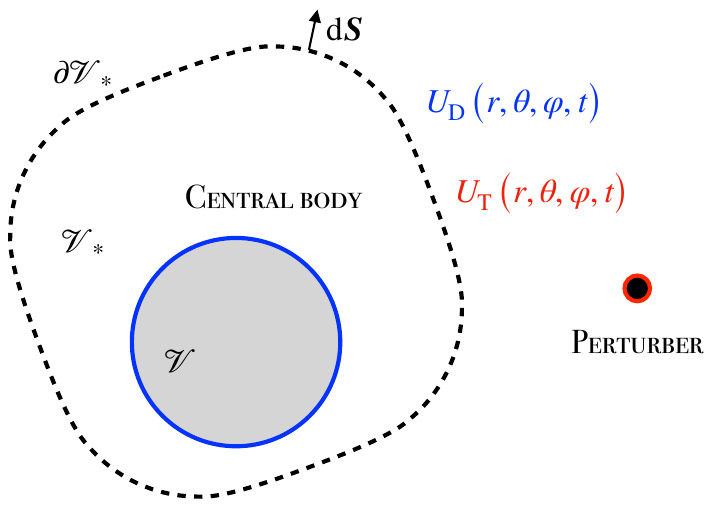}
      \caption{Diagram illustrating the domain $\domainstar$, which includes the central body while excluding the perturber. }
       \label{fig:control_volume}%
\end{figure}

\cite{Ogilvie2013} demonstrates that the integral in \eq{torquev_potentials} can, in fact, be carried out over any volume including the tidally distorted body. We therefore consider a simply connected domain,~$\domainstar$\mynom[S]{$\domainstar$}{Any simply connected domain including the planet and excluding the perturber}, that includes the planet but not the perturber, as illustrated by \fig{fig:control_volume}, and we introduce the vector $\vect{A} = \left( A_\xx, A_\yy, A_\zz \right)$ defined as
\begin{equation}
\label{aintegral}
 \vect{A}  = \integ{ \left( \amoperator \Utide \right) \left( \lap \Uresp  \right) }{\volume}{\domainstar}{}. 
\end{equation}
Since $\domainstar$ excludes the perturber, we have $\lap \Utide = 0$ in $\domainstar$. Furthermore, the angular momentum and Laplacian operator can be interchanged \rev{\citep[e.g.,][Sect.~2.2.2.]{Varshalovich1988}}, giving
\begin{equation}
\label{lap_amoperator}
 \lap \left( \amoperator \Utide \right)  = \amoperator \left( \lap \Utide \right) = 0. 
\end{equation}
As a result, the integral in \eq{aintegral} is rewritten as 
\begin{equation}
 \vect{A} = \integ{\left[ \amoperator \Utide \lap \Uresp - \Uresp \lap \left( \amoperator \Utide \right)   \right] }{\volume}{\domainstar}{}.
\end{equation}
Thus, by virtue of Green's second identity \citep[e.g.,][Sect.~1.11]{Arfken2005}, any component $A_\nuco$ of $\vect{A}$, where $\nuco = \xx, \yy$ or $\zz$, is expressed as 
\begin{equation}
A_\nuco = \ointeg{ \left[ \left( \amopi{\nuco} \Utide \right) \grad  \Uresp - \Uresp \grad \left( \amopi{\nuco} \Utide \right) \right] \cdot }{\surfacev}{\domainstarbd}{}.
\end{equation}
Here, $\infvar{\surfacev}$\mynom[S]{$\infvar{\surfacev}$}{Outward-pointing infinitesimal surface element vector} is an outward-pointing infinitesimal surface element vector, and $\domainstarbd$\mynom[S]{$\domainstarbd$}{Boundary of $\domainstar$} is the boundary of $\domainstar$. This allows the time-averaged torques exerted about the $\xx$, $\yy$ and $\zz$-axes of $\framefix$ to be written as\mynom[S]{$\torquex$}{Time-averaged torque exerted about the $\xx$-axis}\mynom[S]{$\torquey$}{Time-averaged torque exerted about the $\yy$-axis}\mynom[S]{$\torquez$}{Time-averaged torque exerted about the $\zz$-axis}
\begin{align}
\label{torquex}
\torquex & = - \frac{1}{4 \pi \Ggrav} \timeav{ \ointeg{ \! \! \left[ \left( \amopi{\xx} \Utide \right) \grad  \Uresp - \Uresp \grad \left( \amopi{\xx} \Utide \right) \right] \cdot }{\surfacev}{\domainstarbd}{}}, \\
\torquey & = - \frac{1}{4 \pi \Ggrav} \timeav{ \ointeg{ \! \!  \left[ \left( \amopi{\yy} \Utide \right) \grad  \Uresp - \Uresp \grad \left( \amopi{\yy} \Utide \right) \right] \cdot }{\surfacev}{\domainstarbd}{}}, \\
\label{torquez}
\torquez & = - \frac{1}{4 \pi \Ggrav} \timeav{ \ointeg{ \! \!  \left[ \left( \amopi{\zz} \Utide \right) \grad  \Uresp - \Uresp \grad \left( \amopi{\zz} \Utide \right) \right] \cdot }{\surfacev}{\domainstarbd}{}}.
\end{align}

Finally, the rate of energy transferred from the perturber's orbital motion to the planet's rotation is the time-averaged tidal power, $\powertide$\mynom[S]{$\powertide$}{Time-averaged tidal power}, which we express as a function of both the forcing and deformation tidal potentials too \citep[e.g.,][]{Auclair2023}, 
\begin{equation}
\powertide = - \frac{1}{4 \pi \Ggrav} \timeav{\integ{\Utide \lap \left( \dd{\Uresp}{\time} \right)  }{\volume}{\domainstar}{} }.
\end{equation}
By invoking Green's second identity once again, we rewrite this expression as 
\begin{equation}
\label{powertide}
\powertide = - \frac{1}{4 \pi \Ggrav} \timeav{ \ointeg{\left[ \Utide \grad \left( \dd{\Uresp}{\time} \right) - \left( \dd{\Uresp}{\time} \right) \grad \Utide  \right] \cdot}{\surfacev}{\domainstarbd}{} }.
\end{equation}
The time-averaged tidally dissipated power, $\powerdiss$\mynom[S]{$\powerdiss$}{Time-averaged tidally dissipated power}, can be readily deduced from $\torquev$ and $\powertide$, as demonstrated by \cite{Ogilvie2013} and discussed in \sect{ssec:dynamical_eqs}. Since \eqsto{torquex}{torquez} and \eq{powertide} provide the rates of momentum and energy exchanges as functions of the tide-raising and deformation gravitational potentials, $\Utide$ and~$\Uresp$, we shall specify these two potentials in the following.

\subsection{Tidal gravitational potential of the distorted planet}
\label{ssec:gravpot_distorted}

In linear theory, the tidal forcing and deformation potentials in \eqsto{torquex}{torquez} and \eq{powertide} are both series of periodic functions of time since $\Utide$ is composed of oscillatory components. Consequently, it is appropriate to transition from the temporal to the frequency domain by expanding both potentials in Fourier series. These manipulations yield
\begin{align}
\label{Utide}
\Utide \left(  \rr, \col, \lon , \time \right)  & =  \Re \left\{ \sum_{\kk= 0}^{+ \infty}  \Utidensig \left( \rr, \col, \lon  \right) \expo{ \inumber \ftidefixi{\kk} \time} \right\} , \\
\label{Uresp}
\Uresp \left(  \rr, \col, \lon , \time \right)   & =  \Re \left\{\sum_{\kk=0}^{+\infty} \sum_{ \qq , \pp = - \infty}^{+ \infty}  \Urespnsig \left( \rr, \col, \lon  \right) \expo{ \inumber \ftidefixi{\kk,\qq,\pp} \time} \right\} ,
\end{align}
where $\Re$\mynom[S]{$\Re$}{Real part of a complex number} denotes the real part of a complex number (with $\Im$\mynom[S]{$\Im$}{Imaginary part of a complex number} referring to the imaginary part), $\inumber$\mynom[S]{$\inumber$}{Imaginary unit} represents the imaginary unit ($\inumber^2 = -1$)\mynom[S]{$\kk,\qq,\pp$}{Integers}, $\ftidefixi{\kk}$\mynom[S]{$\ftidefixi{\kk}$}{$\kk$-th forcing tidal frequency in $\framefix$} is the $\kk$-th forcing tidal frequency, $\ftidefixi{\kk,\qq,\pp}$\mynom[S]{$\ftidefixi{\kk,\qq,\pp}$}{Frequency of the tidal response defined by the triplet $\left( \kk , \qq , \pp \right)$ in $ \framefix$} is the frequency of the tidal response defined by the triplet $\left( \kk , \qq , \pp \right)$,
\begin{equation}
\ftidefixi{\kk,\qq,\pp} \define \ftidefixi{\kk} + \left(\qq- \pp   \right) \spinrate,
\end{equation}
and $\Utidensig$\mynom[S]{$\Utidensig$}{Complex spatial distribution of the $\kk$-th component of the tide-raising potential} and $\Urespnsig$\mynom[S]{$\Urespnsig$}{Complex spatial distribution of the $\left(\kk,\qq,\pp \right)$-component of the perturbed potential} designate the corresponding complex distributions for each potential. The frequencies $\ftidefixi{\kk}$ are expressed as 
\begin{equation}
\ftidefixi{\kk} = \kk \norb,
\end{equation}
where $\norb$ is the anomalistic mean motion of the perturber in the Galilean frame, and the integer $\kk $ ranges from $0$ and $+ \infty$\footnote{The summation in \eq{Utide} may also be defined with $\kk$ running from $-\infty$ to $+ \infty$ \citep[see e.g.][Eq. (3)]{Ogilvie2014}. In that case, the spherical harmonic orders in \eqs{Utidensig_sph}{Urespnsig_sph} are positive, while they range from $-\llat$ to $\llat$ in the present work.}. It is noteworthy that the summations over $\qq$ and $\pp$ in \eqs{Utide}{Uresp} result from the forward and backward rotations between the Galilean and rotating frames of reference, as will be further discussed. 

Outside the planet, the coefficients $\Utidensig$ and $\Urespnsig$ are expressed as series of spherical harmonics $\Ylm{\llat}{\mm}$ of degree $\llat$ and order $\mm$, as follows:
\begin{align}
\label{Utidensig_sph}
\Utidensig \left( \rr, \col, \lon \right) = & \sum_{\llat=2}^{+ \infty} \sum_{\mm = -\llat}^{\llat}  \Ulmsigj{\llat}{\kk}{\mm} \left( \frac{\rr}{\Rpla} \right)^{\llat} \Ylm{\llat}{\mm} \left( \col , \lon \right) , \\
\label{Urespnsig_sph}
\Urespnsig \left( \rr , \col, \lon \right) = & \sum_{\llat=\abs{\pp}}^{+ \infty} \sum_{\mm = - \llat}^{\llat}  \Uplmsigj{\llat}{\kk,\qq,\pp}{\mm} \left( \frac{\rr}{\Rpla} \right)^{-\left( \llat + 1 \right)} \Ylm{\llat}{\mm} \left( \col , \lon \right) ,
\end{align}
where $\Ulmsigj{\llat}{\kk}{\mm} $\mynom[S]{$\Ulmsigj{\llat}{\kk}{\mm}$}{Complex weighting coefficients of the spherical harmonic expansion of the $\kk$-th component of the tide-raising potential} and $ \Uplmsigj{\llat}{\kk,\qq,\pp}{\mm}$\mynom[S]{$\Uplmsigj{\llat}{\kk,\qq,\pp}{\mm}$}{Complex weighting coefficients of the spherical harmonic expansion of the $\left( \kk, \qq,\pp\right)$-component of the perturbed potential} are complex weighting coefficients associated with the triplet $\left( \kk,  \llat , \mm \right)$ and the quintuplet $\left( \kk,  \llat , \mm, \pp , \qq \right)$, respectively. The spherical harmonics $\Ylm{\llat}{\mm}$ in \eqs{Utidensig_sph}{Urespnsig_sph} are explicitly defined in \append{app:sph}. 

The gravitational potential induced by the planet's tidal response is usually obtained in the rotating frame of reference~($\framepla$), where the equations governing the planet's tidal response are straightforwardly formulated. This requires expressing the tidal components given by \eqs{Utidensig_sph}{Urespnsig_sph} as functions of the coordinates associated with $\framepla$, namely $\left( \colpla, \lonpla \right)$. This step is easily completed by considering the rotation operator of spherical harmonics: the spherical harmonics sharing the same degree in the Galilean and rotated systems of coordinates are linked through the relations \citep[][Sect.~4.1]{Varshalovich1988}
\begin{align}
\label{Ylmpla_Ylmfix}
\Ylm{\llat}{\mm} \left( \colpla , \lonpla \right) & = \sum_{\qq = - \llat}^\llat \wignerD{\qq}{\mm}{\llat} \left( \angalp , \angbet , \anggam \right) \Ylm{\llat}{\qq} \left( \col , \lon \right), \\
\label{Ylmfix_Ylmpla}
\Ylm{\llat}{\mm} \left( \col , \lon \right) &= \sum_{\qq = - \llat}^{\llat} \conj{\wignerD{\mm}{\qq}{\llat}} \left( \angalp , \angbet , \anggam \right) \Ylm{\llat}{\qq} \left( \colpla , \lonpla \right).
\end{align}
In the above equations, the notation $\wignerD{\qq}{\mm}{\llat}$\mynom[S]{$\wignerD{\qq}{\mm}{\llat}$}{Wigner D-functions} refers to the Wigner D-functions  \citep[][Sect.~4.3, Eq. (1)]{Varshalovich1988}, which are detailed in \append{app:wigner_dfunctions}, and $\conj{\wignerD{\mm}{\qq}{\llat}}$ denotes the complex conjugate of $\wignerD{\mm}{\qq}{\llat}$\mynom[S]{$\conj{Z}$}{Conjugate of the complex number $Z$}. Besides, we recall that $\anggam$ is a function of time, as defined in \eq{anggam}.

Using \eq{Ylmpla_Ylmfix} in \eq{Utidensig_sph}, we express the tide-raising potential in the coordinates of the rotating frame of reference, 
\begin{equation}
\label{Utide_rot}
\Utide \left( \rrpla, \colpla, \lonpla, \time \right) = \Re \left\{ \sum_{\kk,\qq } \sum_{\llat = \max \left( \abs{\qq},2\right)}^{+ \infty} \! \! \! \Ulmrotsigj{\llat}{\kk}{\qq} \left( \frac{\rrpla}{\Rpla} \right)^\llat \Ylm{\llat}{\qq} \left( \colpla , \lonpla \right) \expo{\inumber \ftideplai{\kk,\qq} \time} \right\}, 
\end{equation}
where the integers $\kk$ runs from $0$ to $+\infty$ and $\qq$ from $-\infty$ to $+\infty$. In this expression, the tidal frequencies $\ftideplai{\kk,\qq}$ are defined as 
\begin{equation}
\ftideplai{\kk, \qq} \define \qq \spinrate +\ftidefixi{\kk},
\end{equation}
and the complex weighting coefficients $\Ulmrotsigj{\llat}{\qq}{\kk}$\mynom[S]{$\Ulmrotsigj{\llat}{\qq}{\kk}$}{Complex spherical harmonic coefficient of the component of the tide-raising potential associated with the frequency $\ftideplai{\kk,\qq}$ in $\framepla$} as
\begin{equation}
\label{Ulmrot_Ulm}
\Ulmrotsigj{\llat}{\kk}{\qq} \left( \angalp ,\angbet \right) = \sum_{\mm = - \llat}^{\llat} \Ulmsigj{\llat}{\kk}{\mm} \expo{\inumber \mm \angalp} \wignerd{\mm}{\qq}{\llat} \left( \angbet \right). 
\end{equation}
The component of the forcing associated with the triplet $\left( \llat, \qq , \kk \right)$ generates a tidal gravitational potential oscillating with the same frequency, but not necessarily with the same spatial distribution. The spatially dependent factor of this potential is expressed in general as\mynom[S]{$\Uplmrotsigj{\llat}{\kk}{\qq}$}{Spatial distribution function of the component of the perturbed potential associated with the triplet $\left( \llat, \qq , \kk \right)$} 
\begin{equation}
\Uplmrotsigj{\llat}{\kk}{\qq} \left(\rrpla, \colpla , \lonpla \right) =  \sum_{\slon = 0}^{+ \infty} \sum_{\pp = -\slon}^\slon \tidaltfi{\llat}{\slon}{\kk,\qq,\pp} \Ulmrotsigj{\llat}{\kk}{\qq} \left( \frac{\rrpla}{\Rpla} \right)^{- \left( \slon + 1 \right)} \Ylm{\slon}{\pp} \left( \colpla , \lonpla \right) ,
\end{equation} 
with $\tidaltfi{\llat}{\slon}{\kk,\qq,\pp} $ being the transfer function associated with the harmonic of degree $\slon$ and order $\pp$. It is noteworthy that $\tidaltfi{\llat}{\slon}{\kk,\qq,\pp} $\mynom[S]{$\tidaltfi{\llat}{\slon}{\kk,\qq,\pp}$}{Transfer function relating the degree-$\slon$, order-$\pp$ harmonic of the $\kk$-th component of the perturbed potential to the degree-$\llat$, order-$\qq$ harmonic of the same component of the tide-raising potential} stands for a generalised Love number accounting for the fact that the spatial dependence of the tidal response generally differs from that of the tide-raising potential \rev{because of rotational scattering}. For spherically isotropic bodies, this transfer function simplifies to $\tidaltfi{\llat}{\slon}{\kk,\qq,\pp} = \tidaltfi{\llat}{\llat}{\kk,\qq,\qq} \kron{\llat}{\slon} \kron{\qq}{\pp}$,~where $\kron{\qq}{\pp}$\mynom[S]{$\kron{\qq}{\pp}$}{Kronecker delta function, such that $\kron{\qq}{\pp} = 1$ if $\qq = \pp$ and $\kron{\qq}{\pp} = 0$ otherwise} designates the Kronecker delta function, such that $\kron{\qq}{\pp} = 1$ if $\qq = \pp$ and $\kron{\qq}{\pp} = 0$ otherwise. 

In the coordinates of the rotating frame, $\framepla$, the gravitational potential of the tidally distorted body can be expressed as 
\begin{equation}
\Uresp \left( \rrpla, \colpla, \lonpla , \time \right) = \Re \left\{ \sum_{\kk=0}^{+ \infty} \sum_{\qq=-\infty}^{+ \infty} \Urespnrotsigi{\kk,\qq} \left( \rrpla, \colpla, \lonpla \right)  \expo{\inumber \ftideplai{\kk,\qq} \time}  \right\},
\end{equation}
where the spatial distributions associated with the frequencies $\ftideplai{\kk,\qq}$ are given by\mynom[S]{$\Urespnrotsigi{\kk,\qq}$}{Spatial distribution function of the component of the perturbed tidal potential associated with the frequency $\ftideplai{\kk,\qq}$ in $\framepla$}
\begin{equation}
\label{Urespnrotsigi}
\Urespnrotsigi{\kk,\qq}  \left( \rrpla, \colpla, \lonpla \right) = \sum_{\slon=0}^{+ \infty} \sum_{\pp = - \slon}^\slon \Uplmrotsigj{\slon}{\kk,\qq}{\pp} \left(  \frac{\rrpla}{\Rpla} \right)^{-\left( \slon + 1 \right)} \Ylm{\slon}{\pp} \left( \colpla, \lonpla \right)  .
\end{equation}
In this expression, the complex weighting coefficients $\Uplmrotsigj{\slon}{\kk,\qq}{\pp}$\mynom[S]{$\Uplmrotsigj{\slon}{\kk,\qq}{\pp}$}{Complex spherical harmonic coefficients of the component of the perturbed potential associated with the frequency $\ftideplai{\kk,\qq}$ in $\framepla$} are defined as 
\begin{equation}
\Uplmrotsigj{\slon}{\kk,\qq}{\pp} \left( \angalp , \angbet \right) = \sum_{\llat=\max \left( \abs{\qq},2 \right)}^{+ \infty} \tidaltfi{\llat}{\slon}{\kk,\qq,\pp} \Ulmrotsigj{\llat}{\kk}{\qq}.
\label{Uplmrotsig}
\end{equation}
By changing the coordinates from $\left(\rrpla, \colpla , \lonpla \right) $ to $ \left( \rr, \col,\lon\right)$, we obtain the gravitational potential components $\Uplmsigj{\llat}{\kk, \qq, \pp}{\mm}  $, introduced in \eq{Urespnsig_sph}, first as functions of the $\Uplmrotsigj{\slon}{\kk,\qq}{\pp}$ coefficients,
\begin{equation}
\Uplmsigj{\llat}{\kk, \qq, \pp}{\mm} \left( \angalp, \angbet \right) = \Uplmrotsigj{\llat}{\kk,\qq}{\pp} \expo{- \inumber \mm \angalp} \wignerd{\mm}{\pp}{\llat} \left( \angbet \right) , 
\end{equation}
and second, using \eq{Uplmrotsig}, in terms of the $\Ulmrotsigj{\slon}{\kk}{\qq}$,
\begin{equation}
\Uplmsigj{\llat}{\kk, \qq, \pp}{\mm}  \left( \angalp , \angbet \right) = \! \! \sum_{\slon = \max \left( \abs{\qq} , 2 \right) }^{+ \infty} \tidaltfi{\slon}{\llat}{\kk,\qq,\pp} \expo{- \inumber \mm \angalp} \wignerd{\mm}{\pp}{\llat} \left( \angbet \right)  \Ulmrotsigj{\slon}{\kk}{\qq} .
\end{equation}
Finally, substituting \eq{Ulmrot_Ulm} into the above equation, we rewrite $\Uplmsigj{\llat}{\kk, \qq, \pp}{\mm} $ as a function of the weighting coefficients of the forcing tidal potential in the inertial frame $\framefix$,
\begin{equation}
\label{Udlkqpm}
\Uplmsigj{\llat}{\kk, \qq, \pp}{\mm}  \left( \angalp , \angbet \right) = \! \! \! \! \! \!  \sum_{\slon = \max \left( \abs{\qq},2 \right)}^{+ \infty} \sum_{\jj = -\slon}^\slon \tidaltfi{\slon}{\llat}{\kk,\qq,\pp} \Ulmsigj{\slon}{\kk}{\jj}  \expo{\inumber \left( \jj - \mm \right) \angalp} \wignerd{\mm}{\pp}{\llat} \left( \angbet \right) \wignerd{\jj}{\qq}{\slon} \left( \angbet \right) .
\end{equation}
This formulation of the deformation potential accounts for the possible coupling of spherical modes in the tidal response. When the central body is assumed to be spherically isotropic, no such coupling occurs. In that case, each component of the response corresponds to the same spherical harmonic as the forcing component generating it. Therefore, the summation over $\slon$ in \eq{Udlkqpm} vanishes and $\Uplmsigj{\llat}{\kk, \qq, \pp}{\mm}$ simplifies to 
\begin{equation}
\Uplmsigj{\llat}{\kk, \qq, \pp}{\mm} \left( \angalp , \angbet \right) = \sum_{\jj = -\llat}^\llat \tidaltfi{\llat}{\llat}{\kk,\qq,\qq} \Ulmsigj{\llat}{\kk}{\jj}  \expo{\inumber \left( \jj - \mm \right) \angalp} \wignerd{\mm}{\qq}{\llat} \left( \angbet \right) \wignerd{\jj}{\qq}{\llat} \left( \angbet \right) \kron{\qq}{\pp}.
\end{equation}

\subsection{Components of the tidal torque}

As a final step, we express the tidal torque and tidally dissipated power, derived in \eqsto{torquex}{torquez} and \eq{powertide}, in terms of the complex coefficients from the multipole expansions of the forcing and deformation tidal potential. These are the coefficients $\Ulmsigj{\llat}{\kk}{\mm}$ and $ \Uplmsigj{\llat}{\kk,\qq,\pp}{\mm}$ introduced in \eqs{Utidensig_sph}{Urespnsig_sph}, respectively. This step is not straightforward, as the angular momentum operator introduces significant mathematical complexities. However, these difficulties can be elegantly addressed using angular momentum theory, as demonstrated by \cite{Boue2017}. \rev{In this approach, inspired from quantum mechanics,} the problem is simplified by changing to a system of coordinates where the action of the angular momentum operator on spherical harmonics is easily formulated. The terms involving the angular momentum operator in \eqsto{torquex}{torquez} and \eq{powertide} are then directly obtained by converting back to the Cartesian system, $\left( \xx , \yy , \zz \right)$.

 Following \cite{Boue2017}, we introduce the set of complex unit vectors $\left( \eplus, \ezero, \eminus \right)$\mynom[S]{$\eplus,\ezero,\eminus$}{Set of complex unit vectors}, where the coordinates of any vector $\vect{a}$ are represented as $\left( a_+, a_0,a_- \right)$\mynom[S]{$a_+,a_0,a_-$}{Coordinates of any vector $\vect{a}$ in $\left( \eplus,\ezero,\eminus \right)$}. These are related to the Cartesian coordinates $\left( a_\xx , a_\yy, a_\zz \right)$ in the reference frame $\framefix$ by
\begin{equation}
\begin{array}{lll}
\displaystyle a_+ = - \frac{1}{\sqrt{2}} \left( a_\xx + \inumber a_\yy \right), &\displaystyle a_0 = a_\zz, & \displaystyle a_- = \frac{1}{\sqrt{2}} \left( a_\xx - \inumber a_\yy \right). 
\end{array}
\label{change_coord}
\end{equation}
For $\nuco \in \left\{ -, 0 , + \right\} $ (i.e. $\nu = -1,0,1$, respectively), the $\nuco$-component $\amopi{\nuco}$ of the angular momentum operator applied to a spherical harmonic $\Ylm{\llat}{\mm}$ is expressed as 
\begin{equation}
\amopi{\nuco} \Ylm{\llat}{\mm} = \inumber \amopci{\nuco}{\llat}{\mm} \Ylm{\llat}{\mm+\nuco},
\end{equation}
where the coefficients $\amopci{\nuco}{\llat}{\mm} $\mynom[S]{$\amopci{\nuco}{\llat}{\mm} $}{Coefficients of the angular momentum operator in spherical harmonics} are real and defined as \citep[][]{Varshalovich1988}
\begin{equation}
\amopci{\nuco}{\llat}{\mm} \define \left\{
\begin{array}{ll}
\mm & \mbox{if} \ \nuco =0 , \\
\displaystyle - \nuco \sqrt{\frac{\llat \left( \llat + 1 \right) - \mm \left( \mm+\nuco \right)}{2}} &  \mbox{if} \  \nuco = \pm 1. 
\end{array}
\right.
\end{equation}

The forcing and deformation tidal potentials, $\Utide$ and $\Uresp$, are written in terms of the complex potentials, $\Utiden$ and $\Urespn$, as follows:
\begin{align}
& \Utide  = \Re \left\{ \Utiden \right\},  & \Uresp = \Re \left\{ \Urespn \right\}. 
\end{align}
Since the real part and angular momentum operator can be interchanged, we have
\begin{equation} 
\amoperator \Re \left\{ \Utiden \right\}  = \Re \left\{ \amoperator \Utiden \right\}.
\end{equation}
Using the change of coordinates introduced in \eq{change_coord}, we obtain\mynom[S]{$\amopi{\xx}$}{$\xx$-component of the angular momentum operator}\mynom[S]{$\amopi{\yy}$}{$\yy$-component of the angular momentum operator}\mynom[S]{$\amopi{\zz}$}{$\zz$-component of the angular momentum operator} 
\begin{align}
\amopi{\xx} \Re \left\{ \Utiden \right\} & = - \frac{1}{\sqrt{2}} \Re \left\{ \sum_{\nuco=\pm 1} \nuco \amopi{\nuco} \Utiden  \right\}, \\
\amopi{\yy} \Re \left\{ \Utiden \right\} & =  \frac{1}{\sqrt{2}} \Re \left\{ \inumber \sum_{\nuco=\pm 1}  \amopi{\nuco} \Utiden  \right\}, \\
\amopi{\zz} \Re \left\{ \Utiden \right\} & = \Re \left\{ \amopi{0} \Utiden \right\}.
\end{align}
To express the torques defined in \eqsto{torquex}{torquez} in terms of the Fourier components of $\Utiden$ and $\Urespn$, one must compute the conjugates of $\amopi{\xx} \Utiden$, $\amopi{\yy} \Utiden$, and $\amopi{\zz} \Utiden$, as well as the radial component of their gradients and of the gradient of $\Urespn$. The quantity $\amopi{\nuco} \Utiden$ is given by 
\begin{equation}
\amopi{\nuco} \Utiden = \sum_{\kk = 0}^{+ \infty} \sum_{\llat = 2}^{+ \infty} \sum_{\mm = - \llat}^{\llat} \inumber \amopci{\nuco}{\llat}{\mm} \Ulmsigj{\llat}{\kk}{\mm} \left( \frac{\rr}{\Rpla} \right)^\llat \Ylm{\llat}{\mm+\nuco} \left( \col, \lon \right) \expo{\inumber \ftidefixi{\kk} \time}.
\end{equation}
This results in 
\begin{align}
\conj{\amopi{\xx} \Utiden} & = \frac{\inumber}{\sqrt{2}} \sum_{\kk,\nuco} \sum_{\llat=2}^{+ \infty} \sum_{\mm = - \llat}^\llat \nuco \amopci{\nuco}{\llat}{\mm} \conj{\Ulmsigj{\llat}{\kk}{\mm}} \left( \frac{\rr}{\Rpla} \right)^\llat \conj{\Ylm{\llat}{\mm+\nuco}} \left( \col , \lon \right) \expo{-\inumber \ftidefixi{\kk} \time}, \\
\conj{\amopi{\yy} \Utiden} & = -\frac{1}{\sqrt{2}} \sum_{\kk,\nuco} \sum_{\llat=2}^{+ \infty} \sum_{\mm = - \llat}^\llat  \amopci{\nuco}{\llat}{\mm} \conj{\Ulmsigj{\llat}{\kk}{\mm}} \left( \frac{\rr}{\Rpla} \right)^\llat \conj{\Ylm{\llat}{\mm+\nuco}} \left( \col , \lon \right) \expo{-\inumber \ftidefixi{\kk} \time}, \\
\conj{\amopi{\zz} \Utiden} & = - \inumber \sum_{\kk} \sum_{\llat=2}^{+ \infty} \sum_{\mm = - \llat}^\llat  \mm \conj{\Ulmsigj{\llat}{\kk}{\mm}} \left( \frac{\rr}{\Rpla} \right)^\llat \conj{\Ylm{\llat}{\mm}} \left( \col , \lon \right) \expo{-\inumber \ftidefixi{\kk} \time},
\end{align}
with $\kk$ ranging from $0$ to $+ \infty$ and $\nuco = \pm 1$. 

Similarly, the radial component of the gradient of $\Urespn$ is given by 
\begin{equation}
\dd{\Urespn}{\rr} = -   \sum_{\kk,\qq,\pp} \sum_{\llat = \abs{\pp}}^{+ \infty} \sum_{\mm= - \llat}^\llat \frac{\llat + 1}{\rr} \Uplmsigj{\llat}{\kk, \qq, \pp}{\mm}  \left( \frac{\rr}{\Rpla} \right)^{-\left( \llat + 1\right)}   \! \! \Ylm{\llat}{\mm} \left( \col , \lon \right) \expo{\inumber \ftidefixi{\kk,\qq,\pp} \time},
\end{equation}
where $\kk$ runs from $0$ to $+\infty$, and $\qq$ and $\pp$ from $-\infty$ to $+ \infty$. However, only the components associated with the frequencies $\ftidefixi{\kk}$ of the forcing tidal potential contribute to the tidal torque. These correspond to terms where $\pp = \qq$. Thus, the gradient of the tidal potential is rewritten as 
\begin{align}
\dd{\Urespn}{\rr} = & - \sum_{\kk} \sum_{\llat = 0}^{+ \infty} \sum_{\mm= - \llat}^\llat \frac{ \llat + 1 }{\rr} \left(\sum_{\qq= - \llat}^\llat  \Uplmsigj{\llat}{\kk, \qq, \qq}{\mm}  \right) \left( \frac{\rr}{\Rpla} \right)^{-\left( \llat + 1\right)} \! \! \! \! \! \!  \Ylm{\llat}{\mm} \left( \col , \lon \right) \expo{ \inumber \ftidefixi{\kk} \time} \nonumber \\
& + \dd{\mathcal{U}}{\rr},
\end{align}
where $\mathcal{U}$\mynom[S]{$\mathcal{U}$}{Component of $\Urespn$ that does not generate any torque in average} represents the component of $\Urespn$ that does not generate any torque in average. 

The time-averaged components of the torque can be straightforwardly deduced from the preceding derivations. In essence, we consider the function $\ffunc$ defined as 
\begin{equation}
\label{func_power}
\ffunc \left( \time \right) = \integ{A \left(\rvect, \time \right) B  \left(\rvect, \time \right)}{\volume}{\domain}{},
\end{equation}
where $\domain$ denotes the spatial domain over which the integral is evaluated, $\infvar{\volume}$ represents an infinitesimal volume element, and $A$ and $B$\mynom[S]{$A,B$}{Two real functions of spatial coordinates and time} are two functions expressed as
\begin{align}
    & A \left( \rvect , \time \right)  = \Re \left\{ \tilde{A} \left(\rr , \col , \lon , \time \right) \right\},  \\
    & B \left(\rvect , \time \right) = \Re \left\{ \tilde{B} \left( \rr, \col , \lon , \time \right) \right\}.
\end{align}
In the above equations, the complex functions $\tilde{A}$ and $\tilde{B}$\mynom[S]{$\tilde{A},\tilde{B}$}{Two complex-valued time-oscillating functions} are sinusoidal time-oscillating functions of positive frequencies $\ftide_1$ and $\ftide_2$\mynom[S]{$\ftide_1,\ftide_2$}{Frequencies associated with $\tilde{A}$ and $\tilde{B}$, respectively}, respectively, multiplied by complex spatial distributions\mynom[S]{$\tilde{A}_0,\tilde{B}_0$}{Complex spatial components of $\tilde{A}$ and $\tilde{B}$, respectively},
\begin{align}
    & \tilde{A} \left(\rr, \col , \lon, \time \right)  = \tilde{A}_0 \left(\rr, \col , \lon \right) \expo{\inumber \ftide_1 \time},  \\
    & \tilde{B}  \left(\rr, \col , \lon, \time \right) = \tilde{B}_0 \left(\rr,  \col , \lon \right) \expo{\inumber \ftide_2 \time}.
\end{align}

As demonstrated in \append{app:time_averaged_power}, the time-averaged value of~$\ffunc$, defined by \eq{timeav}, can be expressed in terms of the complex spatial distributions $ \tilde{A}_0$ and $ \tilde{B}_0$ as
\begin{equation}
\label{timeav_formula}
\timeav{\ffunc}  = 
\left\{
\begin{array}{ll}
 \displaystyle \Re \left\{ \frac{1}{2}  \integ{ \conj{\tilde{A}_0}  \tilde{B}_0}{\volume}{\domain}{} \right\} & {\rm if} \ \ftide_1 = \ftide_2, \\
 0 & {\rm if} \ \ftide_1 \neq \ftide_2.
\end{array}
\right.
\end{equation}
By applying this identity to the components of the tidal torque (\eqsto{torquex}{torquez}) and tidal power (\eq{powertide}), and following the steps detailed in \append{app:torque_power}, we arrive at 
\begin{align}
\label{torquex_sum}
\torquex & = - \frac{\Ktorque}{\sqrt{2}} \Im  \left\{ \sum_{\kk,\nuco} \sum_{\llat=2}^{+\infty} \sum_{\mm = -\llat}^\llat  \nuco \left( 2 \llat + 1 \right) \amopci{\nuco}{\llat}{\mm} \conj{\Ulmsigj{\llat}{\kk}{\mm} }  \left( \sum_{\qq=-\llat}^{\llat} \Uplmsigj{\llat}{\kk, \qq, \qq}{\mm+\nuco}   \right)   \right\}  , \\
\label{torquey_sum}
\torquey & = - \frac{\Ktorque}{\sqrt{2}} \Re  \left\{ \sum_{\kk,\nuco} \sum_{\llat=2}^{+\infty} \sum_{\mm = -\llat}^\llat   \left( 2 \llat + 1 \right) \amopci{\nuco}{\llat}{\mm} \conj{\Ulmsigj{\llat}{\kk}{\mm} }  \left( \sum_{\qq=-\llat}^{\llat} \Uplmsigj{\llat}{\kk, \qq, \qq}{\mm+\nuco}   \right)   \right\}   , \\
\label{torquez_sum}
\torquez & =  \Ktorque \Im \left\{  \sum_{\kk} \sum_{\llat=2}^{+\infty} \sum_{\mm = -\llat}^\llat \left( 2 \llat  + 1 \right) \mm \conj{\Ulmsigj{\llat}{\kk}{\mm} } \left( \sum_{\qq=-\llat}^{\llat} \Uplmsigj{\llat}{\kk, \qq, \qq}{\mm}  \right)  \right\}, \\
\label{powertide_sum}
\powertide & = - \Ktorque \Im \left\{ \sum_{\kk} \sum_{\llat=2}^{+ \infty} \sum_{\mm = - \llat}^{\llat} \ftidefixi{\kk} \left( 2 \llat + 1 \right) \conj{\Ulmsigj{\llat}{\kk}{\mm} } \left( \sum_{\qq=-\llat}^{\llat} \Uplmsigj{\llat}{\kk, \qq, \qq}{\mm}  \right)   \right\},
\end{align}
where $\kk$ runs from $0$ to $+ \infty$, $\nuco = \pm1$, and $\Ktorque$\mynom[S]{$\Ktorque$}{Constant coefficient} is a constant defined as 
\begin{equation}
\Ktorque = \frac{\Rpla}{8 \pi \Ggrav}.
\end{equation}

The formulae given by \eqsto{torquex_sum}{powertide_sum} enable the calculation of the average rates of angular momentum and energy transfer in a general setting. Notably, these formulae remain valid even if the system does not conform to the spherical isotropy assumptions that typically prevail in the linear theory of bodily tides \citep[see e.g.][Sect.~4]{EW2009} or in the equilibrium tide model. Specifically, (i) the energy dissipation rate of one component is not solely dependent on the associated tidal frequency within this framework, and (ii) the deformation of the planet's shape may differ spatially from the tide-generating potential, though it remains linearly related to it. 

\begin{figure}[t]
   \centering
   \includegraphics[width=0.48\textwidth,trim = 0.cm 0.cm 19.8cm 13.5cm,clip]{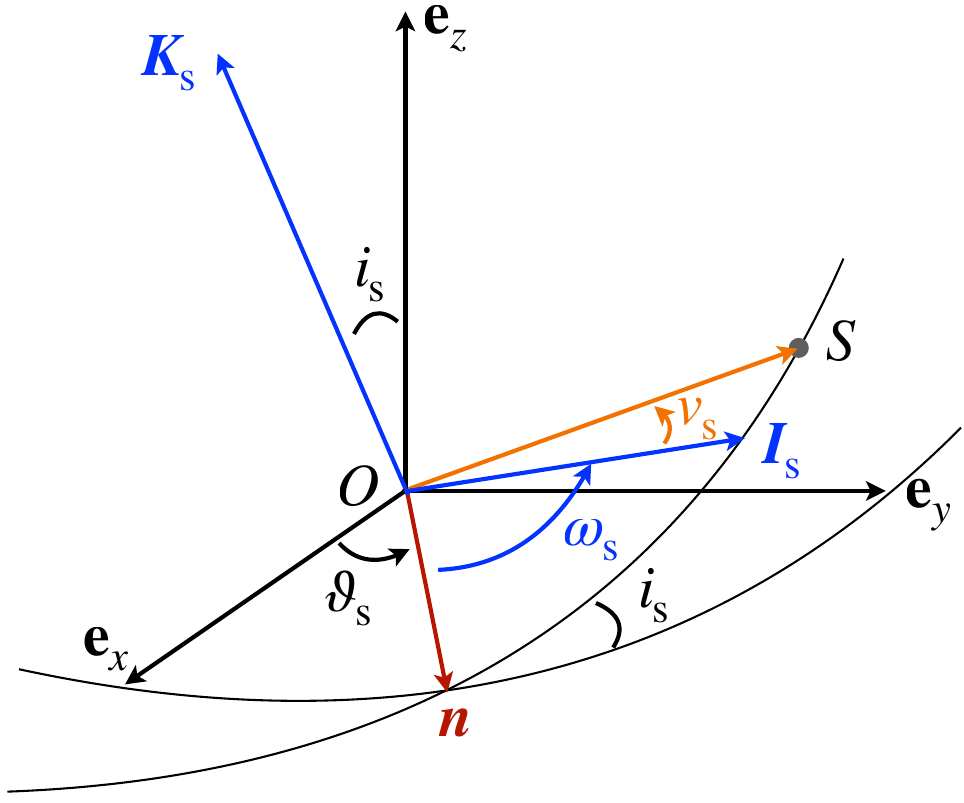}
      \caption{Frame of reference $\rframe{\ipert}{O}{\Ivect}{\Jvect}{\Kvect}$ and Keplerian elements describing the orbit of the perturber.}
       \label{fig:keplerian_elements}%
\end{figure}

\section{Dynamics of a planet-satellite system}
\label{sec:dynamics_system}

\subsection{Tide-raising gravitational potential}

The tidal torque and power, as expressed in \eqsto{torquex_sum}{torquez_sum} and \eq{powertide_sum}, determine the long-term evolution of planetary systems. To demonstrate how they specifically influence the spin and the orbital parameters of celestial bodies, we establish in this section the equations governing the tidal evolution of the Keplerian elements of a planet-perturber system. This is done using the formalism and method outlined in \cite{BE2019}.

We consider a two-body system in which a point-mass satellite orbits a planet that is not spherically isotropic. The satellite's orbit is characterised by the standard Keplerian elements, as illustrated by \fig{fig:keplerian_elements}. We denote the semi-major axis by $\smaxispert$\mynom[S]{$\smaxispert$}{Satellite's semi-major axis}, the eccentricity by $\eccpert$\mynom[S]{$\eccpert$}{Satellite's eccentricity}, the inclination by $\incpert$\mynom[S]{$\incpert$}{Satellite's orbital inclination}, the true anomaly by $\trueapert$\mynom[S]{$\trueapert$}{Satellite's true anomaly}, the longitude of the ascending node by $\lonanpert$\mynom[S]{$\lonanpert$}{Longitude of the ascending node}, and the argument of the pericentre by $\argperipert$\mynom[S]{$\argperipert$}{Argument of the pericentre}. The reference frame associated with the satellite's orbit, $\rframe{\ipert}{O}{\Ivect}{\Jvect}{\Kvect}$\mynom[S]{$\framesymb_{\ipert}$}{Reference frame associated with the satellite's orbit}, is centred on the planet's centre of gravity and inclined relative to the inertial frame $\framefix$. The unit vectors of this orbital frame are defined such that $\Ivect$\mynom[S]{$\Ivect$}{Unit vector pointing towards the pericentre of the satellite's orbit} points towards the pericentre of the orbit, $\Kvect$\mynom[S]{$\Kvect$}{Unit vector aligned with the orbital angular momentum} is aligned with the orbital angular momentum, and $\Jvect = \Kvect \crossp \Ivect$\mynom[S]{$\Jvect$}{Third unit vector defined as $\Jvect = \Kvect \crossp \Ivect$}. Additionally, the vector $\nnodev$ points towards the ascending node, thereby defining the line of nodes. The transformation between the inertial frame $\rframe{\ifix}{O}{\ex}{\ey}{\ez} $ and the orbital frame $ \rframe{\ipert}{O}{\Ivect}{\Jvect}{\Kvect}$ is described by an Euler rotation matrix, characterised by the angles~$\left( \angapert, \angbpert, \anggpert \right)$,\mynom[S]{$ \angapert, \angbpert, \anggpert$}{Euler angles describing the rotation $\rframe{\ifix}{O}{\ex}{\ey}{\ez} \rightarrow \rframe{\ipert}{O}{\Ivect}{\Jvect}{\Kvect}$}
\begin{equation}
\label{euler_angles_orbit}
\begin{array}{lll}
\displaystyle \angapert \define \lonanpert - \frac{\pi}{2} , &\displaystyle  \angbpert \define \incpert , & \displaystyle  \anggpert \define \argperipert + \frac{\pi}{2}.
\end{array}
\end{equation}

In this simplified framework, the forcing tidal potential can be expressed as a function of the mean anomaly of the perturber,~$\meanapert$\mynom[S]{$\meanapert$}{Satellite's mean anomaly}, rather than the true anomaly. This is achieved using the Hansen coefficients $\Hansen{\kk}{\llat}{\mm}$\mynom[S]{$\Hansen{\kk}{\llat}{\mm}$}{Hansen coefficients}, defined as \citep[e.g.,][]{Hughes1981,Laskar2005}
\begin{equation}
\label{Hansen_series}
\left( \frac{\rpert}{\smaxispert} \right)^\llat \expo{\inumber \mm \trueapert} = \sum_{\kk= - \infty}^{ \infty} \Hansen{\kk}{\llat}{\mm} \left( \eccpert \right) \expo{\inumber \kk \meanapert}.
\end{equation}
Here, the mean anomaly $\meanapert$ is defined as $\meanapert \define \npert \time + \meanapertc$, where $\meanapertc$\mynom[S]{$\meanapertc$}{Constant phase angle} is a constant phase angle. After performing the manipulations detailed in \append{app:forcing_potential}, we obtain
\begin{align}
\label{Utidelmkm_gal}
\Ulmsigj{\llat}{\kk}{\mm} = &  \left( 2 - \kron{\kk}{0} \right) \frac{4 \pi}{2 \llat + 1} \frac{\Ggrav \Mpert}{\Rpla} \left( \frac{\Rpla}{\smaxispert} \right)^{\llat +1} \nonumber \\ 
& \times \sum_{\qq=-\llat}^{\llat} \wignerD{\mm}{\qq}{\llat} \left( \angapert , \angbpert, \anggpert \right) \Ylm{\llat}{\qq} \left( \frac{\pi}{2} , 0 \right) \Hansen{\kk}{- \left( \llat + 1 \right)}{- \qq} \left( \eccpert \right).
\end{align}
In this equation, we see the Wigner D-functions previously introduced in \eqs{Ylmpla_Ylmfix}{Ylmfix_Ylmpla}. Notably, the factor $\Ylm{\llat}{\qq} \left( \frac{\pi}{2} , 0 \right) $ in \eq{Utidelmkm_gal} represents a specific value of the spherical harmonics (refer to \append{app:sph} for details).

\subsection{Dynamical equations}
\label{ssec:dynamical_eqs}
The planet's spin angular velocity is typically much greater than the variation rates of the precession and nutation angles introduced in \eq{eulermat} as $\angalp$ and $\angbet$, respectively.\rev{ Mathematically, $\abs{\Dt{\angalp}} \ll \spinrate$ and $\abs{\Dt{\angbet}} \ll \spinrate$.} As a result, the gyroscopic approximation can be applied \citep[see e.g.][Sect.~2.8]{BE2019}, which involves disregarding the contributions of the terms related to $\Dt{\angalp}$ and $\Dt{\angbet}$ in the calculation of the planet's angular momentum, $\angmomplav$\mynom[S]{$\angmomplav$}{Planet's angular momentum vector}. Under this approximation, $\angmomplav$ is simply expressed as  
\begin{equation}
\angmomplav = \Cinertpla \spinvect, 
\end{equation}
where $\Cinertpla$\mynom[S]{$\Cinertpla$}{Planet's principal moment of inertia} designates the principal moment of inertia of the planet, and $\spinvect \define \spinrate \eZ$\mynom[S]{$\spinvect$}{Planet's spin vector} its spin vector. The evolution rate of the planet's angular momentum is the tidal torque established in \eqsto{torquex_sum}{torquez_sum},
\begin{equation}
\Dt{\angmomplav} = \torquev.
\end{equation}
Expanding $\Dt{\spinvect}$ in terms of $\spinrate$, $\angalp$ and $\angbet$, and using the basis unit vectors introduced in \sect{ssec:torque_tidal_potentials} and \fig{fig:euler_angles}, we obtain
\begin{equation}
\label{spinratev_rate_comp}
\Dt{\spinvect} = \Dt{\spinrate} \eZ + \spinrate \Dt{\angalp} \left( \ez \crossp \eZ \right) + \spinrate \Dt{\angbet} \left( \eyprim \crossp \eZ \right),
\end{equation}
which, owing to the orthogonality of the vectors $\eZ$, $\ez \crossp \eZ $, and $\eyprim \crossp \eZ$,  yields the variation rates of $\spinrate$, $\angalp$, and $\angbet$, 
\begin{equation}
\label{rates_vectors_planet}
\begin{array}{lll}
\displaystyle \Dt{\spinrate} = \frac{\torquev \dotp \eZ}{\Cinertpla}, & \displaystyle \Dt{\angalp} = \frac{\torquev \dotp \left(\ez \crossp \eZ \right)}{\Cinertpla \spinrate \sin^2 \angbet}, & \displaystyle \Dt{\angbet} = \frac{\torquev \dotp \left( \eyprim \crossp \eZ \right)}{\Cinertpla \spinrate}. 
\end{array}
\end{equation}
Analogously with $\angmomplav$, the angular momentum of the satellite's orbit, $\angmompertv $\mynom[S]{$\angmompertv$}{Satellite's angular momentum vector}, is defined as 
\begin{equation}
\label{angmompertv}
\angmompertv \define \angmompert \Kvect, 
\end{equation}
and its variation rate is expressed as 
\begin{equation}
\Dt{\angmompertv} = - \torquev. 
\end{equation}
Thus, expanding $\Dt{\angmompertv}$ in terms of the Euler angles describing the inclination of the orbit, $\angapert$ and $\angbpert$ (see \eq{euler_angles_orbit}), we end up with
\begin{equation}
\label{angmompertv_rate_comp}
\Dt{\angmompertv} = \Dt{\angmompert} \Kvect + \angmompert \Dt{\angapert} \left( \ez \crossp \Kvect \right) + \angmompert \Dt{\angbpert} \left( \nnodev \crossp \Kvect \right),
\end{equation}
and the variation rates of $\angmompert$, $\angapert$, and $\angbpert$, 
\begin{equation}
\label{rates_vectors_orbit}
\begin{array}{lll}
\displaystyle \Dt{\angmompert}  = - \torquev \dotp \Kvect , & 
\displaystyle \Dt{\angapert} = - \frac{\torquev \dotp \left( \ez \crossp \Kvect  \right)}{\angmompert \sin^2 \angbpert },  &
\displaystyle \Dt{\angbpert} = - \frac{\torquev \dotp  \left( \nnodev \crossp \Kvect \right)}{\angmompert}. 
\end{array}
\end{equation}
We remark that all the equations established until now are general and do not depend on the choice made for the fixed frame of reference ($\framefix$). 

It is actually appropriate to define $\framefix$ so that $\ez$ is aligned with the total angular momentum of the two-body system,\mynom[S]{$\angmomtotv$}{Total angular momentum vector of the planet-satellite system}
\begin{equation}
\angmomtotv \define \angmomplav + \angmompertv = \angmomtot \ez,
\end{equation}
with the vectors $\ex$ and $\ey$ defining the so-called invariable plane, namely the plane perpendicular to the total angular momentum \citep[e.g.,][]{Tremaine2009,Rubincam2016,Boue2016,BE2019}. In this configuration, $\angalp - \angapert = \pi$, which allows the evolution equations of the set of variables $\left\{ \spinrate , \angmompert, \angbet , \angbpert \right\}$ to be deduced from \eqs{rates_vectors_planet}{rates_vectors_orbit},
\begin{align}
\label{rate_spinrate}
\Dt{\spinrate} & = \Cinertpla^{-1} \left[ \sin \angbet \left( \cos \angalp \torquex + \sin \angalp \torquey \right) + \cos \angbet \torquez  \right] , \\
\label{rate_angmompert}
\Dt{\angmompert} & = - \sin \angbpert \left( \cos \angapert \torquex + \sin \angapert \torquey \right) - \cos \angbpert \torquez, \\
\label{rate_angbet}
\Dt{\angbet} & = \left( \Cinertpla \spinrate \right)^{-1} \left[ \cos \angbet \left( \cos \angalp \torquex + \sin \angalp \torquey \right) - \sin \angbet \torquez  \right],  \\
\Dt{\angbpert} & = \angmompert^{-1} \left[ - \cos \angbpert \left( \cos \angapert \torquex + \sin \angapert \torquey \right) + \sin \angbpert \torquez  \right].
\end{align}

The perturber's semi-major axis and eccentricity can be obtained from energy conservation principles. The total energy of the system, $\enertot$\mynom[S]{$\enertot$}{Total energy of the planet-satellite system}, is formulated as
\begin{equation}
\label{eq_enertot}
\enertot = \frac{1}{2} \Cinertpla \spinrate^2 + \enerpert,
\end{equation}
where $\enerpert$\mynom[S]{$\enerpert$}{Mechanical energy of the perturber's motion} designates the mechanical energy of the perturber's motion; and its variation rate as $\Dt{\enertot} = - \powerdiss$. Since the power transferred from the planet's spin to the planet's orbital motion is expressed as 
\begin{equation}
\label{rate_enerpert}
\Dt{\enerpert} = - \powertide,
\end{equation} 
taking the time derivative of \eq{eq_enertot} yields the relationship
\begin{align}
\label{pdiss_power_torque}
\powerdiss & =  \powertide - \spinvect \dotp \torquev,  \\
                    & =  \powertide - \Cinertpla \spinrate \Dt{\spinrate} . \nonumber
\end{align}
The semi-major axis of the perturber is readily deduced from $\enerpert$ \citep[e.g.,][]{MacDonald1964},
\begin{equation}
\label{smaxispert}
\smaxispert = - \frac{\Ggrav \Mpert \Mpla}{2\enerpert}.
\end{equation}
Analogously, the angular momentum of the perturber is given by \citep[e.g.,][]{MacDonald1964}
\begin{equation}
\angmompert = \frac{\Mpla \Mpert }{ \Mpla +\Mpert} \sqrt{\Ggrav \left( \Mpla + \Mpert \right)  \smaxispert \left( 1 - \eccpert \right)^2}, 
\end{equation}
which allows the eccentricity to be expressed as a function of the other quantities,
\begin{equation}
\label{eccpert}
\eccpert =  1 - \frac{\left( \Mpla + \Mpert \right) \angmompert^2}{\left( \Mpla \Mpert \right)^2  \Ggrav \smaxispert}. 
\end{equation}

We remark that the tangential component of $\angmomtotv $ is zero by construction, given that $\angmomtotv $ is aligned with $\ez$ in the adopted frame of reference. Consequently, the tangential components of $\angmomplav$ and $\angmompertv$ exactly compensate each other, which is formulated as
\begin{equation}
\Cinertpla \spinrate \sin \angbet = \angmompert \sin \angbpert.
\end{equation}
The equatorial plane of the planet and the orbital plane of the perturber are thus inclined in two opposite directions, and the angle $\angbpert$ can be deduced from $\angbet$, $\spinrate$, and $\angmompert$, 
\begin{equation}
\label{angbpert}
\angbpert = \arcsin \left( \frac{\Cinertpla \spinrate}{\angmompert} \sin \angbet \right). 
\end{equation}
Similarly, $\angmomtot$ appears to be a first integral of the problem, as no external torque acts on the system. The constancy of $\angmomtot$ over time ($\Dt{\angmomtot}=0$) imposes an additional constraint on $\spinrate$, $\angmompert$, $\angbet$, and~$\angbpert$, expressed by the equation
\begin{equation}
\Cinertpla \spinrate \cos \angbet + \angmompert \cos \angbpert = \angmomtot. 
\end{equation}
This constraint can be used to verify the consistency of the system's dynamical evolution over time, as the left-hand side must remain unchanged despite variations in $\spinrate$, $\angmompert$, $\angbet$, and $\angbpert$. The system's evolution is then determined by integrating \eqsto{rate_spinrate}{rate_angbet} and \eq{rate_enerpert} for $\left\{ \spinrate , \angbet, \angmompert , \enerpert \right\}$. The quantities $\smaxispert$, $\eccpert$, and $\angbpert$ are subsequently derived from the former using the relations provided in \eqsthree{smaxispert}{eccpert}{angbpert}, respectively. 

\section{Application to the Earth-Moon system}
\label{sec:application_earth_moon}

\subsection{Physical setup of tidal solutions}

In the previous sections, we derived the general expressions of tidal torque and power as functions of the forcing and deformation tidal potentials, incorporating the non-isotropic effects that are usually ignored. This formalism now enables us to investigate, as a next step, how deviations from the isotropic assumption may alter the estimates of these quantities and potentially lead to inaccurate predictions regarding the evolution of planetary systems. 

\begin{table}[h]
\centering
\caption{\label{tab:param_refcase} Values of parameters used in our case study. }
\begin{small}
\begin{tabular}{lccc} 
 \hline 
 \hline 
Parameter & Symbol & Value & Ref.  \\ 
 \hline \\[-0.3cm]
  \multicolumn{4}{c}{\textit{Planet's solid part}} \\
  Planet mass & $\Mpla$ & $5.9722 \times 10^{24}$~kg & 1 \\  
  Planet radius & $\Rpla$ & $6371.0 $~km & 1\\  
  Planet's moment of inertia & $\Cinertpla$ & $0.3308 \Mpla \Rpla^2$ & 1  \\
  Effective shear modulus & $\mupla$ & 25.1189~GPa & 2 \\ 
  Maxwell time & $\tauM$ & 685~yr & 2\\ 
  Andrade time & $\tauA$ & 12897.1~yr & 2 \\ 
  Rheological parameter  & $\alphaA$ & 0.25 & 2 \\[0.1cm]
  \multicolumn{4}{c}{\textit{Planet's ocean}} \\
  Oceanic depth & $\Hoc$ & 4.0~km & \\
  Rayleigh drag frequency & $\fdrag$ & $1.0 \times 10^{-5} \ {\rm s^{-1}}$ &  \\ 
  Seawater density & $\rhowater$ & $1035 \ {\rm kg \ m^{-3}}$  & 3 \\[0.1cm]
   \multicolumn{4}{c}{\textit{Satellite}} \\
   Mass of the satellite & $\Mpert$ & $ 7.346 \times 10^{22}  $~kg  & 1 \\
   Orbital period of the satellite & $\Porb$ & $ 27.3217 \ {\rm days}$ & 1\\ 
\hline
 \end{tabular}
 \end{small}
 \tablebib{(1)~\citet{Fienga2021}; (2) \citet{Bolmont2020}; (3) \citet{Ray2001}.}
 \end{table}

For a didactical purpose, we apply the described methods to an idealised Earth-Moon system. The Earth is modelled as a two-layer, spherically symmetric body with a massive solid part and a thin, uniformly deep liquid water ocean, with a depth of $\Hoc \ll \Rpla$\mynom[S]{$\Hoc$}{Ocean's depth}. The formalism used to compute the linear tidal response of such a planet has been extensively detailed in previous works \citep[][]{ADLML2018,Auclair2019,Farhat2022ellip,Auclair2023}. Therefore we will only briefly outline the key aspects of the approach followed here and direct readers to these studies for further details. 

As outlined in \cite{Auclair2023}, the tidal response of the Earth's solid part is modelled using Andrade rheology, with parameter values prescribed by \cite{Bolmont2020} for the actual Earth. Andrade rheology captures both visco-elastic and inelastic deformations of the solid regions. Although initially developed from lab experiments on metals \citep[][]{Andrade1910}, later studies demonstrated its applicability to silicate rocks and ices \citep[see e.g.,][and references therein]{Efroimsky2012}. In this framework, the tidal response of the solid Earth is governed by the mantle's effective shear modulus, $\mupla$\mynom[S]{$\mupla$}{Mantle's effective shear modulus}, the Maxwell time, $\tauM$\mynom[S]{$\tauM$}{Maxwell time}, which parametrises viscous friction, and two additional parameters linked to inelastic elongation: the Andrade time, $\tauA$\mynom[S]{$\tauA$}{Andrade time}, and a dimensionless parameter, $\alphaA$\mynom[S]{$\alphaA$}{Dimensionless parameter characterising the unrecoverable creep}, which characterises the unrecoverable creep \citep[e.g.,][]{Castelnau2008,CR2011,Efroimsky2012}. The latter parameter scales the frequency-dependent energy dissipation in the high-frequency regime.  

The Andrade model is believed to accurately represent the tidal response of rocky bodies such as terrestrial planets and icy satellites, particularly in the high-frequency range where inelasticity dominates \citep[e.g.,][]{EL2007}. However, alternative models can also be applied \citep[see e.g.,][]{Henning2009}. The detailed computation of the solid Love numbers is presented in \cite{Auclair2023}, Appendix~H. This approach is based on Kelvin's closed-form solution for an incompressible, homogeneous sphere \citep[e.g.,][Chapter~5]{MM1960book}. Here, we employ this simplified model as a general method to describe the contribution of solid tides to energy dissipation in Earth-like scenarios, where oceanic tides are predominant. Nevertheless, the calculation of the solid Love numbers can be refined later using more sophisticated methods based on elasto-gravitational theory \citep[e.g.,][]{TS1972,Tobie2005}, along with realistic radial profiles of material properties \citep[e.g.,][]{Sotin2007}. 

The oceanic tidal theory used here follows the approach detailed in \cite{Auclair2023}. The ocean's tidal response is obtained by solving the linear Laplace tidal equations (hereafter LTEs)\mynom[A]{LTEs}{Laplace tidal equations} over a spherical surface. Originally formulated by Laplace in his `Traité de Mécanique Céleste' \citep[][Book~IV]{Laplace1798}, these equations were given a modern formulation in the foundational work by \cite{LH1968}. We consider the LTEs in the shallow water approximation, where tidal variables are depth-averaged \citep[][]{Vallis2006}. In this approach, the tidal solution reduces to barotropic tidal flows -- those independent of vertical structure -- which are dominant on Earth \citep[][]{Gerkema2019}. It represents planetary-scale surface gravity waves, restored by variations in self-attraction due to local ocean surface elevation \citep[][]{Vallis2006}. These waves are directly forced by the tide-raising potential, with their typical phase speed given by~$\sqrt{\ggravi \Hoc}$ \citep[e.g.,][]{Cartwright1977book}. 

Consequently, vertical oceanic structure effects on tidal dynamics are neglected, meaning the contribution of internal gravity waves \citep[i.e. waves restored by the Archimedean force; e.g.,][]{Gerkema2008} is formally excluded. However the impact of these waves is implicitly accounted for in the effective damping caused by dissipative processes. Using high-resolution TOPEX/Poseidon altimetric data, \cite{ER2000} and \cite{ER2001} demonstrated that energy dissipation on Earth primarily arises from bottom drag in shallow seas (${\sim}70\%$) and barotropic-to-baroclinic tidal conversion in deep oceans ($25{-}35\%$). This conversion results from internal gravity waves generated by the interaction between tidal flows and ocean floor topography \citep{Bell1975,JL2001,GK2007}.   

The LTEs consist of the horizontal momentum and continuity equations of fluid dynamics linearised for tidal perturbations, which are assumed to be small disturbances relative to the background fields \citep[see, e.g.,][for an exhaustive formulation of the LTEs including nonlinear terms]{HM2019}. These equations are expressed, respectively, as \citep[e.g.,][]{Hendershott1972,Matsuyama2014,Matsuyama2018,Auclair2023}
\begin{align}
\label{momentum}
\dd{\Vvect}{\time} + \fdrag \Vvect + \fcorio \crossp \Vvect + \ggravi \grad \left( \tiltopd \zetaoc - \tiltopg \zetaeq \right)  &= 0, \\
\label{continuity}
\dd{\zetaoc}{\time} + \div  \left( \Hoc \Vvect \right) &= 0, 
\end{align}
where $\ggravi \define \Ggrav \Mpla/ \Rpla^2$\mynom[S]{$\ggravi$}{Planet's surface gravity} is the planet's surface gravity, $\fdrag$\mynom[S]{$\fdrag$}{Rayleigh drag frequency} is the Rayleigh drag frequency accounting for the efficiency of dissipative processes, and $\fcorio \define 2 \spinrate \cos \colpla \, \er $\mynom[S]{$\fcorio$}{Coriolis parameter} denotes the Coriolis parameter, with $\er$ being the outward-pointing radial unit vector. Here, $\Vvect$\mynom[S]{$\Vvect$}{Horizontal velocity field} represents the horizontal velocity field, $\zetaoc$\mynom[S]{$\zetaoc$}{Vertical displacement of the ocean's surface relative to the ocean's floor} the vertical displacement of the ocean's surface relative to the ocean floor, and $\zetaeq = \Utide/\ggravi $\mynom[S]{$\zetaeq$}{Equilibrium surface elevation in the zero-frequency limit} the equilibrium displacement caused by the tidal gravitational potential in the zero-frequency limit. 

For simplification, mean flows are ignored in \eq{momentum}, treating the ocean as a static fluid layer. This assumption is based on the separation of spatial and time scales between tidal and mean flows, with the latter characterised by smaller spatial structures and longer time scales, reducing the likelihood of significant coupling. However, this assumption does not always hold for thick fluid envelopes, where large-scale circulation patterns, such as differential rotation or latitudinal shear, can affect tidal flows by inducing Doppler-shift effects \citep[e.g.,][]{Boyd1978,Ortland2005a,Ortland2005b,BR2013,Guenel2016}. While permanent oceanic currents are generally thought to have minimal impact on tidal dynamics, they are influenced by tidal energy dissipation and internal waves, which interact with oceanic mesoscale eddies \citep[see e.g.,][and references therein]{Arbic2022}.

The Rayleigh drag frequency, $\fdrag$, plays a crucial role as it governs the frequency-dependent rates of energy and momentum exchanges within the planet-perturber system. As $\fdrag$ decreases, both energy and momentum transfer rates become more sensitive to tidal frequency \citep[e.g.,][]{ADLML2018}. This behaviour is linked to the phenomenon of the dynamical tide, which refers to the component formed by forced tidal waves propagating through a fluid layer \citep[][]{Zahn1975}. In this context, the dynamical tide results from long-wavelength surface gravity waves, which can be resonantly excited by the tide-raising potential, leading to significant variations in the tidally dissipated energy \citep[e.g.,][]{Webb1980,Tyler2014,Tyler2021,Kamata2015,ADLML2018,Auclair2019,Auclair2023,Matsuyama2018,HM2019,Farhat2022ellip,Rovira-Navarro2023}. As $\fdrag$ diminishes, the strength of the dynamical tide increases, with the height of resonant peaks scaling as $\scale \fdrag^{-1}$ in linear theory \citep[see e.g.,][Eq.~(58)]{Auclair2019}. 

This relationship holds generally for fluid bodies. The frequency dependence of the tidal response, driven by forced wave propagation in planetary fluid layers and stars, becomes more pronounced as the damping coefficient for frictional forces (or other dissipative mechanisms) decreases. This can result in energy dissipation rates that span several orders of magnitudes in stars and giants planets \citep[e.g.,][]{SW2002,OL2004,OL2007,Ogilvie2009,Ogilvie2013,ADMLP2015,Mathis2016,AB2023}, or in the subsurface oceans of icy moons \citep[][]{Tyler2011,Tyler2014,Chen2014,Matsuyama2014,Matsuyama2018,Beuthe2016}. Rayleigh drag frequency is typically unconstrained and can vary widely. For instance, \cite{Matsuyama2018} reports values as low as $\fdrag \sim 10^{-11} \ {\rm s^{-1}}$ for icy moons in the Solar System with $\fdrag \sim 10^{-9} \ {\rm s^{-1}}$ for obliquity tides on Europa. In the present study, we adopt $\fdrag = 10^{-5} \ {\rm s^{-1}}$, in line with estimates reckoned from high-precision measurements of Earth's tidal energy dissipation and advanced tidal models \citep[e.g.,][]{Webb1980,ER2001,ER2003,Farhat2022ellip}. Given the strong sensitivity of tidal dissipation to $\fdrag$, this value represents a conservative estimate for examining the impact of anisotropy on tidal dissipation. Any smaller value would amplify the frequency dependence of the oceanic tidal response.

The operators $\tiltopd$ and $\tiltopg$\mynom[S]{$\tiltopd,\tiltopg$}{Operators accounting for ocean loading, self-attraction, and the deformation of the planet's solid regions} in \eq{momentum} account for ocean loading, self-attraction, and the deformation of the planet's solid regions, which slightly modify the oceanic tidal response \citep[e.g.,][]{Hendershott1972,Ray1998,ER2004}. These operators are expressed in the frequency domain as functions of the solid tidal and load Love numbers, as detailed in \cite{Auclair2023}. Also, it is important to note that the three-dimensional gradient operator $\grad$, introduced in \eq{torquev_def}, is applied over the spherical surface in \eq{momentum}, along with the divergent operator $\div$ in \eq{continuity}. These operators are explicitly formulated in spherical coordinates in \append{app:vectorial_operators}. Equations~(\ref{momentum}) and~(\ref{continuity}) are solved for the velocity field $\Vvect$ and ocean surface displacement $\zetaoc$ in the frequency domain for every component of the tide-raising gravitational potential expressed in the rotating frame of reference, as provided by \eq{Utide_rot}. This potential is computed using \eq{Utidelmkm_gal}, with the degree-2 truncation ($\llat\leq 2$) corresponding to the classical quadrupolar approximation \citep[e.g.,][Sect.~7.2.2.5]{Mathis2013}. 

Following the approach of \cite{Auclair2023}, we numerically solve the LTEs by employing a spectral method that expands the tidal quantities in spherical harmonics series. Specifically, using the notations introduced in \sect{ssec:gravpot_distorted}, the variations in ocean depth are expressed as
\begin{equation}
\zetaoc \left( \colpla, \lonpla , \time \right) = \Re \left\{ \sum_{\kk=0}^{+ \infty} \sum_{\qq=-\infty}^{+ \infty} \zetaocrotsigi{\kk,\qq} \left(  \colpla, \lonpla \right)  \expo{\inumber \ftideplai{\kk,\qq} \time}  \right\}.
\end{equation}
The spatial functions $\zetaocrotsigi{\kk,\qq} \left(  \colpla, \lonpla \right)$ are further expanded as
\begin{equation}
\label{zetaocrotsigi}
\zetaocrotsigi{\kk,\qq}  \left(\colpla, \lonpla \right) = \sum_{\slon=0}^{\lmax} \sum_{\pp = - \slon}^\slon \zetalmrotsigi{\slon}{\kk,\qq}{\pp} \Ylm{\slon}{\pp} \left( \colpla, \lonpla \right)  ,
\end{equation}
where $\zetalmrotsigi{\slon}{\kk,\qq}{\pp} $\mynom[S]{$\zetalmrotsigi{\slon}{\kk,\qq}{\pp} $}{Complex coefficients of the multipole expansion of ocean's surface elevation in $\framepla$} are the complex coefficients of the multipole expansion and $\lmax$ denotes the truncation degree of the series. As discussed in \append{app:convergence_tests} from a convergence analysis, the truncation degree in \eq{zetaocrotsigi} is set to $\lmax = 30$\mynom[S]{$\lmax$}{Truncation degree of the spherical harmonic series in numerical solutions}. \rev{This value empirically appears to be sufficiently large to account for the complex mapping between the ocean's forced modes and spherical harmonics.} 

Finally, from the solution, we derive the Fourier coefficients of the deformation tidal potential needed to evaluate the tidal torque and power, as formulated in \eqsto{torquex_sum}{powertide_sum}. These coefficients, introduced in \eq{Urespnrotsigi}, are expressed in terms of the components of the tide-raising potential and the variation in ocean depth as shown in \citep[e.g.,][Eq.~(72)]{Auclair2023}\footnote{Equations~(72) and~(75) of \cite{Auclair2023} contain a typo that we correct here: a missing $\ggravi$ factor in the numerator of the second term on the right-hand side. This factor is included in the corrected formula given by \eq{Uresp_solution}.}
\begin{equation}
\label{Uresp_solution}
\Uplmrotsigj{\slon}{\kk,\qq}{\pp} = \klsigi{\slon}{\ftideplai{\kk,\qq}} \Ulmrotsigj{\slon}{\kk}{\qq} + \left( 1 + \kloadlsigi{\slon}{\ftideplai{\kk,\qq}}  \right) \frac{3 }{2 \slon +1} \frac{\rhowater}{\rhocore} \ggravi \zetalmrotsigi{\slon}{\kk,\qq}{\pp}, 
\end{equation}
where $\rhocore \define  3 \Mpla / \left( 4 \pi \Rpla^3 \right)$\mynom[S]{$\rhocore$}{Planet's mean density} denotes the planet's mean density, $\rhowater \approx 1035 \ {\rm kg \ m^{-3}}$\mynom[S]{$\rhowater$}{Average seawater density} represents the average density of seawater \citep[e.g.,][]{Ray2001}, and $\klsigi{\slon}{\ftideplai{\kk,\qq}} $\mynom[S]{$\klsigi{\slon}{\ftideplai{\kk,\qq}}$}{Complex solid tidal Love numbers} and $ \kloadlsigi{\slon}{\ftideplai{\kk,\qq}}$\mynom[S]{$\kloadlsigi{\slon}{\ftideplai{\kk,\qq}}$}{Complex solid load Love numbers} are the complex solid tidal and load Love numbers that describe the visco-elastic deformation of solid regions at the frequency $\ftideplai{\kk,\qq}$. The first term on the right-hand side of \eq{Uresp_solution} accounts for the solid elongation caused by the forcing tidal potential. The second term captures the contribution of oceanic tides to self-attraction variations, consisting of two components: the gravitational potential created by surface elevation changes and the deformation of solid regions due to ocean loading. 

\def\hraisebox{0.05\textwidth}
\def\wpanel{0.40\textwidth}
\def\wkey{0.0881\textwidth}
\begin{figure*}[t]
   \centering
   \RaggedRight \twolabelsat{5.5cm}{\textsc{Dry}}%
                           {13cm}{\textsc{Global ocean}}\\[0.1cm]
  \raisebox{\hraisebox}[1cm][0pt]{%
   \begin{minipage}{1.4cm}%
   \textsc{Full}
\end{minipage}}
\centering
   \includegraphics[width=\wpanel,trim = 0.cm 2.4cm 9cm 2.3cm,clip]{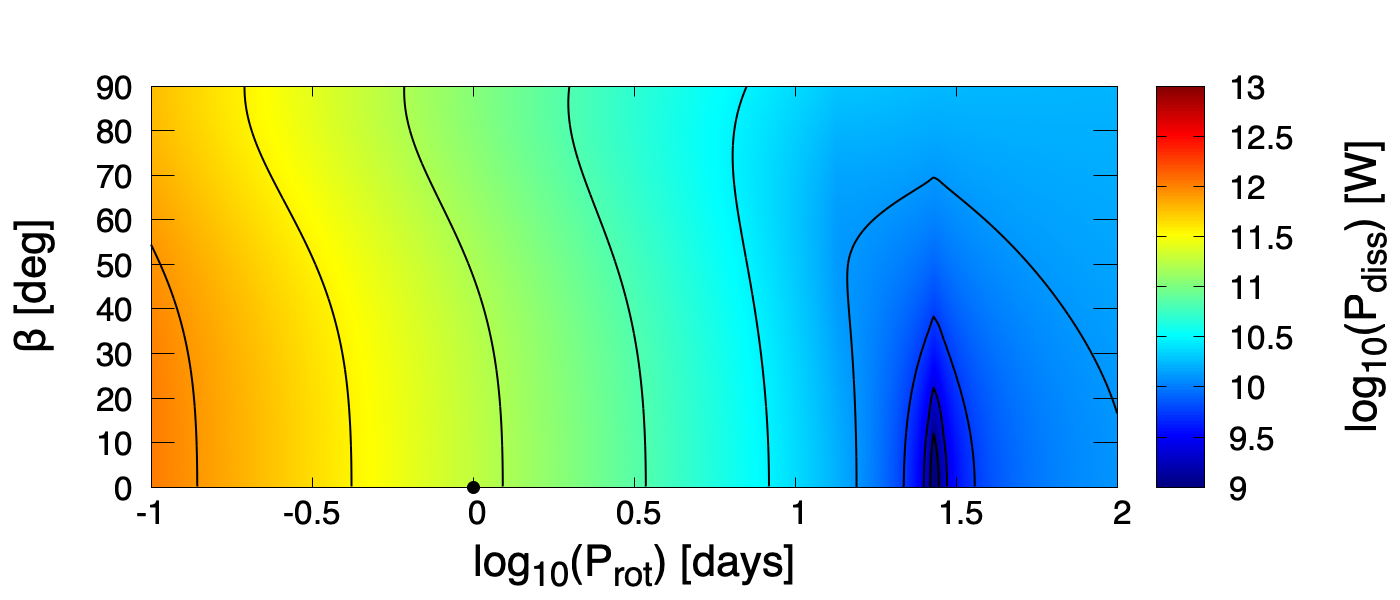}
   \includegraphics[width=\wpanel,trim = 0.cm 2.4cm 9cm 2.3cm,clip]{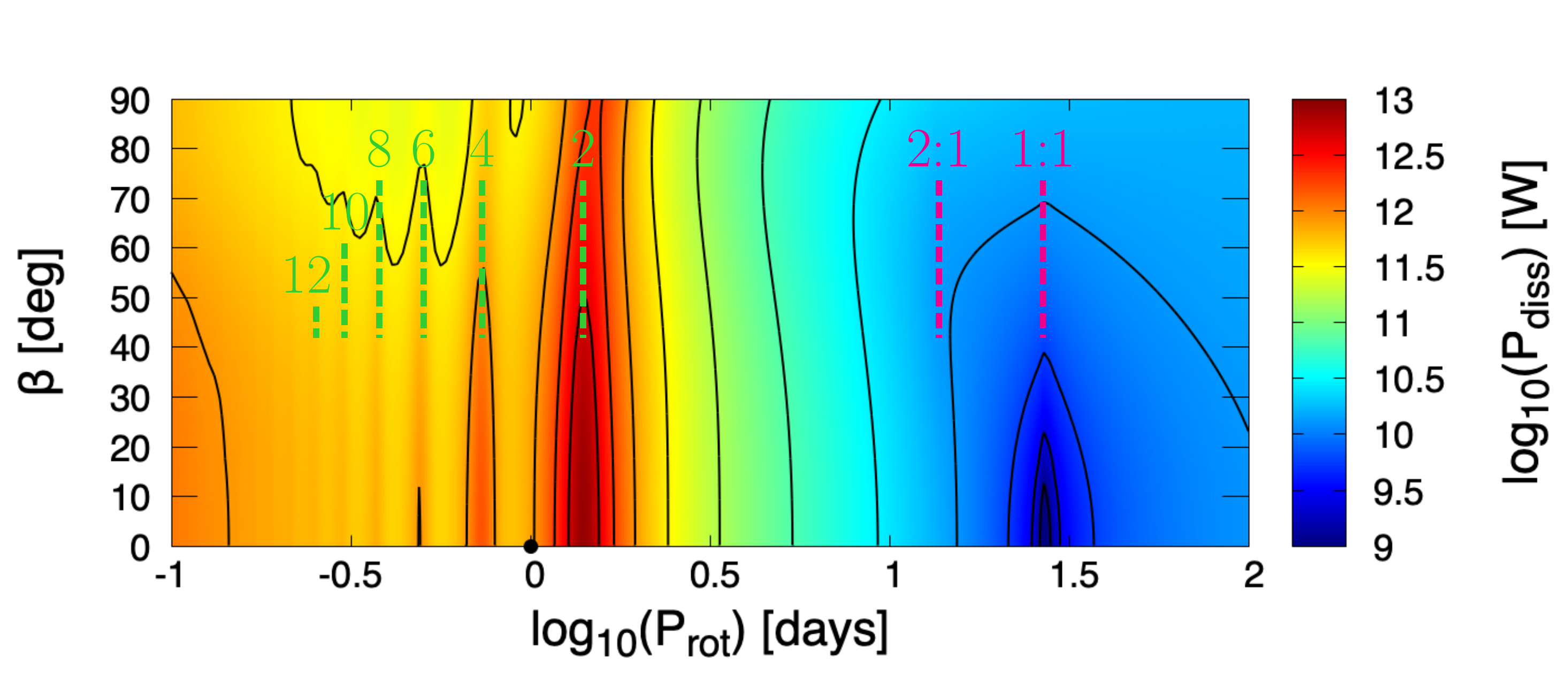} \hspace{\wkey}~ \\
   \raisebox{\hraisebox}[1cm][0pt]{%
   \begin{minipage}{1.4cm}%
   \textsc{Approx}
\end{minipage}}
   \includegraphics[width=\wpanel,trim = 0.cm 2.4cm 9cm 2.3cm,clip]{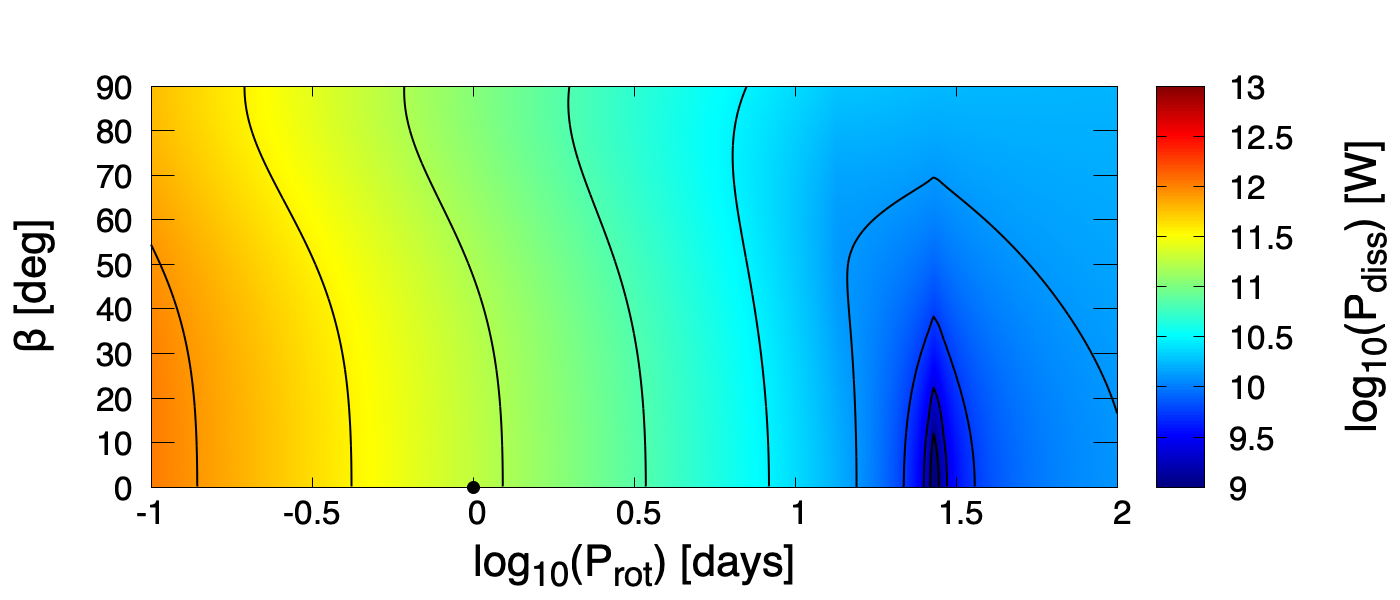}
   \includegraphics[width=\wpanel,trim = 0.cm 2.4cm 9cm 2.3cm,clip]{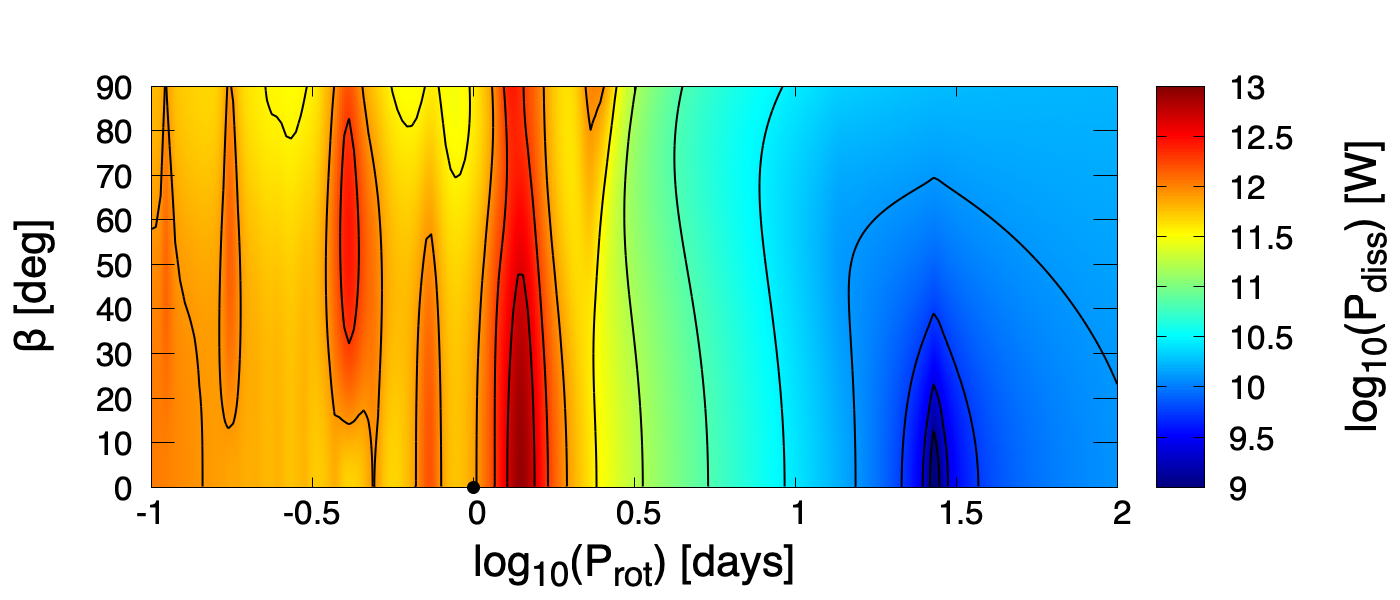}
   \includegraphics[width=\wkey,trim = 40.5cm 2.4cm 0cm 2.3cm,clip]{auclair-desrotour_fig4d.png} \\
   \raisebox{\hraisebox}[1cm][0pt]{%
   \begin{minipage}{1.4cm}%
   \textsc{Diff.}
\end{minipage}}
   \includegraphics[width=\wpanel,trim = 0.cm 0.1cm 9cm 2.3cm,clip]{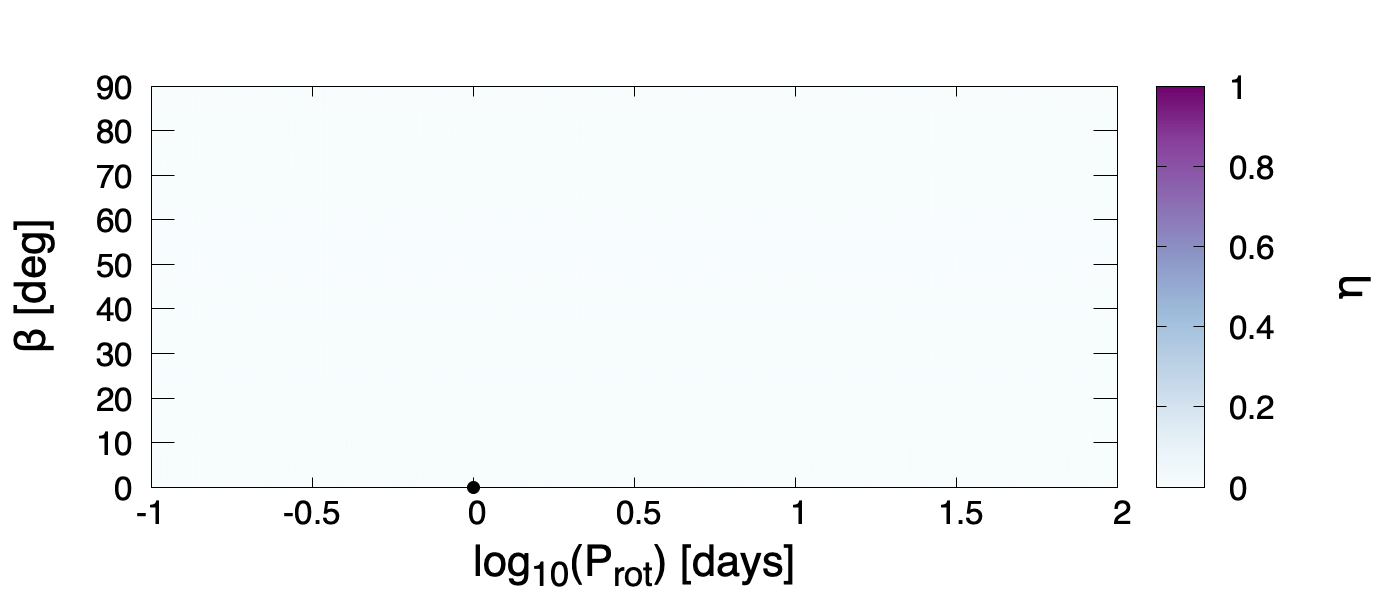}
   \includegraphics[width=\wpanel,trim = 0.cm 0.1cm 9cm 2.3cm,clip]{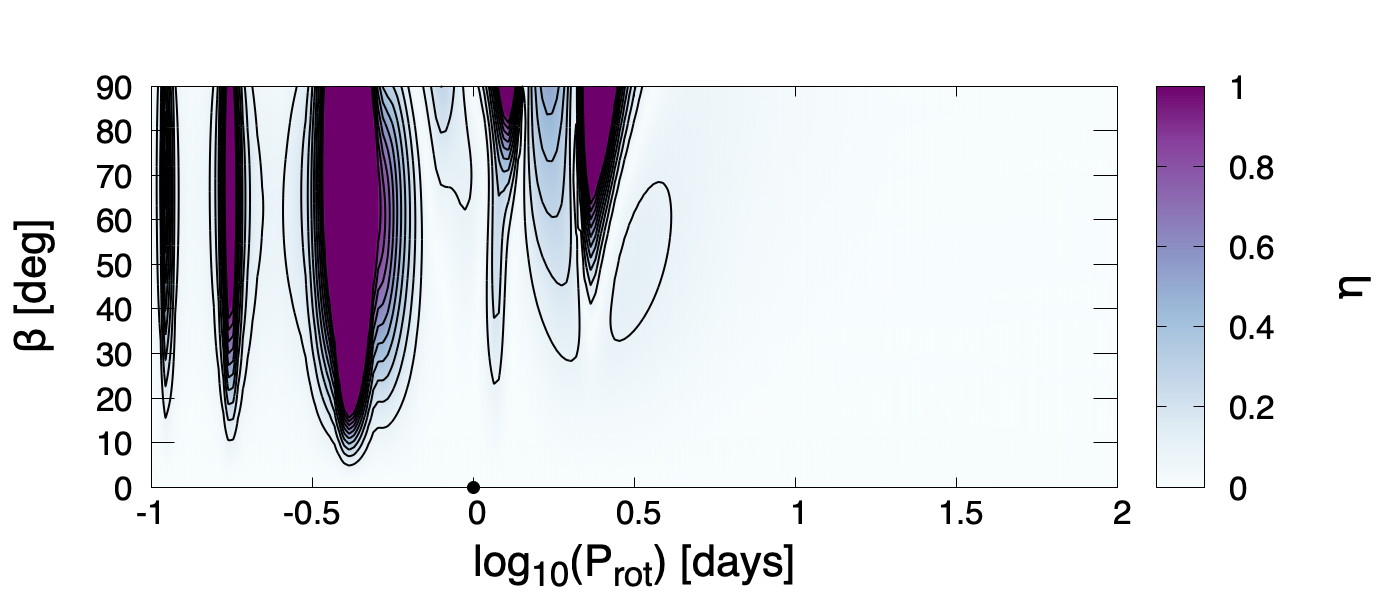} 
   \includegraphics[width=\wkey,trim = 40.5cm 0.1cm 0cm 2.3cm,clip]{auclair-desrotour_fig4f.png} 
      \caption{Tidally dissipated power ($\powerdiss$) as a function of the planet's spin rotation period (horizontal axis) and obliquity (vertical axis) for Earth-sized dry (left panels) and global-ocean (right panels) planets. {\it Top:} Full calculation of the planet's tidal response (\textsc{Full}). {\it Middle:} Calculation using, for all tidal forcing terms, the degree-2 Love number computed in the equatorial plane assuming the coplanar-circular configuration (\textsc{Approx}). {\it Bottom:} Relative difference between the \textsc{Full} and \textsc{Approx} cases, defined by \eq{eta_diff}. The tidally dissipated power is computed from the tidal power and torque using \eq{pdiss_power_torque}. The dashed green lines indicate the resonances associated with the oceanic surface gravity modes (see \eq{periodn} and Table~\ref{tab:period_res}), and the dashed magenta lines the $1{:}1$ and $2{:}1$ spin-orbit resonances. The black dot at $\left( \Prot , \obli \right) = \left( 1 \ {\rm day}, 0^\circ \right)$ designates the actual Earth-Moon system in the coplanar-circular approximation.}
       \label{fig:pdiss_solid_globoc}%
\end{figure*}

\begin{figure*}[t]
   \RaggedRight \twolabelsat{5.5cm}{\textsc{Dry}}%
                           {13cm}{\textsc{Global ocean}}\\[0.1cm]
  \raisebox{\hraisebox}[0.5cm][0pt]{%
   \begin{minipage}{1.4cm}%
   \textsc{Full}
\end{minipage}}
\centering
   \includegraphics[width=\wpanel,trim = 0.cm 2.4cm 9cm 2.3cm,clip]{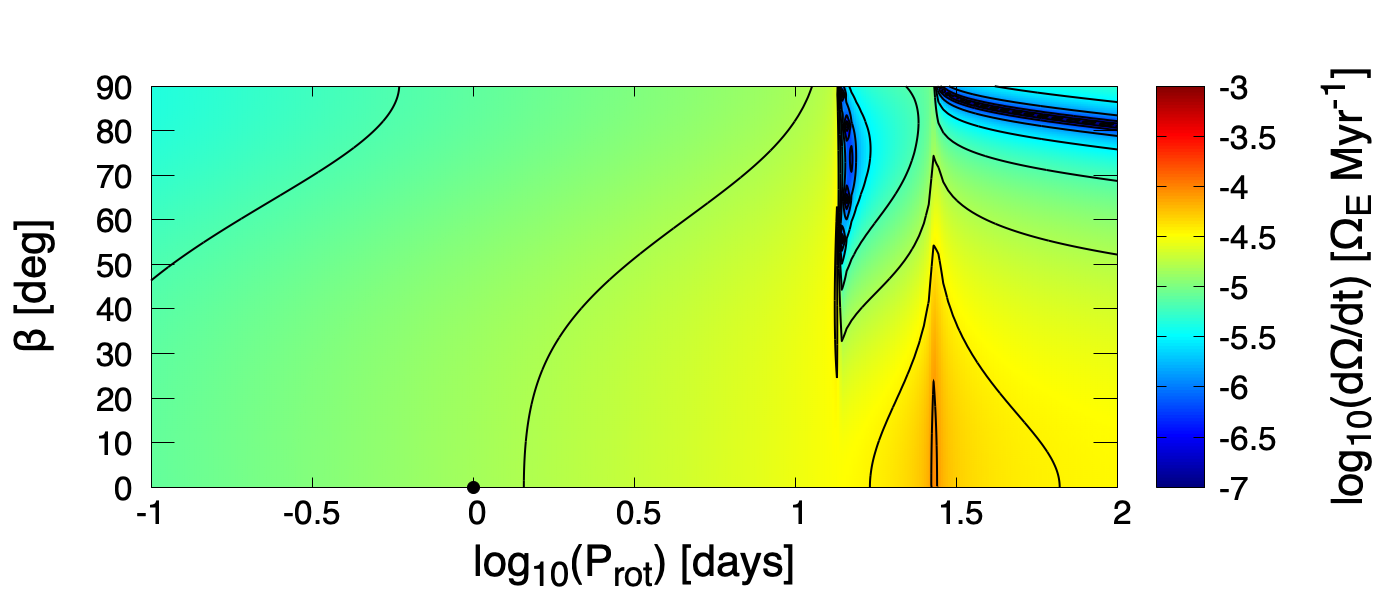}
   \includegraphics[width=\wpanel,trim = 0.cm 2.4cm 9cm 2.3cm,clip]{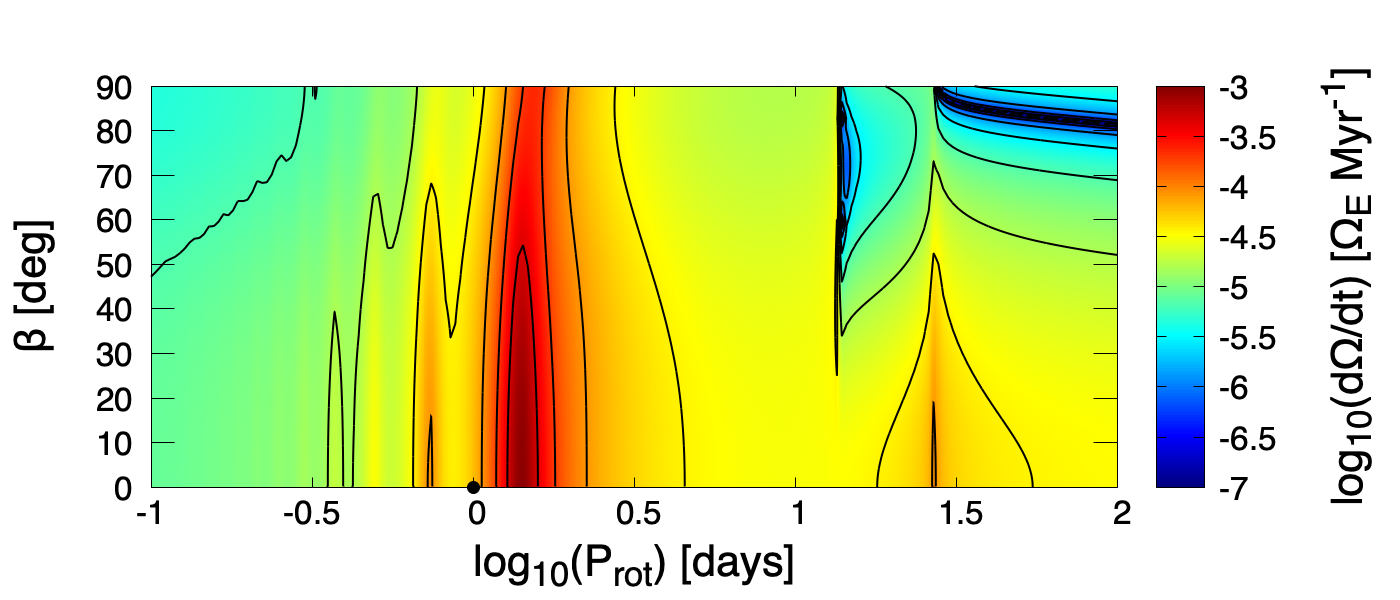} \hspace{\wkey}~  \\
   \raisebox{\hraisebox}[1cm][0pt]{%
   \begin{minipage}{1.4cm}%
   \textsc{Approx}
\end{minipage}}
   \includegraphics[width=\wpanel,trim = 0.cm 2.4cm 9cm 2.3cm,clip]{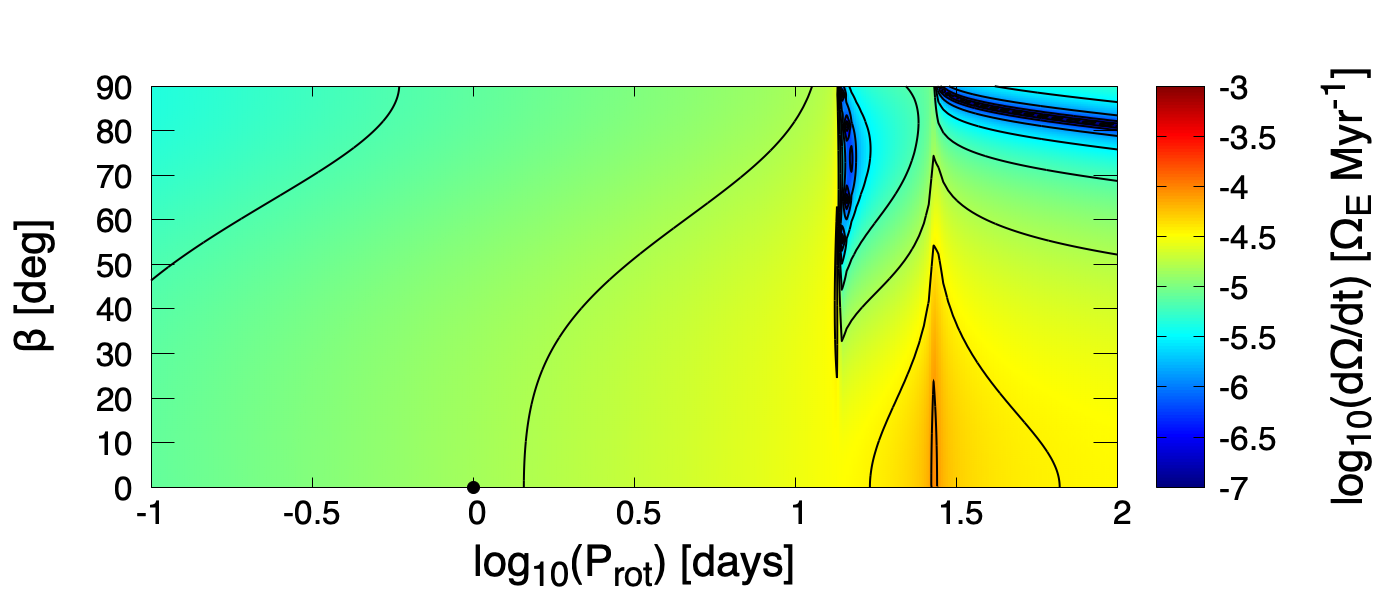}
   \includegraphics[width=\wpanel,trim = 0.cm 2.4cm 9cm 2.3cm,clip]{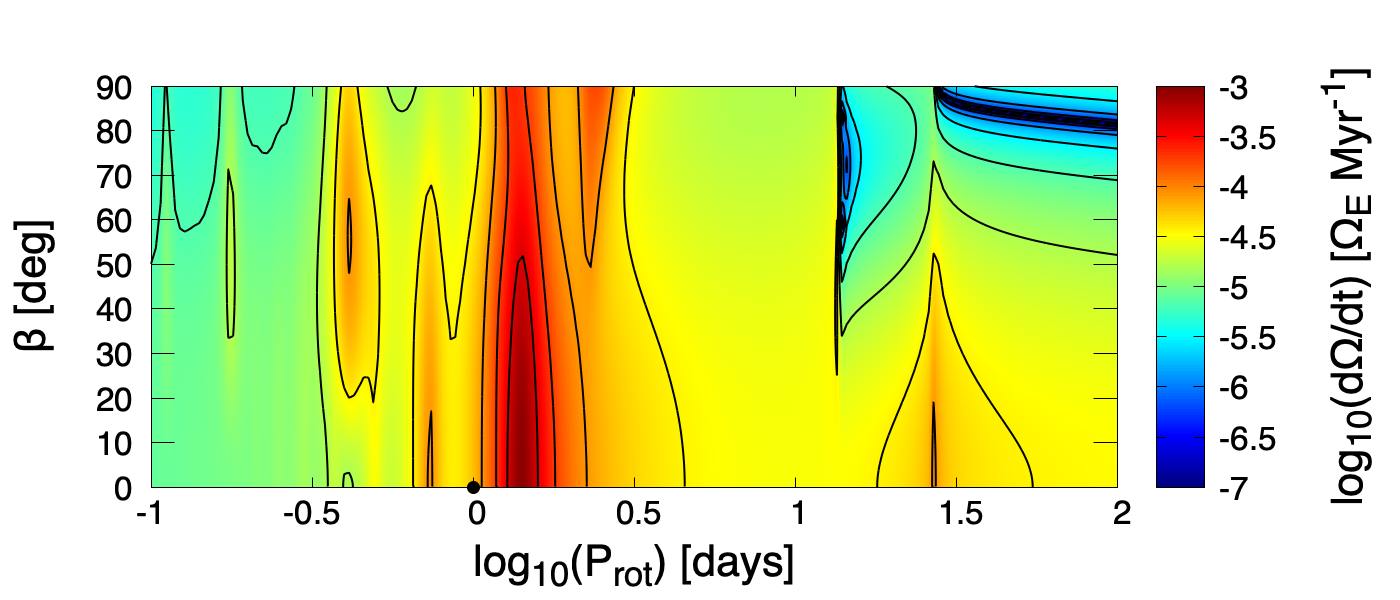} 
     \includegraphics[width=\wkey,trim = 40.5cm 2.4cm 0cm 2.3cm,clip]{auclair-desrotour_fig5d.png} \\
   \raisebox{\hraisebox}[1cm][0pt]{%
   \begin{minipage}{1.4cm}%
   \textsc{Diff.}
\end{minipage}}
   \includegraphics[width=\wpanel,trim = 0.cm 0.1cm 9cm 2.3cm,clip]{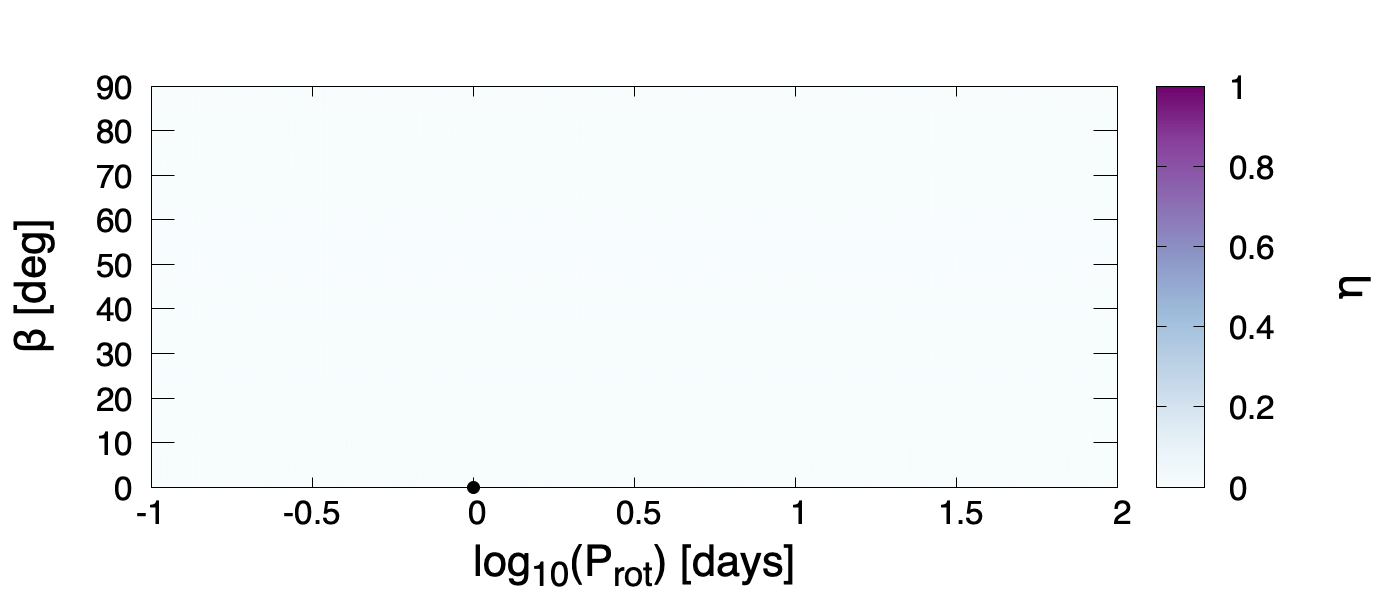}
   \includegraphics[width=\wpanel,trim = 0.cm 0.1cm 9cm 2.3cm,clip]{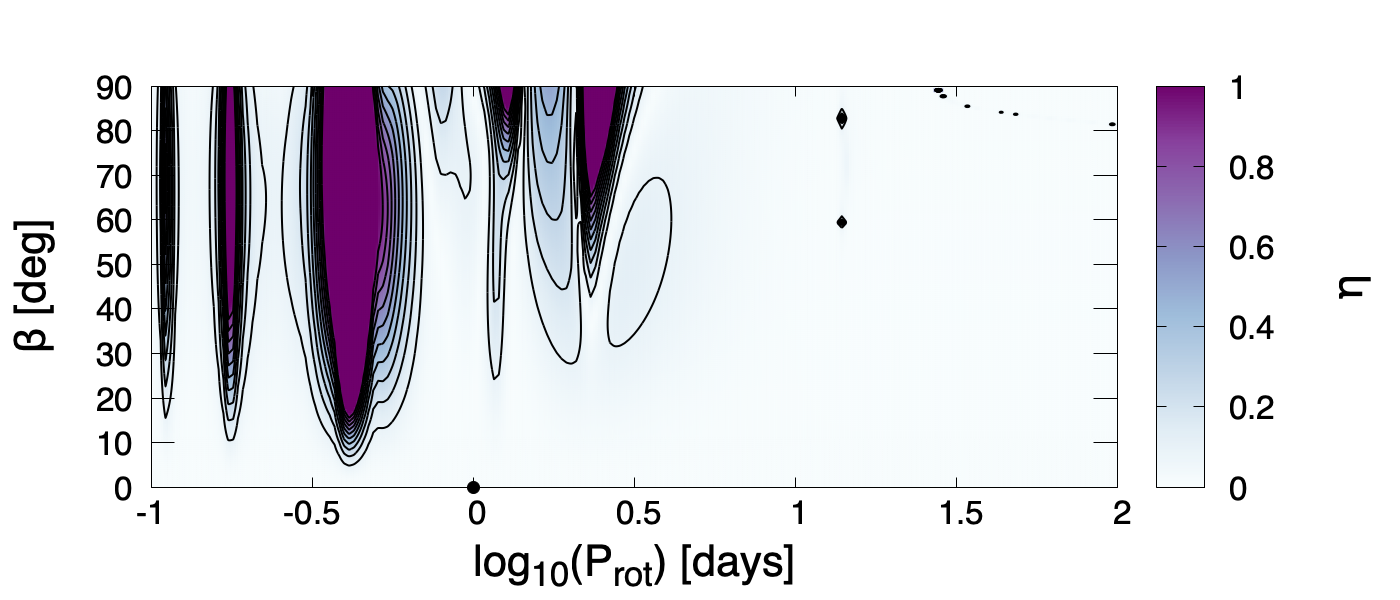} 
   \includegraphics[width=\wkey,trim = 40.5cm 0.1cm 0cm 2.3cm,clip]{auclair-desrotour_fig5f.png} 
      \caption{Variation rate of the planet's spin angular velocity ($\Dt{\spinrate}$) as a function of the planet's spin rotation period (horizontal axis) and obliquity (vertical axis) for Earth-sized dry (left panels) and global-ocean (right panels) planets. {\it Top:} Full calculation of the planet's tidal response (\textsc{Full}). {\it Middle:} Calculation using, for all tidal forcing terms, the degree-2 Love number computed in the equatorial plane assuming the coplanar-circular configuration (\textsc{Approx}). {\it Bottom:} Relative difference between the \textsc{Full} and \textsc{Approx} cases, defined by \eq{eta_diff}. The variation rate of the planet's spin angular velocity is computed from the three-dimensional tidal torque using \eq{rate_spinrate}. The black dot at $\left( \Prot , \obli \right) = \left( 1 \ {\rm day}, 0^\circ \right)$ designates the actual Earth-Moon system in the coplanar-circular approximation.}
       \label{fig:spinvel_solid_globoc}%
\end{figure*}

\begin{figure*}[t]
   \RaggedRight \twolabelsat{5.5cm}{\textsc{Dry}}%
                           {13cm}{\textsc{Global ocean}}\\[0.1cm]
  \raisebox{\hraisebox}[1cm][0pt]{%
   \begin{minipage}{1.4cm}%
   \textsc{Full}
\end{minipage}}
\centering
   \includegraphics[width=\wpanel,trim = 0.cm 2.4cm 9cm 2.3cm,clip]{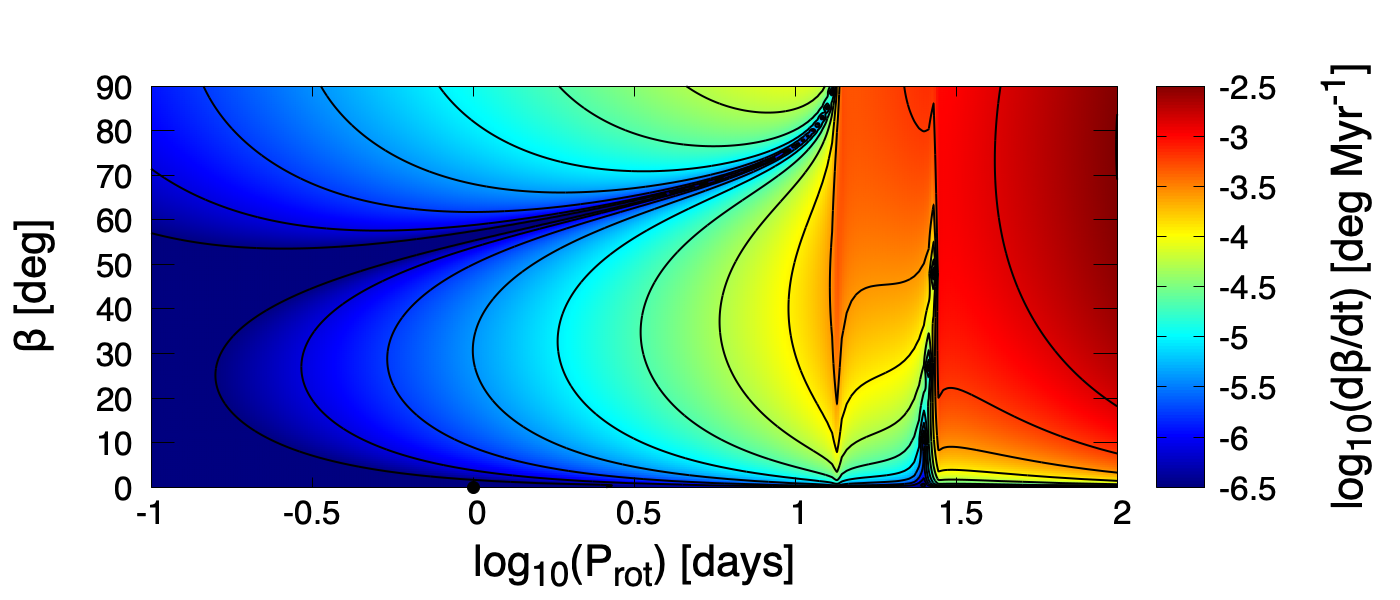}
   \includegraphics[width=\wpanel,trim = 0.cm 2.4cm 9cm 2.3cm,clip]{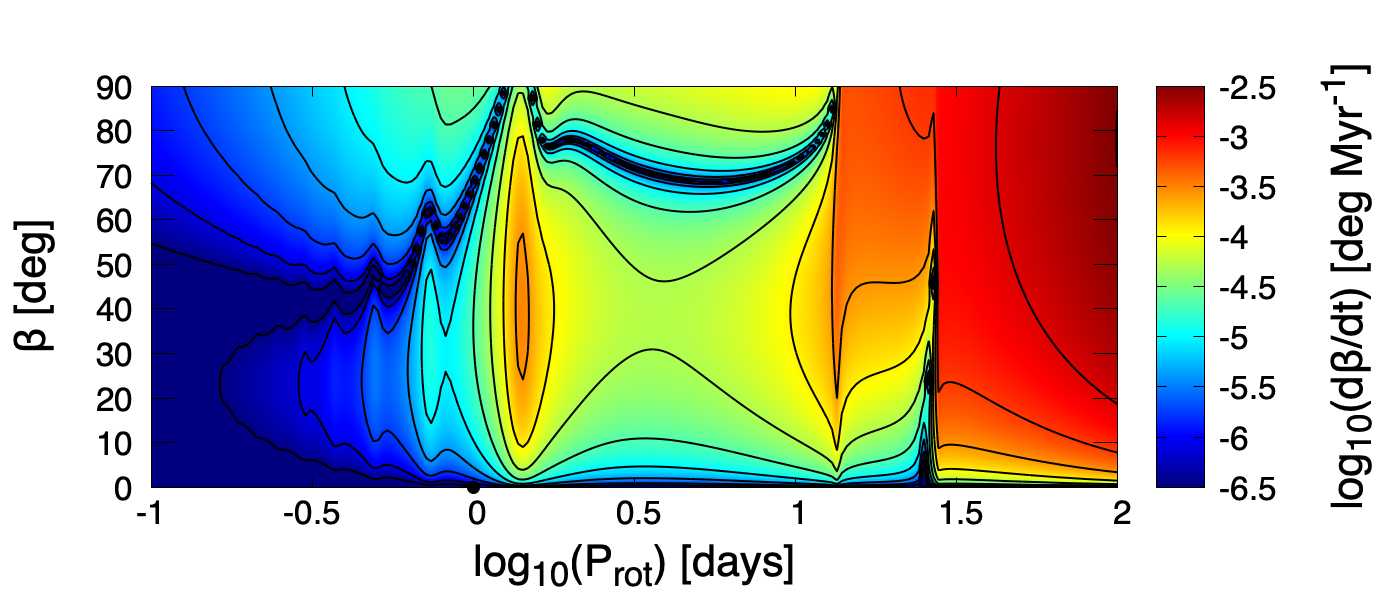} \hspace{\wkey}~ \\
   \raisebox{\hraisebox}[1cm][0pt]{%
   \begin{minipage}{1.4cm}%
   \textsc{Approx}
\end{minipage}}
   \includegraphics[width=\wpanel,trim = 0.cm 2.4cm 9cm 2.3cm,clip]{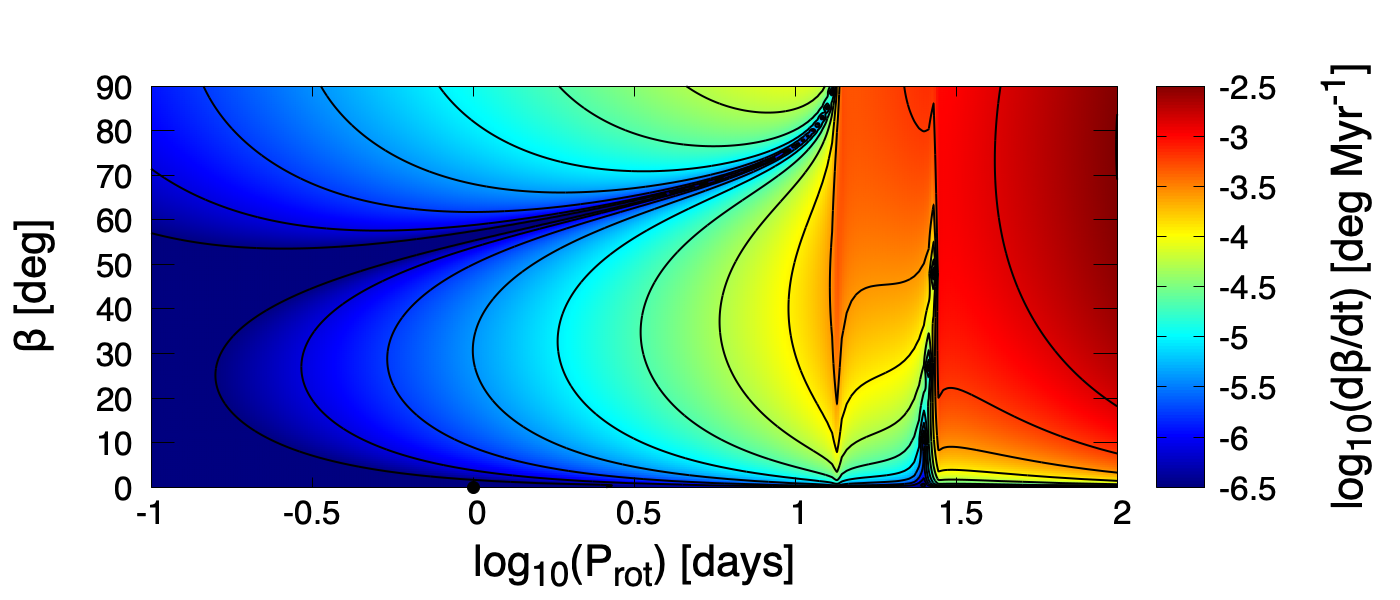}
   \includegraphics[width=\wpanel,trim = 0.cm 2.4cm 9cm 2.3cm,clip]{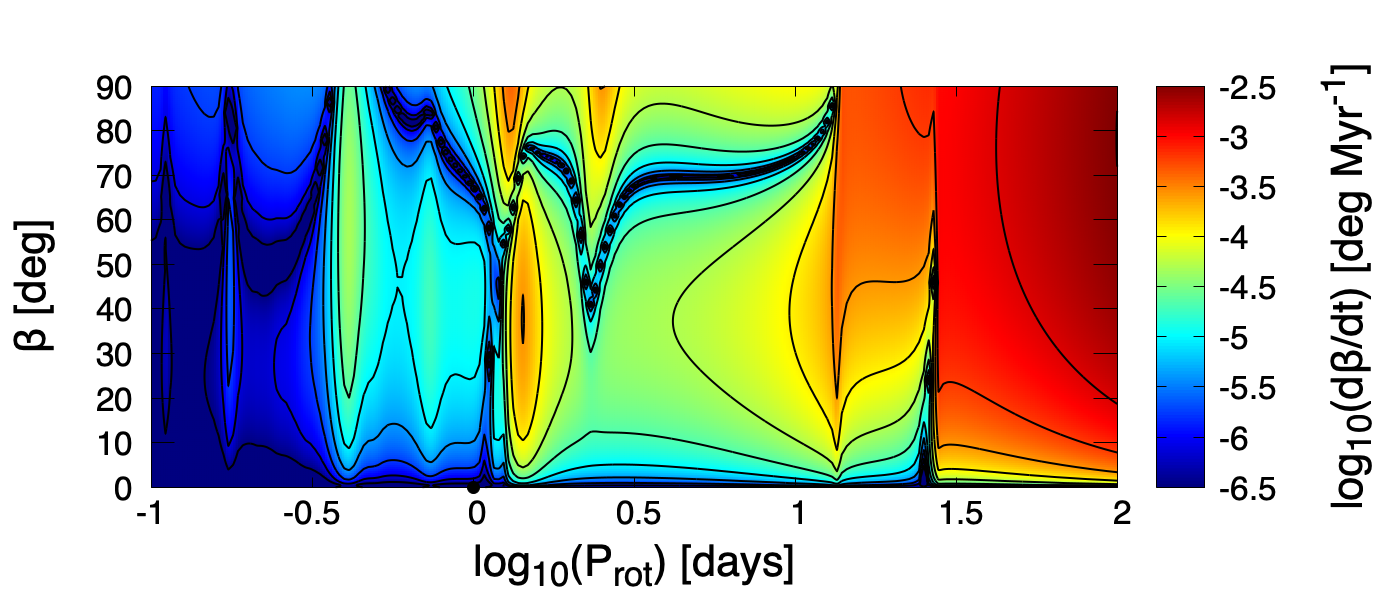} 
   \includegraphics[width=\wkey,trim = 40.5cm 2.4cm 0cm 2.3cm,clip]{auclair-desrotour_fig6d.png}  \\
   \raisebox{\hraisebox}[1cm][0pt]{%
   \begin{minipage}{1.4cm}%
   \textsc{Diff.}
\end{minipage}}
   \includegraphics[width=\wpanel,trim = 0.cm 0.1cm 9cm 2.3cm,clip]{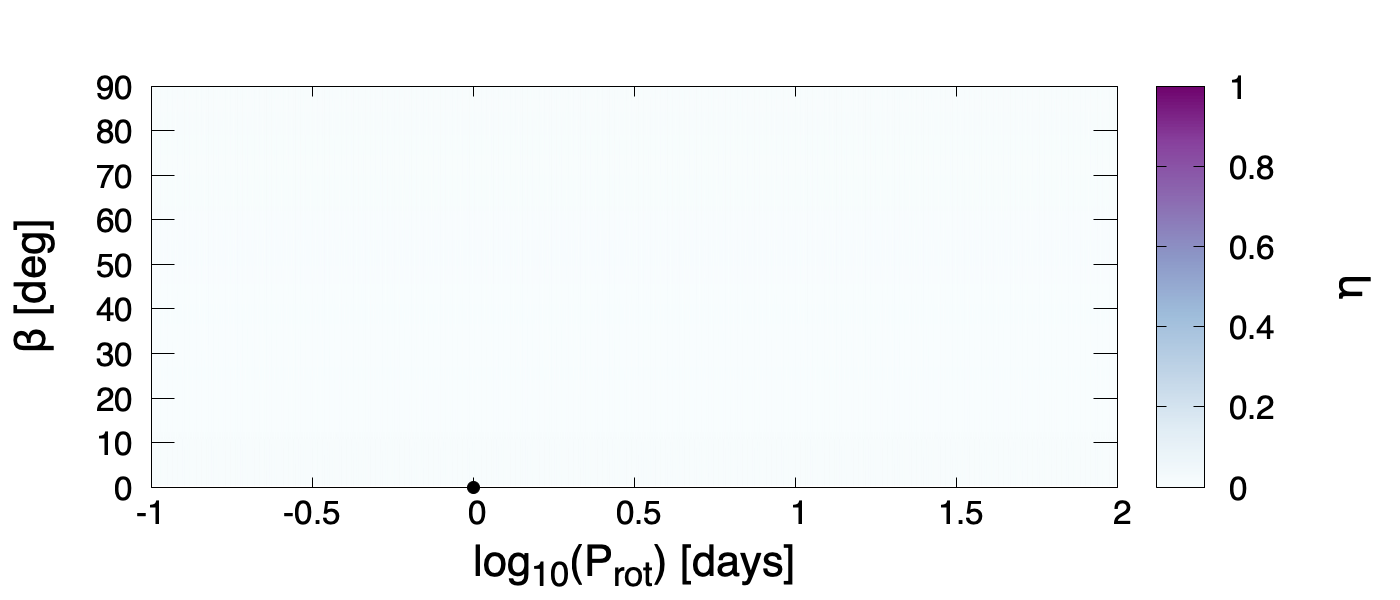}
   \includegraphics[width=\wpanel,trim = 0.cm 0.1cm 9cm 2.3cm,clip]{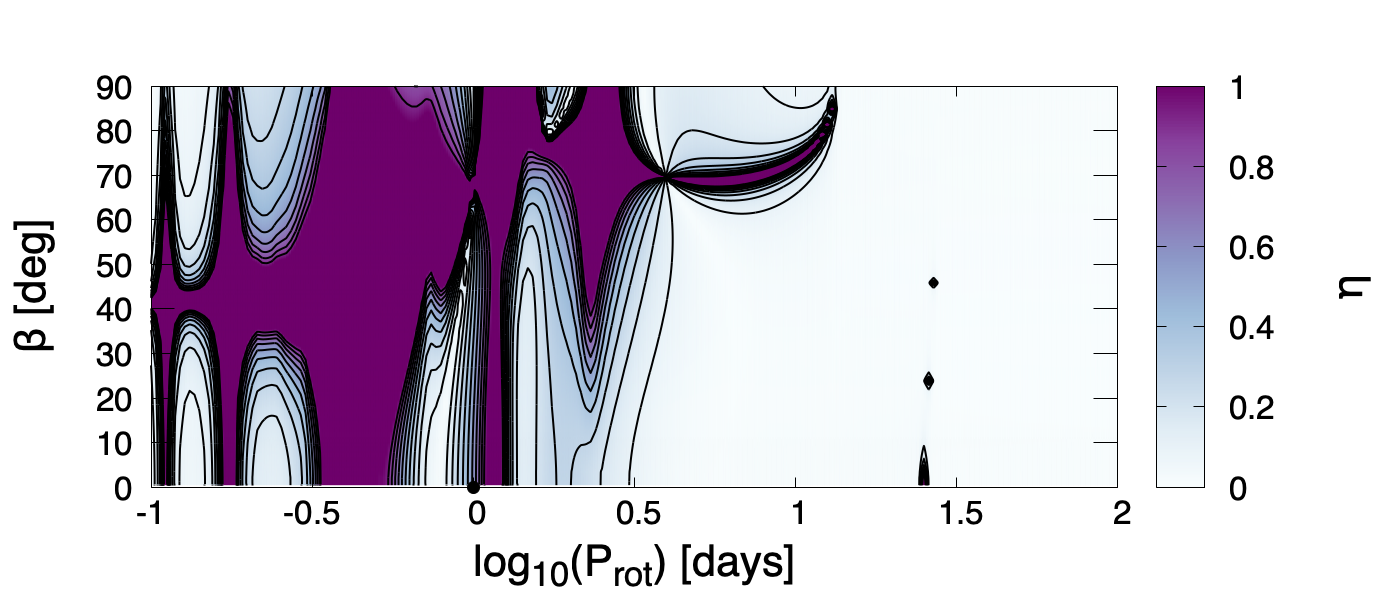} 
   \includegraphics[width=\wkey,trim = 40.5cm 0.1cm 0cm 2.3cm,clip]{auclair-desrotour_fig6f.png} 
      \caption{Variation rate of the planet's obliquity ($\Dt{\angbet}$) as a function of the planet's spin rotation period (horizontal axis) and obliquity (vertical axis) for Earth-sized dry (left panels) and global-ocean (right panels) planets. {\it Top:} Full calculation of the planet's tidal response (\textsc{Full}). {\it Middle:} Calculation using, for all tidal forcing terms, the degree-2 Love number computed in the equatorial plane assuming the coplanar-circular configuration (\textsc{Approx}). {\it Bottom:} Relative difference between the \textsc{Full} and \textsc{Approx} cases, defined by \eq{eta_diff}. The variation rate of the planet's obliquity is computed from the three-dimensional tidal torque using \eq{rate_angbet}. The black dot at $\left( \Prot , \obli \right) = \left( 1 \ {\rm day}, 0^\circ \right)$ designates the actual Earth-Moon system in the coplanar-circular approximation.}
       \label{fig:obli_solid_globoc}%
\end{figure*}

\def\hraisebox{0.15\textwidth}
\def\wpanel{0.40\textwidth}
\def\wkey{0.0495\textwidth} 
\def\hsval{0.2cm}
\begin{figure*}[t]
   \RaggedRight \twolabelsat{5.5cm}{\textsc{Full}}%
                           {13cm}{\textsc{Approx}}\\[0.1cm]
  \raisebox{\hraisebox}[1cm][0pt]{%
   \begin{minipage}{1.4cm}%
  $\obli = 90^\circ$
\end{minipage}}
\centering
   \includegraphics[width=\wpanel,trim = 0.cm 0.cm 9.cm 2.8cm,clip]{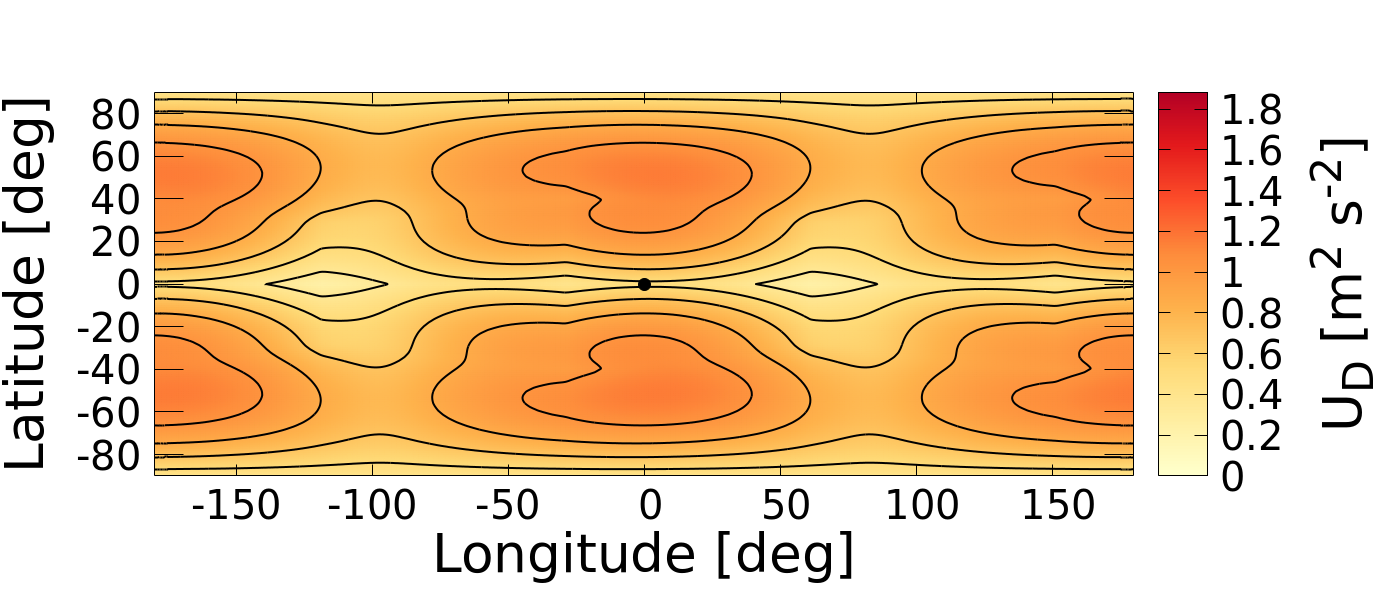} \hspace{\hsval}
   \includegraphics[width=\wpanel,trim = 0.cm 0.cm 9.cm 2.8cm,clip]{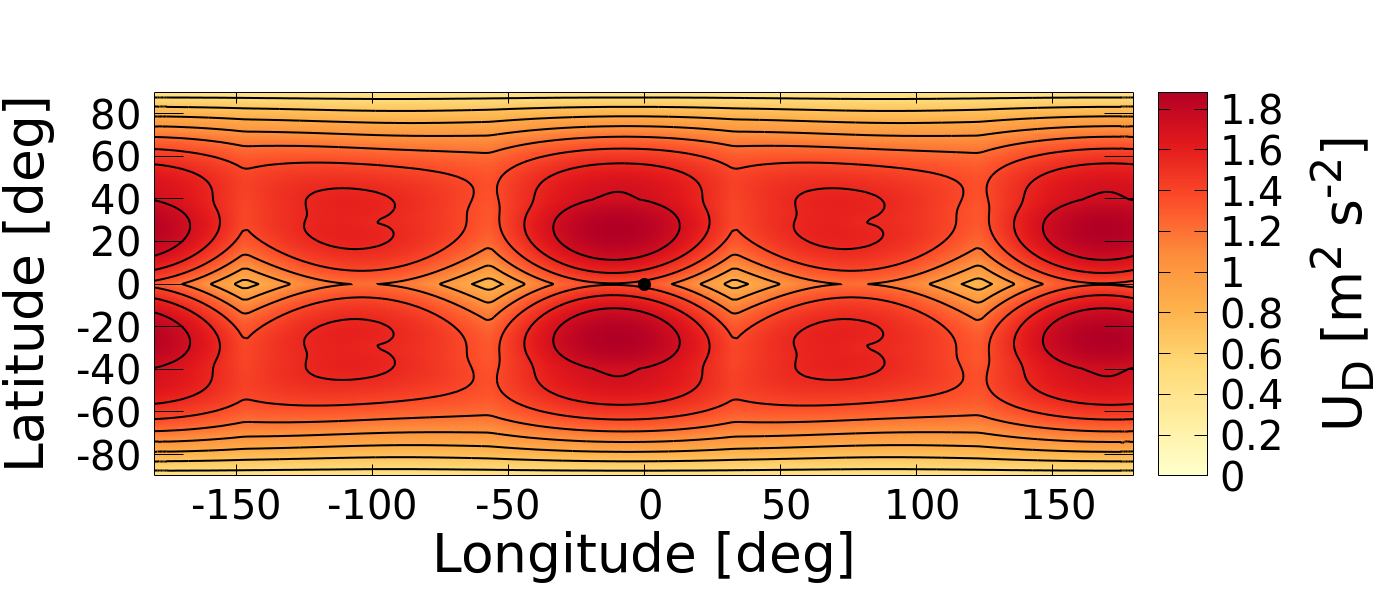} \hspace{0.36cm}  \hspace{\wkey}~ \\
   \raisebox{\hraisebox}[1cm][0pt]{%
   \begin{minipage}{1.4cm}%
   $\obli = 60^\circ$
\end{minipage}}
   \includegraphics[width=\wpanel,trim = 0.cm 0.cm 9.cm 2.8cm,clip]{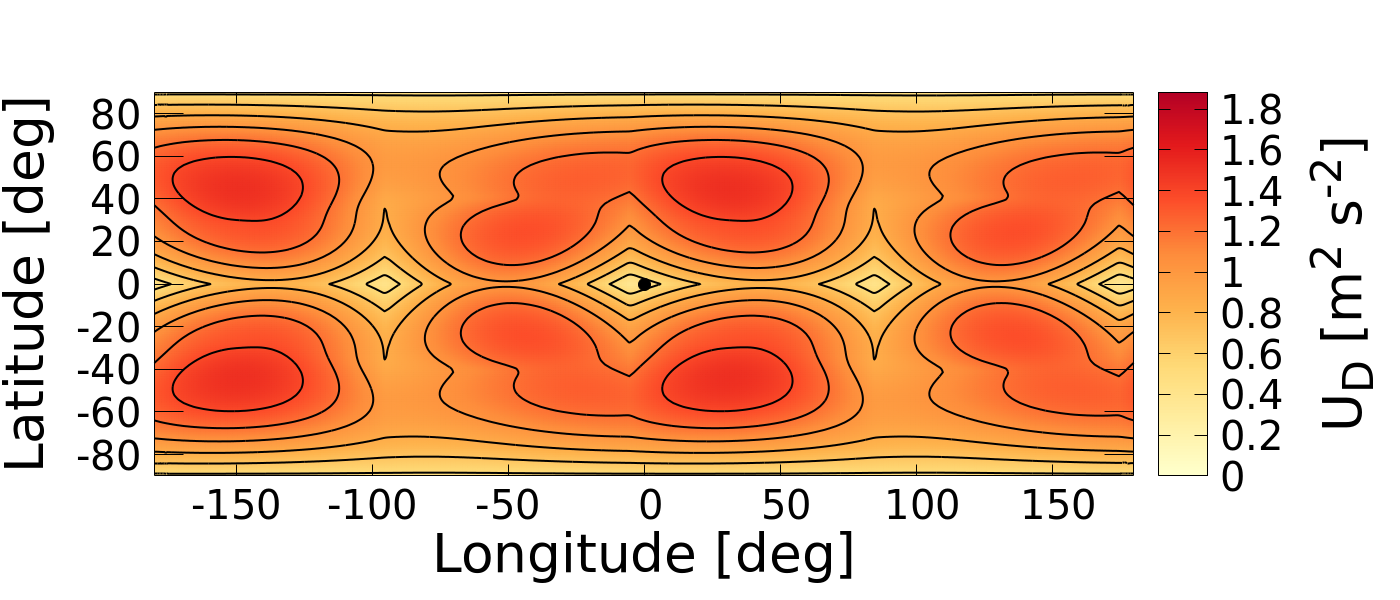} \hspace{\hsval}
   \includegraphics[width=\wpanel,trim = 0.cm 0.cm 9.cm 2.8cm,clip]{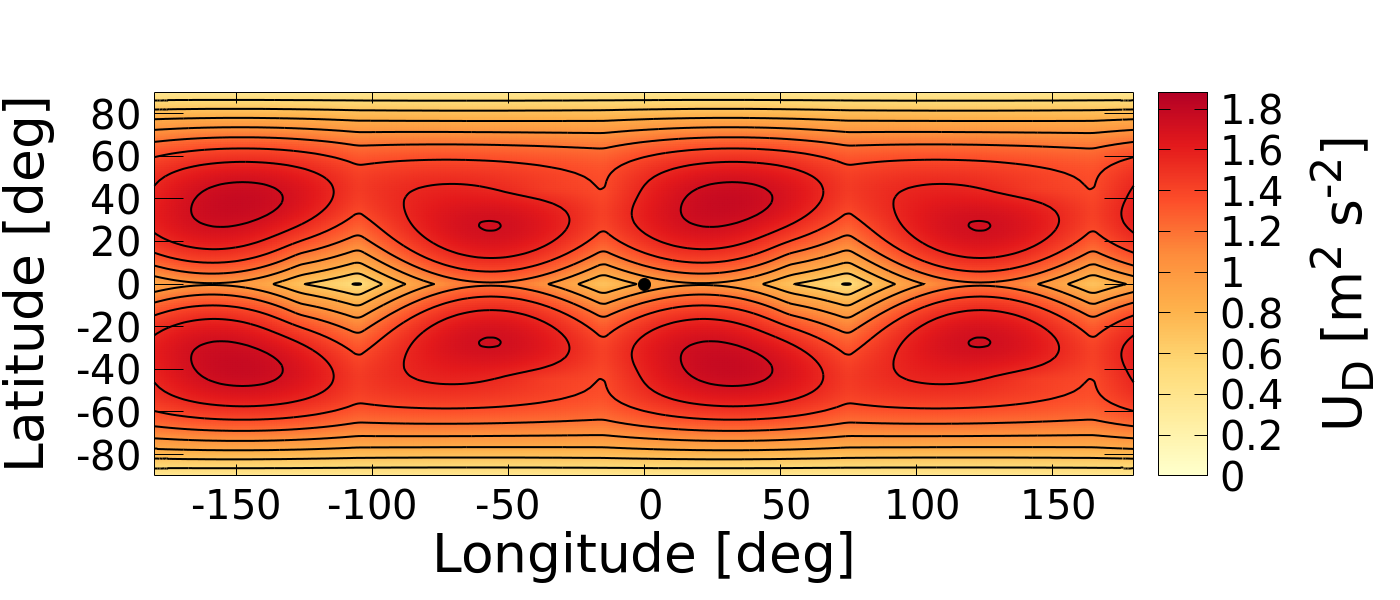} \hspace{0.36cm}  \hspace{\wkey}~ \\
   \raisebox{\hraisebox}[1cm][0pt]{%
   \begin{minipage}{1.4cm}%
   $\obli = 30^\circ$
\end{minipage}}
   \includegraphics[width=\wpanel,trim = 0.cm 0.cm 9.cm 2.8cm,clip]{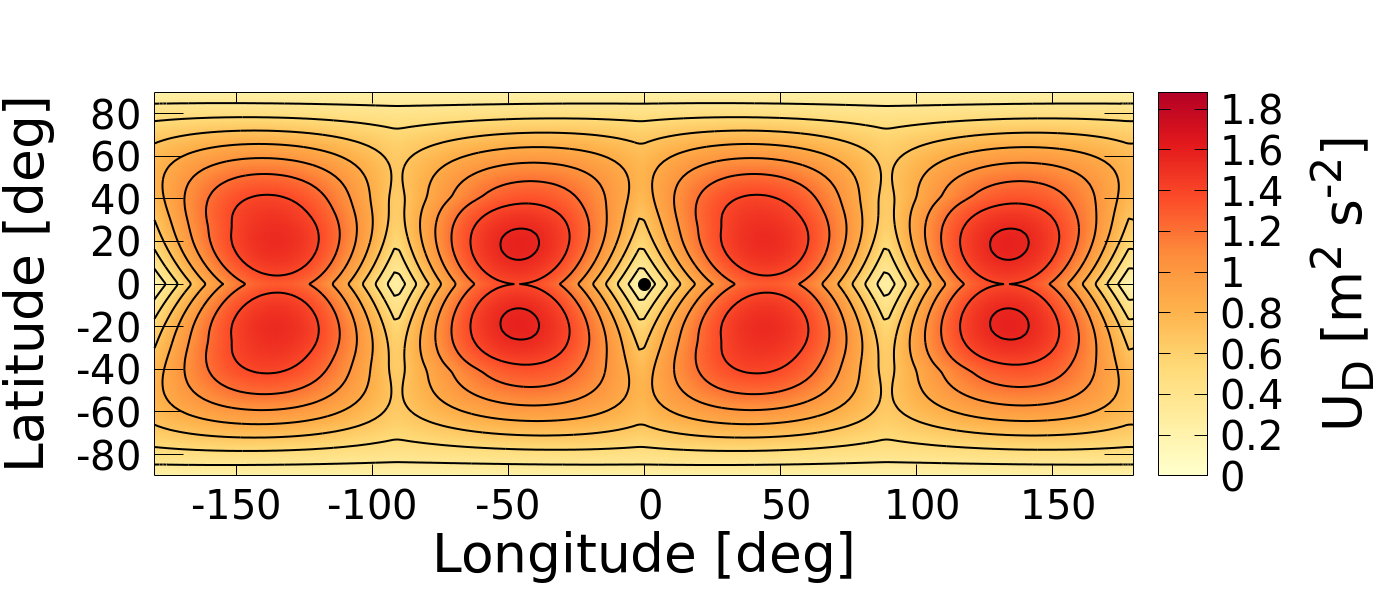} \hspace{\hsval}
   \includegraphics[width=\wpanel,trim = 0.cm 0.cm 9.cm 2.8cm,clip]{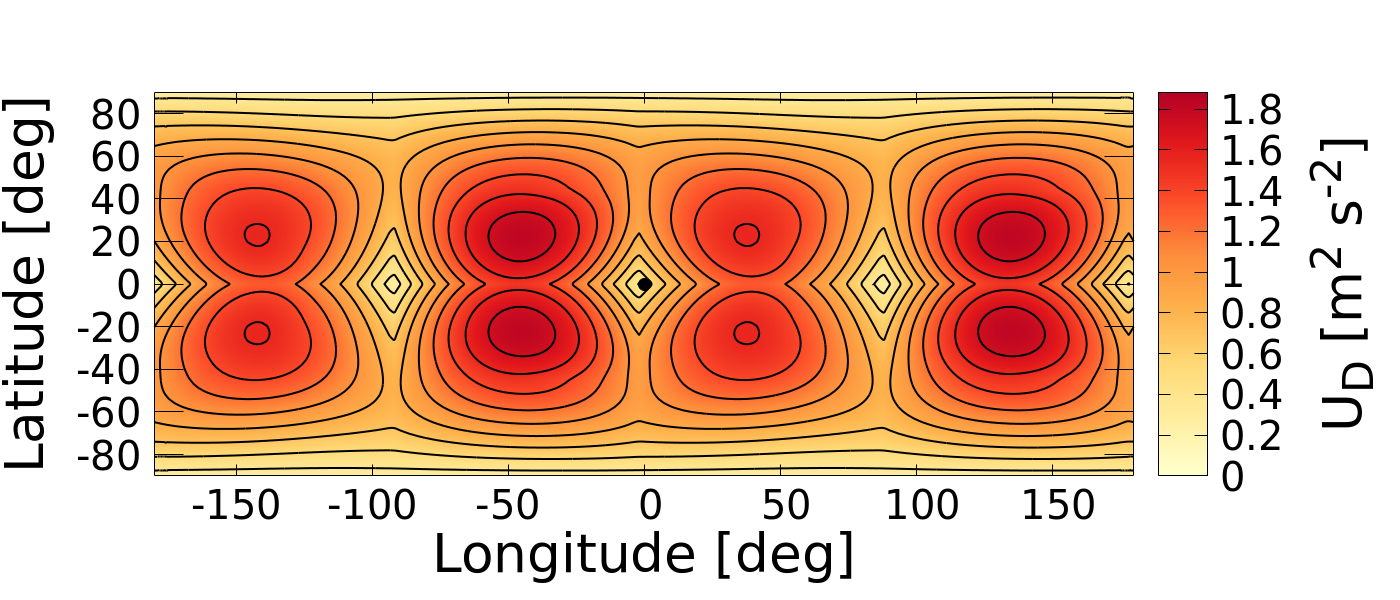} \hspace{0.36cm} \hspace{\wkey}~ \\
   \raisebox{\hraisebox}[1cm][0pt]{%
   \begin{minipage}{1.4cm}%
   $\obli = 0^\circ$
\end{minipage}}
   \includegraphics[width=\wpanel,trim = 0.cm 0.cm 9.cm 2.8cm,clip]{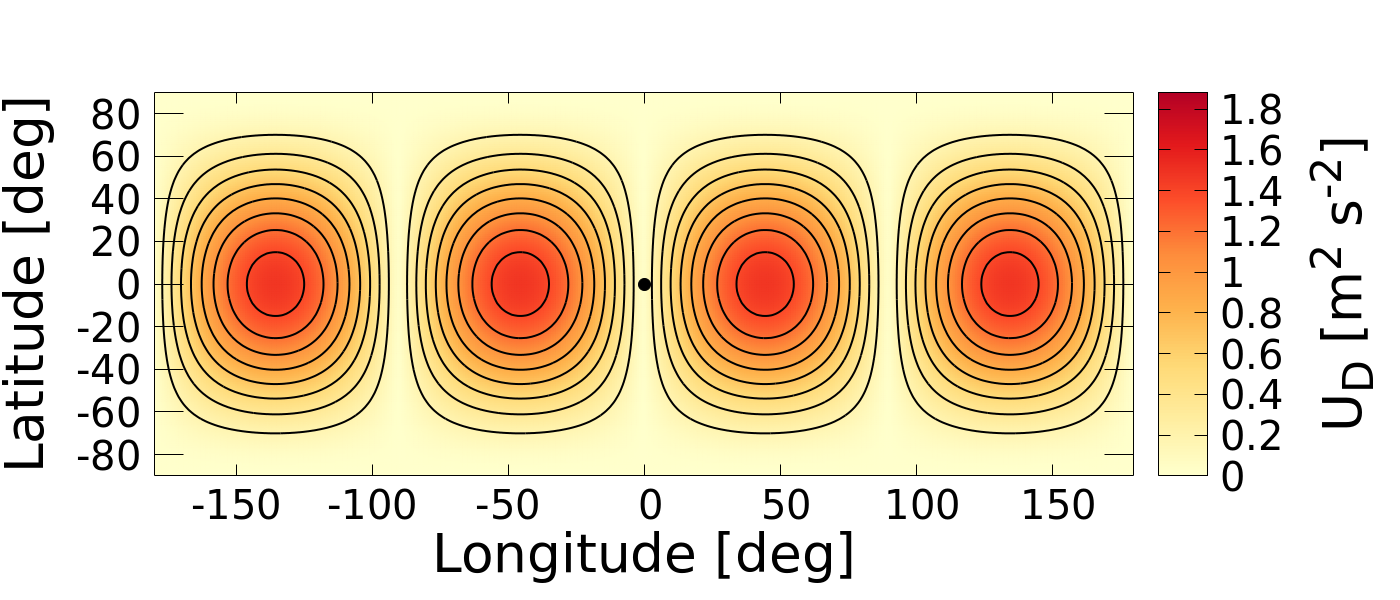} \hspace{\hsval}
   \includegraphics[width=\wpanel,trim = 0.cm 0.cm 9.cm 2.8cm,clip]{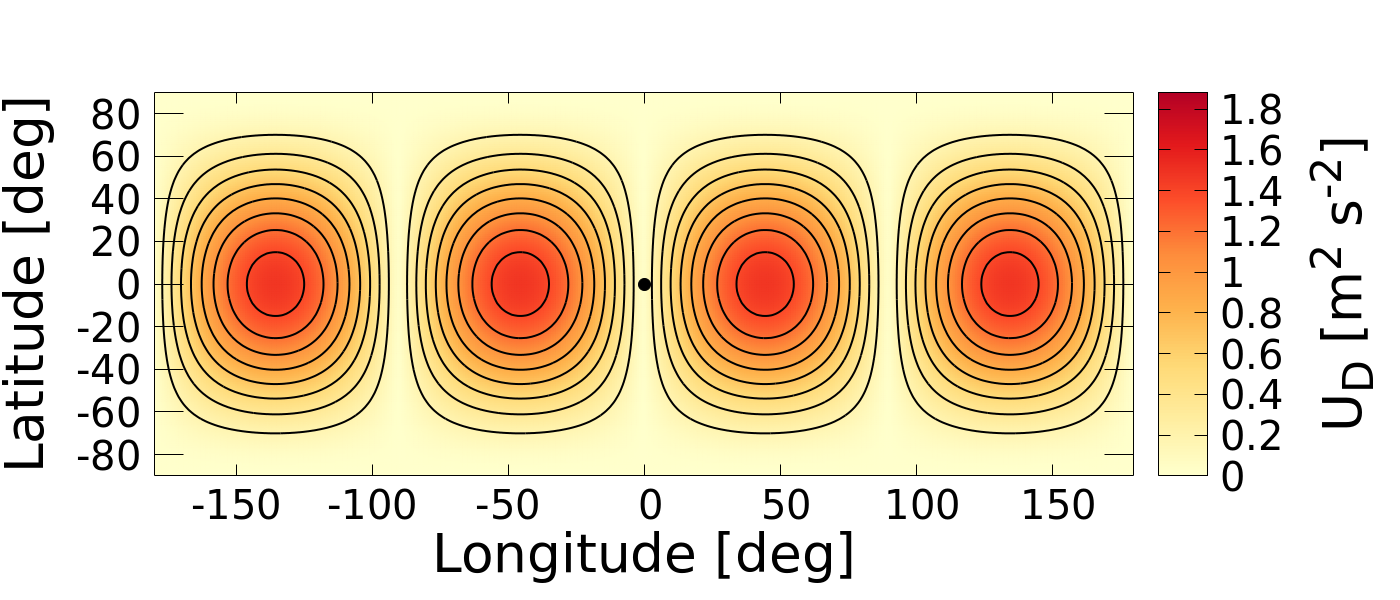} 
   \includegraphics[width=\wkey,trim = 40.5cm 0.cm 3.9cm 2.8cm,clip]{auclair-desrotour_fig7h.png} 
   \raisebox{0.9cm}{\rotatebox{90}{$ \Urespub \  {\rm [m^2 \ s^{-2}]} $}}
      \caption{Gravitational potential induced by the tidal response of an ocean planet with rigid solid regions in the system of coordinates rotating with the perturber for $\Prot = 33$~hr and obliquity values ranging between $0^\circ$ and $90^\circ$. {\it Left:} Maximum amplitude of the tidal potential obtained from the full calculation (\textsc{Full}). {\it Right:} Maximum amplitude of the tidal potential obtained with the standard approximation based on the equatorial degree-2 Love number (\textsc{Approx}). Amplitudes are plotted as functions of longitude, $\lonrot$ (horizontal axis), and latitude, $\colrot^\prime = 90^\circ - \colrot$ (vertical axis). Red areas indicate large amplitudes, and yellow areas small amplitudes. The black dot at $\left( \colrot^\prime, \lonrot \right) = \left( 0^\circ , 0^\circ \right)$ designates the sub-satellite point. }
       \label{fig:map_prot_33h}%
\end{figure*}

\begin{figure*}[t]
   \RaggedRight \twolabelsat{5.5cm}{\textsc{Full}}%
                           {13cm}{\textsc{Approx}}\\[0.1cm]
  \raisebox{\hraisebox}[1cm][0pt]{%
   \begin{minipage}{1.4cm}%
  $\obli = 90^\circ$
\end{minipage}}
\centering
   \includegraphics[width=\wpanel,trim = 0.cm 0.cm 9.cm 2.8cm,clip]{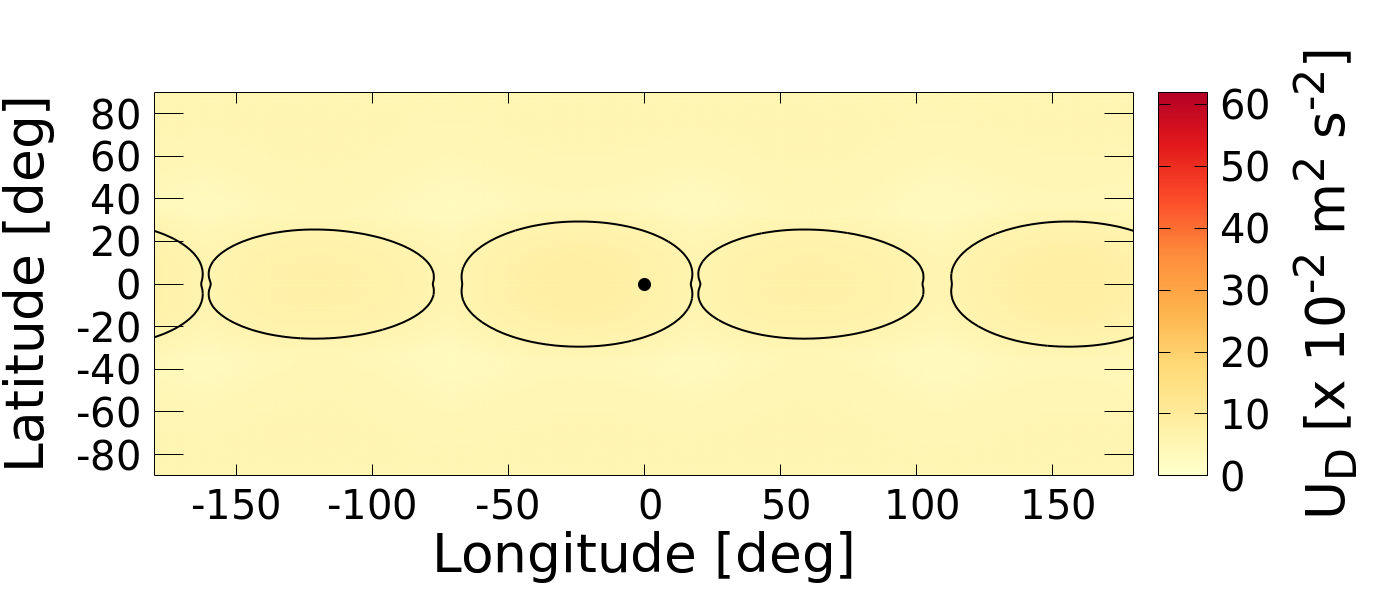} \hspace{\hsval}
   \includegraphics[width=\wpanel,trim = 0.cm 0.cm 9.cm 2.8cm,clip]{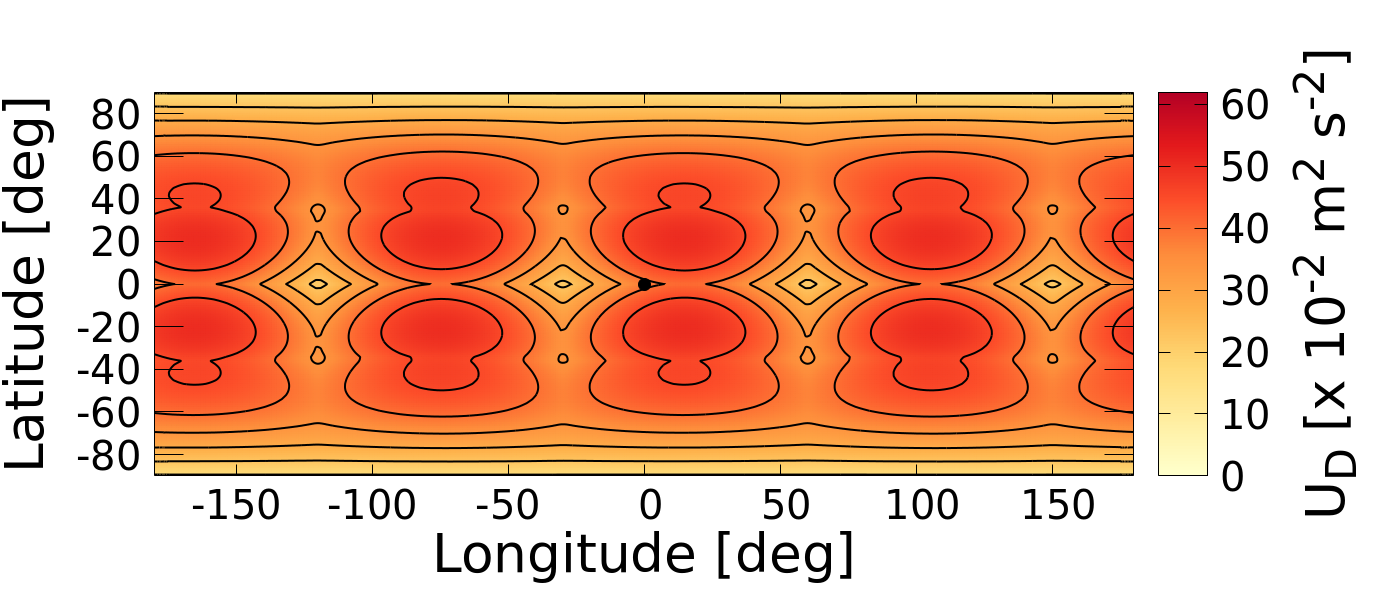} \hspace{0.36cm}  \hspace{\wkey}~ \\
   \raisebox{\hraisebox}[1cm][0pt]{%
   \begin{minipage}{1.4cm}%
   $\obli = 60^\circ$
\end{minipage}}
   \includegraphics[width=\wpanel,trim = 0.cm 0.cm 9.cm 2.8cm,clip]{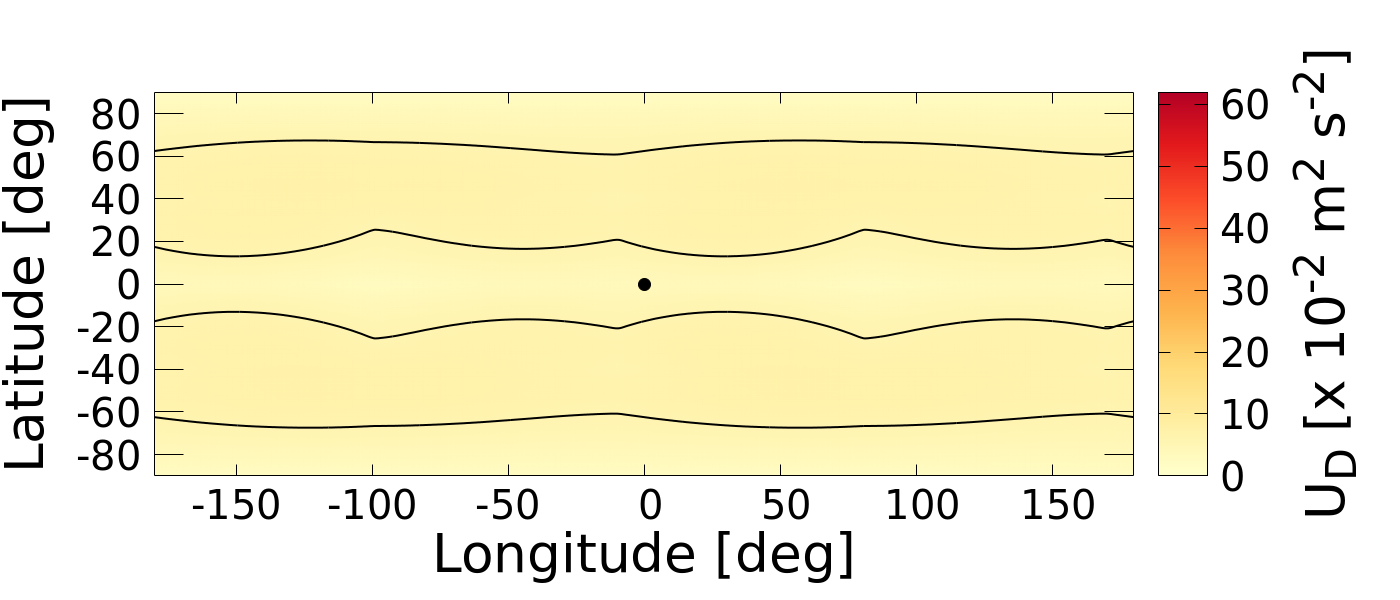} \hspace{\hsval}
   \includegraphics[width=\wpanel,trim = 0.cm 0.cm 9.cm 2.8cm,clip]{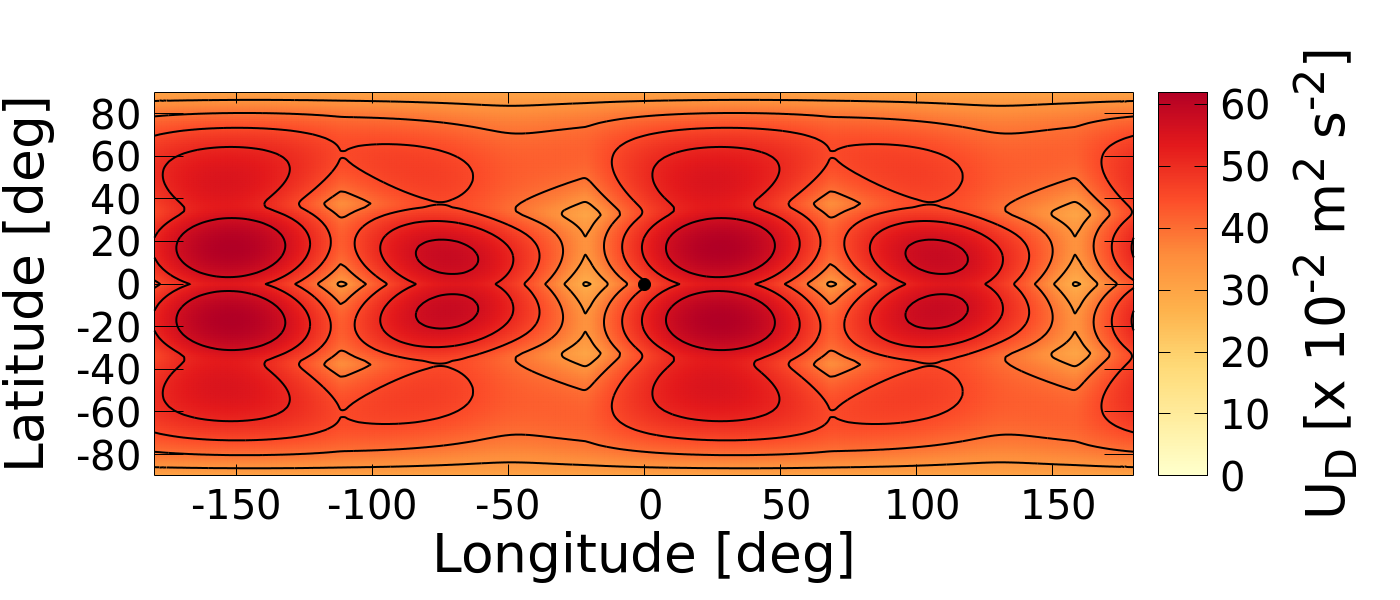} \hspace{0.36cm}  \hspace{\wkey}~ \\
   \raisebox{\hraisebox}[1cm][0pt]{%
   \begin{minipage}{1.4cm}%
   $\obli = 30^\circ$
\end{minipage}}
   \includegraphics[width=\wpanel,trim = 0.cm 0.cm 9.cm 2.8cm,clip]{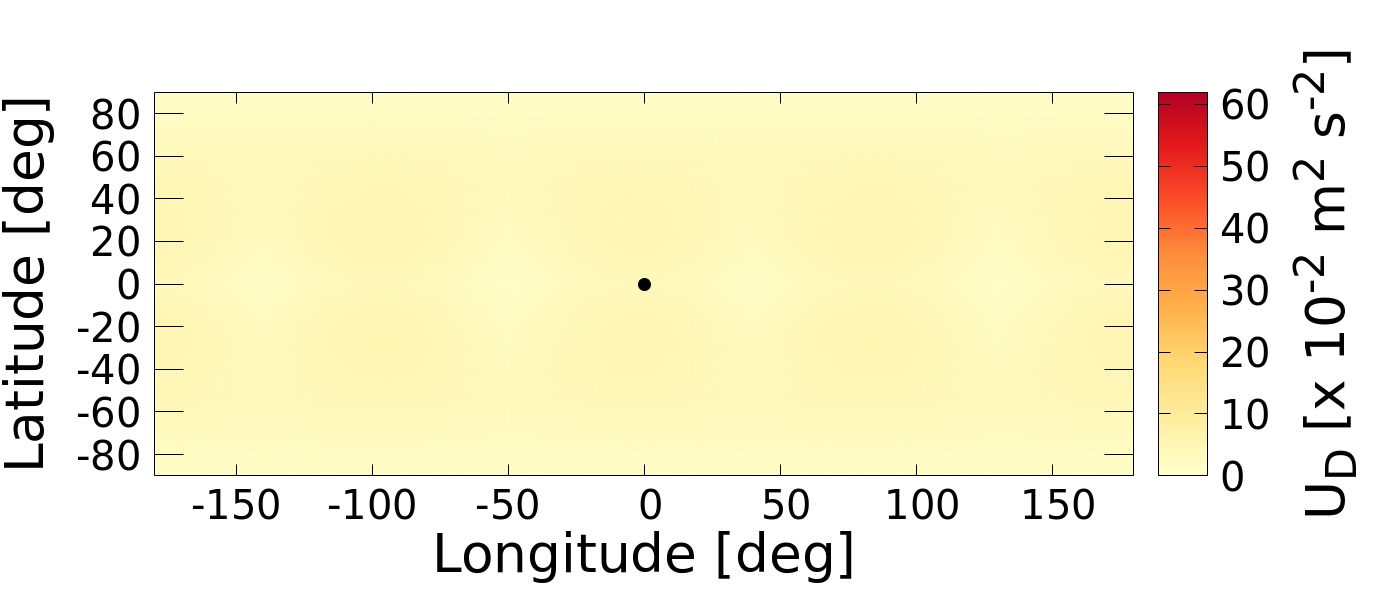}  \hspace{\hsval}
   \includegraphics[width=\wpanel,trim = 0.cm 0.cm 9.cm 2.8cm,clip]{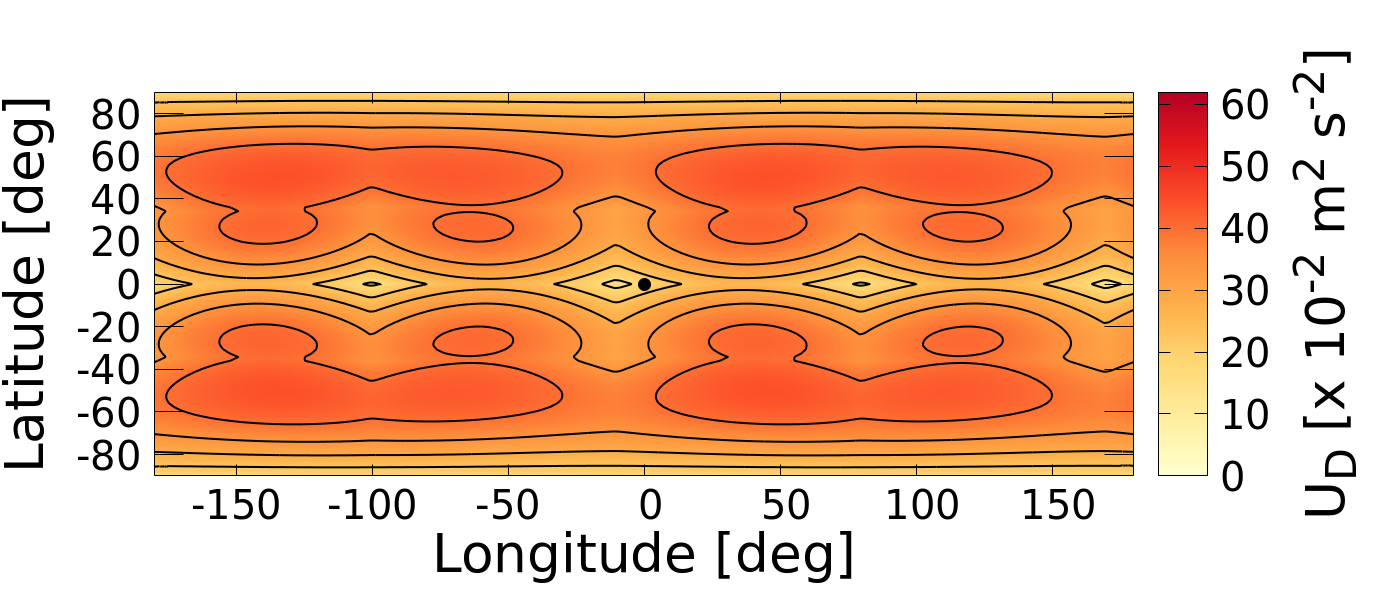} \hspace{0.36cm}  \hspace{\wkey}~ \\
   \raisebox{\hraisebox}[1cm][0pt]{%
   \begin{minipage}{1.4cm}%
   $\obli = 0^\circ$
\end{minipage}}
   \includegraphics[width=\wpanel,trim = 0.cm 0.cm 9.cm 2.8cm,clip]{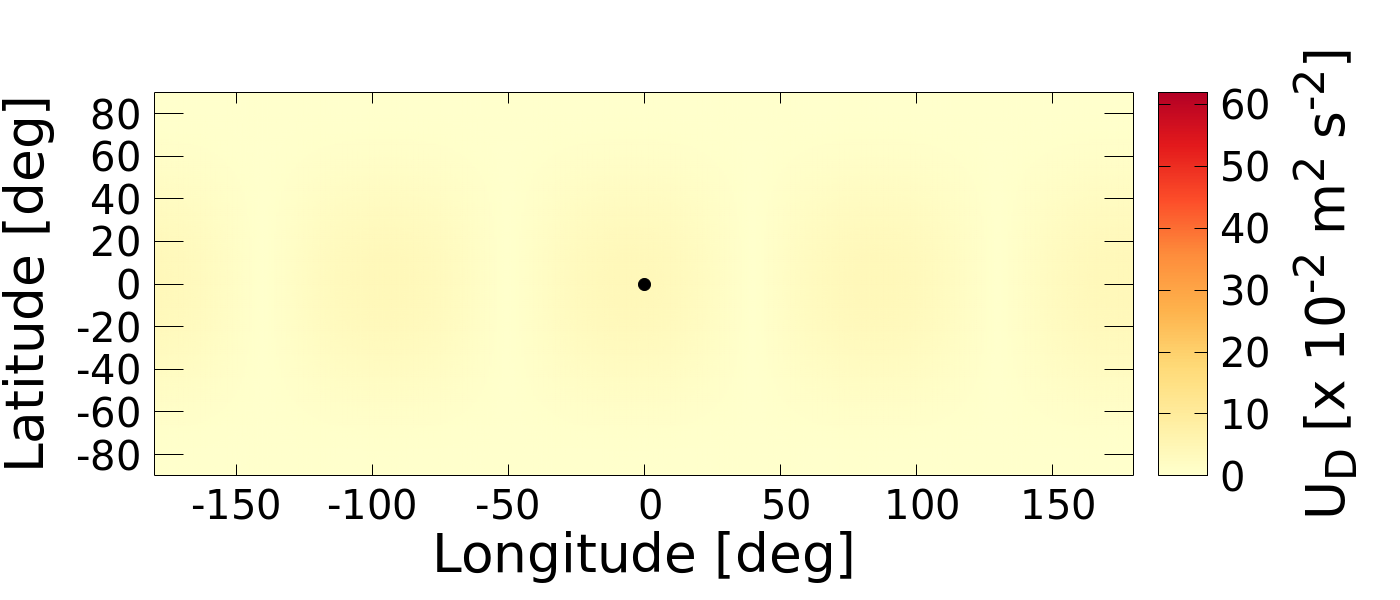} \hspace{\hsval}
   \includegraphics[width=\wpanel,trim = 0.cm 0.cm 9.cm 2.8cm,clip]{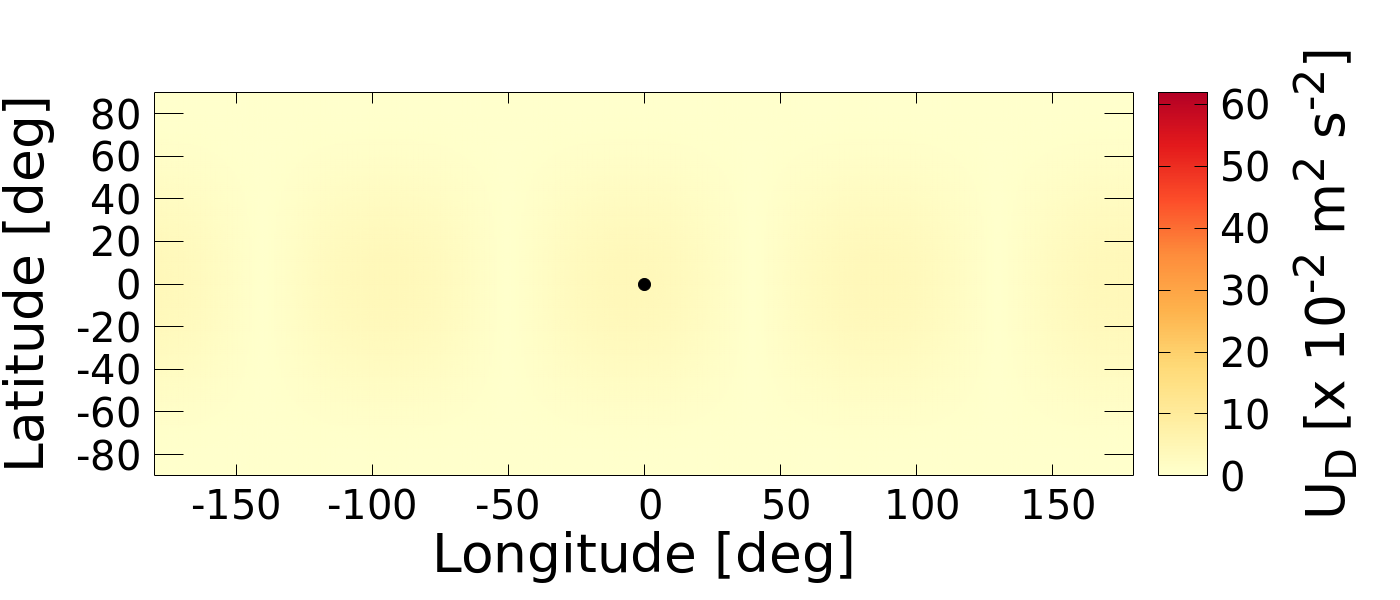} 
   \includegraphics[width=\wkey,trim = 40.5cm 0.cm 3.9cm 2.8cm,clip]{auclair-desrotour_fig8h.png} 
   \raisebox{0.1cm}{\rotatebox{90}{$ \Urespub \ [ \times 10^{-2} \ {\rm m^2 \ s^{-2}} ] $}}
      \caption{Gravitational potential induced by the tidal response of an ocean planet with rigid solid regions in the system of coordinates rotating with the perturber for $\Prot = 10$~hr and obliquity values ranging between $0^\circ$ and $90^\circ$. {\it Left:} Maximum amplitude of the tidal potential obtained from the full calculation (\textsc{Full}). {\it Right:}  Maximum amplitude of the tidal potential obtained with the standard approximation based on the equatorial degree-2 Love number (\textsc{Approx}). Amplitudes are plotted as functions of longitude, $\lonrot$ (horizontal axis), and latitude, $\colrot^\prime = 90^\circ - \colrot$ (vertical axis). Red areas indicate large amplitudes, and yellow areas small amplitudes. The black dot at $\left( \colrot^\prime, \lonrot \right) = \left( 0^\circ , 0^\circ \right)$ designates the sub-satellite point. }
       \label{fig:map_prot_10h}%
\end{figure*}

\subsection{Sensitivity of tidal dissipation to planet's spin rotation and obliquity}

In our calculations, we assume that the perturber causing the tidal forcing is a Lunar-mass satellite in a circular orbit around the planet ($\eccpert = 0$), with the same semi-major axis as the current Earth-Moon system, and a fixed orbital period $\Porb$\mynom[S]{$\Porb$}{Satellite's orbital period}. The planet's rotation period, $\Prot$\mynom[S]{$\Prot$}{Planet's rotation period}, is varied from $0.1$ to $ 100$~days, and its obliquity, $\obli$, from $0^\circ$ to~$90^\circ$. We investigate two configurations. In the first configuration ('\textsc{Dry}'), the ocean is neglected, and only the solid tide is considered. In the second configuration ('\textsc{Global ocean}'), the full tidal response, including the ocean, is evaluated. For each configuration, the calculations are performed in two scenarios: (i) using the full formalism introduced in \sect{sec:torque_power}, where the planet's tidal response is computed comprehensively and self-consistently ('\textsc{Full}'), and (ii) under the classical isotropic approximation, where all tidal components are derived from the equatorial degree-2 Love number defined in the circular-coplanar configuration for the semidiurnal tide ('\textsc{Approx}'). We compute the tidal solutions using the numerical tools of the \texttt{TRIP}  software \citep[][]{GL2011}\footnote{\url{https://www.imcce.fr/trip/}}. 

In the \textsc{Approx} case, the equatorial degree-2 Love number is calculated for every tidal frequency of the forcing potential in the planet's rotating frame. Each tidal frequency, $\ftide$, is treated as a semidiurnal tidal frequency, $\ftide = 2 \left( \spinrate - \norb \right)$, where the fictitious orbital frequency is defined as $\norb = \spinrate - \ftide/2$. The corresponding tidal response is then evaluated using the spatial distribution for the semidiurnal quadrupolar tidal potential, represented by the spherical harmonic of degree 2 and order 2 (denoted by $ \Ylm{2}{2}$). From the resulting solution, the degree-2 Love number is derived and used to compute the deformation potential for the relevant tidal component. This potential is simply the product of the degree-2 Love number and the tide-raising potential, following the theory of bodily tides. It is important to note that the Love number depends solely on $\ftide$ in the dry configuration due to the spherical isotropy of the solid part. However, in the global ocean configuration, it also depends on the planet's spin period because of the influence of Coriolis forces, as described earlier. 

Figures~\ref{fig:pdiss_solid_globoc}, \ref{fig:spinvel_solid_globoc} and \ref{fig:obli_solid_globoc} present the results for tidally dissipated power ($\powerdiss$), the variation rate of the planet's spin angular velocity ($\Dt{\spinrate}$), and the variation rate of the planet's obliquity ($\Dt{\obli}$), respectively. Additionally, the relative difference between the two cases (\textsc{Full} and \textsc{Approx}), denoted by $\eta$\mynom[S]{$\eta$}{Relative difference between the $\textsc{Full}$ and \textsc{Approx} cases for a quantity}, is shown for each configuration and for the three quantities (bottom panels). This difference is normalised by the values obtained in the full calculation. For a given tidal quantity, $f$, the relative difference is defined mathematically as 
\begin{equation}
\label{eta_diff}
\eta = \abs{\frac{f_{\rm approx} - f_{\rm full}}{f_{\rm full}}},
\end{equation} 
where $f_{\rm full}$ and $f_{\rm approx} $ represent the values obtained in the \textsc{Full} an \textsc{Approx} cases, respectively.

We first examine the solid-body configuration (\figsto{fig:pdiss_solid_globoc}{fig:obli_solid_globoc}, left panels). In this scenario, the three quantities -- $\powerdiss$, $\Dt{\spinrate}$ and $\Dt{\obli}$ -- vary smoothly with changes in the planet's rotation period and obliquity. In the coplanar case ($\obli = 0^\circ$), both the tidally dissipated power and the variation rate of spin velocity approach zero as the planet crosses the $1{:}1$ spin-orbit resonance ($\Prot = \Porb$). As obliquity increases, obliquity tides become more dominant, shifting the equilibrium states. Due to the spherical isotropy of the solid part, the results from the \textsc{Full} and \textsc{Approx} calculations are identical, leading to a relative difference of zero between the two approaches. 

At higher obliquities, obliquity tides may induce an additional equilibrium state corresponding to the \rev{$2{:}1$ spin-obit resonance ($\Prot =  \Porb/2$)}, which is particularly visible in \fig{fig:spinvel_solid_globoc}. In this configuration, the frequency of the quadrupolar obliquity component of the tidal potential \rev{($\ftide =   \spinrate - 2 \norb$, and $\mm=1$)} becomes zero. Both the $1{:}1$ and $2{:}1$ resonances are indicated by dashed magenta lines in \fig{fig:pdiss_solid_globoc} (top right panel). It is worth noting that the solid part's ability to dissipate energy via viscous friction is maximised at very low tidal frequencies, given that typical Maxwell timescales for rocky planets reach several centuries \citep[][]{Peltier1974,KS1990,Bolmont2020}. Consequently, any change in the sign of a tidal frequency results in a sharp variation in the associated tidal component. 

\begin{table}
\centering
\caption{\label{tab:period_res} Features of the global ocean modes that are resonantly excited by the semidiurnal tidal forces.}
\begin{tabular}{rrrr} 
 \hline 
 \hline 
$\nn$ & $\Houghval{\nn}{2}{1}$ & $\periodi{\nn}$~(hours) & $\log_{10} \left( \periodi{\nn}  \right)$~(days)   \\ 
 \hline \\[-0.3cm]
$2$ & $      11.129$ & $      33.653$ & $       0.147$ \\ 
$4$ & $      41.333$ & $      17.462$ & $      -0.138$ \\ 
$6$ & $      91.060$ & $      11.765$ & $      -0.310$ \\ 
$8$ & $     160.424$ & $       8.864$ & $      -0.433$ \\ 
$10$ & $     249.469$ & $       7.108$ & $      -0.528$ \\ 
$12$ & $     358.213$ & $       5.932$ & $      -0.607$ \\ 
$14$ & $     486.671$ & $       5.089$ & $      -0.674$ \\ 
$16$ & $     634.847$ & $       4.456$ & $      -0.731$ \\ 
$18$ & $     802.747$ & $       3.962$ & $      -0.782$ \\ 
$20$ & $     990.374$ & $       3.567$ & $      -0.828$ \\ 
\hline 
 \end{tabular}
 \tablefoot{From left to right are provided the degrees of the ocean modes, their associated eigenvalues, and their resonant rotation periods, given in both hours and days (in logarithmic scale).}
\end{table}

When the global ocean is included in the tidal response, long-wavelength surface gravity modes can be resonantly excited by tidal forces. This greatly increases the sensitivity of both dissipated power and tidal torque to the planet's rotation period. The resonant peaks of these modes occur at specific frequencies, given by \citep[e.g.,][]{LH1968,LS1997,Auclair2023} 
\begin{equation}
\focn^{\mm,\spinpar} = \frac{\sqrt{\Houghval{\nn}{\mm}{\spinpar} \ggravi \Hoc}}{\Rpla}.
\end{equation}
In this equation, $\spinpar \define 2 \spinrate / \ftide$\mynom[S]{$\spinpar$}{Spin parameter} is the spin parameter, and $\Houghval{\nn}{\mm}{\spinpar} $\mynom[S]{$\Houghval{\nn}{\mm}{\spinpar} $}{Eigenvalues associated with Hough functions} represents the eigenvalues associated with the Hough functions that describe the horizontal structure of the forced oceanic eigenmodes. Hough functions are essentially spherical harmonics modified by the planet's rotation \citep[e.g.,][]{LH1968,LC1969,LS1997,Wang2016}. These functions arise from the LTEs and were first introduced by Hough in his pioneering work \citep[][]{Hough1898}. 

The most prominent peaks observed in \figsto{fig:pdiss_solid_globoc}{fig:obli_solid_globoc} are caused by the semidiurnal tide (i.e. $\ftide =  2 \left( \spinrate - \norb \right)$ with $\mm = 2$). These peaks appear as long as the tidal frequency exceeds the eigenfrequency corresponding to the longest-wavelength gravity mode ($\nn=2$). In the regime where oceanic modes become resonant, $\norb \ll \spinrate$, so the semidiurnal tidal frequency simplifies to $\ftide \approx 2 \spinrate$, implying that $\spinpar \approx 1$. Consequently, the spin rotation periods at which resonances occur can be expressed as\mynom[S]{$\periodi{\nn}$}{Rotation periods at which oceanic resonances occur}
\begin{equation}
\label{periodn}
\periodi{\nn} \approx \frac{4\pi \Rpla}{\sqrt{\Houghval{\nn}{2}{1} \ggravi \Hoc}},
\end{equation}
with $\nn = 2, 4, 6, \ldots, + \infty$. The eigenvalues and corresponding rotation periods for the main resonances are listed in Table~\ref{tab:period_res}, and these resonances are indicated by dashed green lines in \fig{fig:pdiss_solid_globoc} (top right panel). 

In the resonant regime, the oceanic dynamical tide greatly enhances the energy dissipated through tidal forces, as well as the rates of change of spin angular velocity and obliquity. The amplification factor associated with a resonance is directly related to the efficiency of tidal flow drag against the ocean floor. Typically, the maximum tidal torque during a resonance crossing scales as $\scale 1/\fdrag$ in the adopted linear model \citep[see e.g.,][Eqs.~(55) and (58)]{Auclair2019}. Using the drag frequency inferred from Earth's present tidal energy dissipation, the torque can be resonantly amplified by approximately an order of magnitude \citep[][]{Webb1980,Auclair2023}. The strength of these resonances also depend on the overlap coefficients between Hough functions and spherical harmonics, which vary with the planet's obliquity and the orbital inclination of the perturbing body. Furthermore, the geometry of the ocean basin influences the coupling between the excited oceanic modes and the tidal forcing potential. Continental coastlines introduce more resonant peaks compared to the global ocean scenario studied here, as discussed in \cite{Auclair2023}. The interactions between tidal flows and coastlines also enable the resonant excitation of modes with lower eigenfrequencies.

Next, we consider the divergences between the \textsc{Full} and \textsc{Approx} cases when oceanic tides are included. These discrepancies are evident in the plots showing the evolution of tidally dissipated power and the rate of change of spin angular velocity (\figs{fig:pdiss_solid_globoc}{fig:spinvel_solid_globoc}, top and middle right panels). In the \textsc{Full} case, the peaks associated with oceanic equatorial gravity modes tend to diminish as the obliquity increases,  consistent with the decreasing coupling between these modes and the tidal forcing. \rev{Gravity modes are progressively replaced by non-resonant polar waves.} 

Conversely, in the \textsc{Approx} case, new peaks emerge as the obliquity increases. Notably, the two solutions diverge significantly for a planet with a 10-hour rotation period ($\log \left( \Prot \right) = -0.380 $ where $\Prot$ is in days). These new peaks are artefacts arising from the assumption that the planet's tidal response is independent of its orientation relative to the perturber's orbit. This assumption artificially reproduces resonances linked to equatorial gravity modes in scenarios where these modes are, in reality, subdominant. As a result, the \textsc{Approx} model can overestimate the tidal torque on the planet by several orders of magnitude, leading to regions of the parameter space with substantial errors ($\eta \geq 1$), as shown by \figs{fig:pdiss_solid_globoc}{fig:spinvel_solid_globoc} (bottom panels). It should be noted that in these regions, $\eta$ may exceed~1 by a considerable margin, as it is directly related to the resonant amplification factor, which is a function of $\fdrag$. The error introduced by this approximation, particularly regarding the degree-2 Love number, is even more pronounced for the rate of change of obliquity. This is evident from \fig{fig:obli_solid_globoc} (bottom right panel), where the inaccurate predictions of the \textsc{Approx} model extend over a broad area of the parameter space, including configurations with near-zero obliquity.

\subsection{Spatial distributions of tidal self-attraction variations}
\label{ssec:mapping_self_attraction}

To visualise how the solution is altered in the \textsc{Approx} case, we consider the gravitational potential induced by the ocean's elevation in the rigid-body limit, where solid regions are non-deformable. This potential is generally expressed by \eq{Uresp}. However, terms where $\qq \neq \pp$ can be discarded since the frequencies associated with these terms ($\ftidefixi{\kk,\qq,\pp} $) differ from those of the tide-raising potential ($\ftidefixi{\kk} $), as given in \eq{Utide}, preventing their coupling with the forcing components on average. Therefore, only components of $\Uresp$ with $\qq = \pp$ are considered. Furthermore, the Galilean frame of reference is not the most suitable for plotting $\Uresp$, as tidal deformations follow the perturber, which moves in this frame. Instead, we plot the deformation tidal potential in a frame that rotates with the perturber, with the planet's centre of gravity as its origin. This rotating frame uses spherical coordinates $\left( \rrrot, \colrot, \lonrot \right)$, where $\rrrot=\rr$, and $\colrot=0$ corresponds to the direction of the angular momentum vector of the satellite's orbit, $\angmompertv$, as introduced in \eq{angmompertv}. In this system, $\left( \colrot, \lonrot \right) = \left( 90^\circ, 0^\circ \right)$ marks the sub-satellite point. It is important to note that the coordinate notations used here to describe the frame following the perturber are the same as those for the planet-rotating frame in \sect{ssec:gravpot_distorted}, so the two systems should not be confused.

The steps for transitioning from $\left( \rr, \col , \lon \right) $ to $\left( \rrrot, \colrot, \lonrot \right)$ are detailed in \append{app:gravpot_perturber_frame}. This yields the expression 
\begin{equation}
\label{Uresp_framepert}
\Uresp \left( \rrrot, \colrot, \lonrot, \time \right) = \Re \left\{ \sum_{\qq = - \infty}^{+ \infty} \Urespnrotsigi{\qq} \left( \rrrot, \colrot, \lonrot \right) \expo{\inumber \qq \npert \time}   \right\},
\end{equation}
where the spatial functions $\Urespnrotsigi{\qq}$ are given by
\begin{equation}
\Urespnrotsigi{\qq}  \left( \rrrot , \colrot, \lonrot \right) = \sum_{\llat=0}^{+ \infty} \sum_{\mm = -\llat}^{\llat} \Uplmrotsigj{\llat}{\qq }{\mm} \left( \frac{\rrrot}{\Rpla} \right)^{-\left(\llat+ 1 \right)} \Ylm{\llat}{\mm} \left( \colrot , \lonrot \right).
\end{equation}
In this equation, the complex coefficients $\Uplmrotsigj{\llat}{\qq }{\mm}$ are expressed using the weighting coefficients introduced in \eq{Urespnsig_sph} as
\begin{equation}
\Uplmrotsigj{\llat}{\qq }{\mm} = \sum_{\kk=0}^{+\infty} \sum_{\jj = -\llat}^{+\llat}   \Hansen{\qq - \kk}{0}{\mm} \left( \eccpert \right) \conj{\wignerD{\jj}{\mm}{\llat}} \left( \angapert , \angbpert , \anggpert  \right) \left( \sum_{\pp=-\llat}^{+\llat}  \Uplmsigj{\llat}{\kk,\pp,\pp}{\jj} \right).
\end{equation}
We remark that the dominant component in \eq{Uresp_framepert} is static ($\qq=0$), while the other components travel either eastwards ($\qq<0$) or westwards ($\qq>0$) relative to the sub-satellite point. At the planet's surface ($\rrrot=\Rpla$), the amplitude of $\Uresp$ is upper-bounded (superscript $^{\rm UB}$) by\mynom[S]{$\Urespub$}{Upper bound of the perturbed potential at planet's surface}
\begin{equation}
\label{Urespub}
\Urespub \left( \colrot, \lonrot \right) = \abs{\Re \left\{ \Urespnrotsigi{0} \left( \Rpla, \colrot, \lonrot \right)   \right\}} + \sum_{\qq \neq 0} \abs{\Urespnrotsigi{\qq} \left( \Rpla, \colrot, \lonrot \right)} .
\end{equation}

We calculate $\Urespub$ using both the \textsc{Full} and \textsc{Approx} models, considering obliquities ranging between $0^\circ$ and $90^\circ$ and two spin rotation periods: $\Prot = 33$~hr and $\Prot = 10$~hr. The first period corresponds to the rotation rate at which the primary oceanic mode is resonantly excited by the semidiurnal tide (see Table~\ref{tab:period_res}, $\nn=2$). The second falls within the parameter space where the \textsc{Full} and \textsc{Approx} models significantly diverge. The remaining physical parameters are set to the values used for generating the solutions in \figsto{fig:pdiss_solid_globoc}{fig:obli_solid_globoc}, as detailed in Table~\ref{tab:param_refcase}. The upper bound of the tidal potential, as defined in \eq{Urespub}, is plotted against the coordinates $\left( \colrot , \lonrot \right)$ in \fig{fig:map_prot_33h} for $\Prot = 33$~hr, and in \fig{fig:map_prot_10h} for $\Prot=10$~hr. We note that the tidal potential is truncated at $\llat = 2$ (quadrupolar approximation), as the contribution from higher-degree terms to tidal dissipation is negligible when $\Rpla \ll \smaxispert$. Consequently, the smaller horizontal structures in the tidal potential are not visible in the colour maps, although they are accounted for in the tidal solutions. 

For $\Prot = 33$~hr (\fig{fig:map_prot_33h}), the dominant pattern arises from the resonantly excited equatorial gravity mode in both the \textsc{Full} and \textsc{Approx} cases, resulting in the two solutions appearing largely similar. The slight differences observed are due to errors in calculating the oceanic obliquity tides within the \textsc{Approx} model, but these discrepancies remain minor compared to the semidiurnal tidal component. This component is represented in the colour maps generated for the coplanar-circular configuration, where the two solutions are identical (see \fig{fig:map_prot_33h}, bottom panels). At $\Prot=10$~hr (\fig{fig:map_prot_10h}), the semidiurnal oceanic tide is weak, as no equatorial gravity mode is excited by the associated forcing potential. As a result, the tidal response is more sensitive to obliquity tides compared to $\Prot=33$~hr. Moreover, the \textsc{Approx} model artificially reproduces the resonance of an equatorial gravity mode for an obliquity tidal component, leading to an amplification of this component by an order of magnitude. Consequently, the \textsc{Full} and \textsc{Approx} solutions show substantial divergence for non-zero obliquities, consistent with the large errors observed in estimates of tidally dissipated power, and the variation rates of spin angular velocity and obliquity (\figsto{fig:pdiss_solid_globoc}{fig:obli_solid_globoc}, bottom right panels). 

As discussed in \append{app:sensitivities_bulge_torque}, the spatial distribution of the deformation tidal potential, shown in \figs{fig:map_prot_33h}{fig:map_prot_10h}, is strongly influenced by the elasticity of solid regions. In the rigid limit considered here, variations in self-attraction are solely generated by oceanic tides, which display complex horizontal structures due to the excitation of modes of various degrees. This helps to clarify the causes of the discrepancies observed in the tidally dissipated power and the variation rates of the planet's spin angular velocity and obliquity between the \textsc{Full} and \textsc{Approx} cases. When solid elasticity is taken into account, as seen in \figsto{fig:pdiss_solid_globoc}{fig:obli_solid_globoc}, the tidal bulge is primarily shaped by the degree-2 visco-elastic deformation of solid regions, driven by the density contrast between rocks and liquid water. Consequently, the deformation potential exhibits similar patterns in both the \textsc{Full} and \textsc{Approx} models. Although this may seem counterintuitive, the sensitivity of the planet's tidal distortion to solid elasticity has a minimal effect on the tidally dissipated power and torque. This is because the angular lag induced by viscous friction in solid regions is much smaller than that of the oceanic tide in the configurations studied. This point is further explored in \append{app:sensitivities_bulge_torque}, where we use a toy model to illustrate the underlying mechanisms. 

The primary limitations of our approach originate in the simplified geometry used to compute tidal solutions. Our model assumes a global surface layer with uniform depth, which overlooks the complex coupling between shallow seas and deep oceans that help damping resonances on Earth \citep[e.g.,][]{Arbic2009,AB2010}. Furthermore, the intricate interactions between tidal flows and coastlines, which significantly impact tidal dissipation, are not considered. Additionally, nonlinear processes generally inhibit the dynamical tide. For example, bottom friction between tidal flows and the ocean floor, which dominates on Earth, is a quadratic function of the velocity field \citep[][]{ER2001,ER2003}. This tends to mitigate the predictions of linear tidal theory by smoothing out resonant peaks. 

\rev{Nevertheless, even in the presence of nonlinearities, which may significantly attenuate tidal energy transfer rates, dissipation remains strongly dependent on tidal frequency in stars and the fluid envelopes of giant planets due to the dynamical tide \citep[e.g.,][]{AB2022,AB2023}}. As a result, the resonant amplification mechanism highlighted is equally applicable to fluid bodies beyond ocean planets. Finally, it is important to note that any symmetry-breaking effect, in addition to Coriolis forces, would \rev{increase scattering and} make the tidal response more sensitive to the orientation of both the central body and the perturber's orbit, causing the system to deviate further from the \rev{isotropic assumption}. 


\section{Conclusions}

In this study, we explored how anisotropy influences the tidal response of fluid bodies, focusing specifically on Earth-sized rocky planets with global oceans. Building on previous research  that sought to deepen our understanding of oceanic tides and their role in the long-term evolution of the Earth-Moon system and exoplanets with liquid surface layers \citep[][]{ADLML2018,Auclair2019,Auclair2023,Farhat2022ellip}, we examined the implications of a commonly used approximation. This approximation involves using the equatorial degree-2 Love number of the semidiurnal tide to quantify energy and angular momentum exchange rates across all components of the tidal response. By assuming spherical isotropy, it presumes that the Love number's dependence on tidal frequency is uniform, regardless of the body's orientation relative to the perturber's orbit. Given that \rev{the isotropic assumption} does not hold for stars and planetary fluid layers due to Coriolis forces our primary goal was to assess the impact of this approximation on planetary evolution models, where it is frequently employed. 

Our calculations are grounded in linear theory and the conventional framework of two-body tidal interactions. Using angular momentum theory, we first derived expressions for the time-averaged rate of energy exchange and the three-dimensional tidal torque as functions of the complex coefficients that describe the spherical harmonic expansions of tidal forcing and deformation potentials (\eqsto{torquex_sum}{powertide_sum}). These expressions are general and apply even to \rev{non-isotropic bodies}, enabling the computation of the long-term tidal evolution of planet-satellite or star-planet systems in three dimensions. For completeness, we also provided the variation rates of the planet's spin angular velocity and the orbital elements of the perturber within this formalism. 

We then applied the theory to an idealised Earth-Moon system. In this setup, Earth is treated a rocky planet with a thin, uniform-depth ocean at its surface, while the Moon is modelled as a point-mass satellite, following \cite{Auclair2023}. We calculated the evolution of tidally dissipated energy as a function of the planet's spin period and obliquity, both with and without the aforementioned Love number approximation. By comparing the two approaches, we quantified the error induced by this approximation in tidal models. Our results indicate that the error can be significant for ocean planets, as resonances associated with tidally forced oceanic modes greatly amplify it. In Earth-like configurations, the approximated model can overestimate tidally dissipated energy and angular momentum exchange rates by an order of magnitude, especially when tidal frequencies exceed the eigenfrequency of the ocean mode with the largest wavelength.

Interestingly, the Love number approximation does not uniformly affect the variation rates that describe the planet-satellite's dynamical evolution. For instance, the variation rate of obliquity is significantly altered even in quasi-coplanar configurations (where the perturber's orbit lies nearly in the planet's equatorial plane), whereas the tidally dissipated energy and the planet's spin rate remain largely unaffected for obliquities under approximately ${\sim}10^\circ$. However, the extensive region of parameter space where the approximation fails for rotation periods shorter than one day suggests that it is unsuitable for studying the past evolution of the Earth-Moon system. In such cases, the tidal response should be computed self-consistently for each tidal component to properly account for \rev{scattering} effects of Coriolis forces and continental coastlines, which enable the resonant excitation of lower-frequency modes. 

These conclusions can be extended to stars and the fluid envelopes of gaseous giants, where the dynamical tide is typically the dominant component of the tidal response. However, the approximation remains valid in the non-wavelike regime, where fluid waves cannot be resonantly excited by tidal forces. Additionally, the model used in this study, which is based on simplified geometry and linear tidal theory, has its limitations. Specifically, the propagation of forced tidal modes may be hindered by nonlinear processes such as bottom friction, as well as by the complexity of land-ocean distributions and bathymetry. Therefore, while our findings provide valuable qualitative insights, they should be refined in future studies depending on the specific problem at hand.

\begin{acknowledgements}
\rev{The authors are thankful to the referee, Robert Tyler, for his thoughtful comments, which helped to improve the manuscript.} This research has made use of NASA's Astrophysics Data System.
\end{acknowledgements}

%
%

\bibliographystyle{aa} 
\bibliography{references} 


\begin{appendix}

\section{Nomenclature}
\label{app:nomenclature}

The notations introduced in the main text are listed below in order of appearance.

\vspace{-1.2cm}

\printnomenclature

\section{Vectorial operators in spherical coordinates}
\label{app:vectorial_operators}

Throughout this study, we use the standard spherical coordinate system $\left( \rr, \col , \lon \right)$, where $\rr$ denotes the radial coordinate, $\col$ is the colatitude (the angle measured from the $\zz$-axis), and $\lon$ represents the longitude. In this coordinate system, the gradient of any scalar quantity $f$ is expressed as \citep[e.g.,][Sect.~2.5]{Arfken2005}
\begin{equation}
\grad f = \dd{f}{\rr} \er + \frac{1}{\rr} \dd{f}{\col} \etheta + \frac{1}{\rr \sin \col} \dd{f}{\lon} \ephi,
\end{equation}
where $\left( \er , \etheta, \ephi \right)$ form the orthogonal basis of unit vectors corresponding to the spherical coordinates $\left( \rr ,\col , \lon \right)$. The Laplacian of $f$ is given by
\begin{equation}
\lap f = \frac{1}{\rr} \ddd{\left( \rr f \right)}{\rr}{\rr}  + \frac{1}{\rr^2 \sin \col} \dd{\left( \sin \col \dd{f}{\col} \right)}{\col} + \frac{1}{\rr^2 \sin^2 \col} \ddd{f}{\lon}{\lon}. 
\end{equation}
Similarly, the divergence of a vector field $\Vvect = \Vr \er + \Vtheta \etheta + \Vphi \ephi$ is expressed as
\begin{equation}
\div \Vvect = \frac{1}{\rr^2} \dd{\left( \rr^2 \Vr \right)}{\rr} + \frac{1}{\rr \sin \col} \dd{\left( \sin \col \Vtheta \right)}{\col} + \frac{1}{\rr \sin \col} \dd{\Vphi}{\lon}. 
\end{equation}
When applying the gradient and divergence operators over a 2D spherical surface, as in \eqs{momentum}{continuity}, the radial terms $\dd{}{\rr}$ and $\ddd{}{\rr}{\rr}$ vanish. As a result, the gradient of $f$ and the divergence of $\Vvect = \left( \Vtheta , \Vphi \right)$ simplify as follows:
\begin{align}
\grad f  & =  \frac{1}{\rr} \dd{f}{\col} \etheta + \frac{1}{\rr \sin \col} \dd{f}{\lon} \ephi, \\
\div \Vvect & =  \frac{1}{\rr \sin \col} \dd{ \left( \sin \col \Vtheta \right)}{\col} + \frac{1}{\rr \sin \col} \dd{\Vphi}{\lon}. 
\end{align}

\section{Time-averaged power}
\label{app:time_averaged_power}

We demonstrate here how the time-averaged power, as defined by \eqs{timeav}{func_power}, can be related to the complex Fourier coefficients of the relevant quantities, specifically using the formula given in \eq{timeav_formula}. We consider two fields, $A$ and $B$, which vary as functions of time and spatial coordinates and are expressed as the real parts of the complex fields $\tilde{A}$ and $\tilde{B}$,
\begin{align}
    & A \left( \col , \lon , \time \right)  = \Re \left\{ \tilde{A} \left( \col , \lon , \time \right) \right\}, & B \left( \col , \lon , \time \right) = \Re \left\{ \tilde{B} \left( \col , \lon , \time \right) \right\}.
\end{align}
We assume that both $\tilde{A}$ and $\tilde{B}$ can be written as sinusoidal time-oscillations multiplied by complex spatial functions,
\begin{align}
    & \tilde{A} = \tilde{A}_0 \left( \col , \lon \right) \expo{\inumber \ftide_1 \time}, & \tilde{B} = \tilde{B}_0 \left( \col , \lon \right) \expo{\inumber \ftide_2 \time}.
\end{align}
From \eq{timeav}, the time-average of the product $AB$ is defined as 
\begin{equation}
    \timeav{AB} =\lim_{\period \rightarrow + \infty} \frac{1}{\period} \integ{A B  \, }{\time }{0}{\period},
\end{equation}
where $\period$ is the time interval over which the average is calculated. Expanding the real parts of $\tilde{A}$ and $\tilde{B}$, we write
\begin{align}
    & \Re \left\{ \tilde{A} \right\} = \Re \left\{ \tilde{A}_0 \right\} \cos \left( \ftide_1 \time \right) - \Im \left\{ \tilde{A}_0  \right\} \sin \left( \ftide_1 \time \right), \\
    & \Re \left\{ \tilde{B} \right\} = \Re \left\{ \tilde{B}_0 \right\} \cos \left( \ftide_2 \time \right) - \Im \left\{ \tilde{B}_0  \right\} \sin \left( \ftide_2 \time \right).
\end{align}
Now, expanding the product $AB$ yields
\begin{equation}
AB  = \Re \left\{ \tilde{A}  \right\}  \Re \left\{ \tilde{B} \right\} ,
\end{equation}
which can be written as 
\begin{align}
   AB & = \frac{1}{2} \left( \Re \left\{ \tilde{A}_0 \right\} \Re \left\{ \tilde{B}_0 \right\} - \Im \left\{ \tilde{A}_0  \right\} \Im \left\{ \tilde{B}_0  \right\} \right) \cos \left( \left( \ftide_1 + \ftide_2 \right) \time \right) \nonumber \\
    & + \frac{1}{2} \left( \Re \left\{ \tilde{A}_0 \right\} \Re \left\{ \tilde{B}_0 \right\} + \Im \left\{ \tilde{A}_0  \right\} \Im \left\{ \tilde{B}_0  \right\} \right) \cos \left( \left( \ftide_1 - \ftide_2 \right) \time \right) \nonumber \\
    & - \frac{1}{2} \left( \Im \left\{ \tilde{A}_0 \right\} \Re \left\{ \tilde{B}_0 \right\} + \Re \left\{ \tilde{A}_0  \right\} \Im \left\{ \tilde{B}_0  \right\} \right) \sin \left( \left( \ftide_1 + \ftide_2 \right) \time \right) \nonumber \\
    & + \frac{1}{2} \left( \Re \left\{ \tilde{A}_0 \right\} \Im \left\{ \tilde{B}_0 \right\} - \Im \left\{ \tilde{A}_0  \right\} \Re \left\{ \tilde{B}_0  \right\} \right) \sin \left( \left( \ftide_1 - \ftide_2 \right) \time \right) . \nonumber
\end{align}
If $\ftide_1 \neq \ftide_2 $ and $\ftide_2 \neq - \ftide_1 $, we obtain
\begin{equation}
   \lim_{\period \rightarrow + \infty} \frac{1}{\period} \integ{AB  \, }{\time }{0}{\period} = 0,
\end{equation}
indicating that there is no coupling between $A$ and $B$. However, if $\ftide_1 = \ftide_2 = \ftide$, then 
\begin{equation}
   \lim_{\period \rightarrow + \infty}  \frac{1}{\period} \integ{ AB \, }{\time}{0}{\period} = \frac{1}{2} \left( \Re \left\{ \tilde{A}_0 \right\} \Re \left\{ \tilde{B}_0 \right\} + \Im \left\{ \tilde{A}_0 \right\} \Im \left\{ \tilde{B}_0 \right\} \right).
\end{equation}
We recognise this as the real part of the product $\conj{\tilde{A}_0} \tilde{B}_0$, expanded as
\begin{align}
    \conj{\tilde{A}_0} \tilde{B}_0 = & \Re \left\{ \tilde{A}_0 \right\} \Re \left\{ \tilde{B}_0 \right\} + \Im \left\{ \tilde{A}_0 \right\} \Im \left\{ \tilde{B}_0 \right\} \nonumber \\
    & + \inumber \left( \Im \left\{ \tilde{A}_0 \right\} \Re \left\{ \tilde{B}_0 \right\} - \Im \left\{ \tilde{B}_0 \right\} \Re \left\{ \tilde{A}_0 \right\} \right) .
\end{align}
Thus, we have
\begin{equation}
   \timeav{AB} = \lim_{\period \rightarrow + \infty} \frac{1}{\period} \integ{ A B \, }{\time}{0}{\period} = \frac{1}{2} \Re \left\{  \conj{\tilde{A}_0} \tilde{B}_0 \right\},
\end{equation}
which, when integrated over the spatial coordinates, gives \eq{timeav_formula}.


\section{Legendre polynomials and spherical harmonics}
\label{app:sph}

The spherical harmonics $\Ylm{\llat}{\mm}$ in \eqs{Utidensig_sph}{Urespnsig_sph} are defined as \citep[][Sect.~5.2, Eq.~(1)]{Varshalovich1988}
\begin{equation}
\Ylm{\llat}{\mm} \left( \col , \lon \right) = \sqrt{\frac{\left( 2 \llat + 1 \right) \left( \llat - \mm \right) ! }{4 \pi \left( \llat + \mm \right) !}} \LegF{\llat}{\mm} \left( \cos \col \right) \expo{\inumber \mm \lon},
\end{equation}
where $\LegF{\llat}{\mm}$ denotes the associated Legendre functions (ALFs), defined for $ - 1 \leq \xx \leq 1$, with $\llat$ and $\mm$ as integers such that $\abs{\mm} \leq \llat$, as follows:
\begin{equation}
\label{alf}
\LegF{\llat}{\mm} \left( \xx \right) \define \left( -1 \right)^\mm \left( 1 - \xx^2 \right)^{\mm/2} \DDn{}{\xx}{\mm} \LegP{\llat} \left( \xx \right).
\end{equation}
In this equation, $\LegP{\llat}$ represents the degree-$\llat$ Legendre polynomial \citep[][Eq. (8.6.18)]{AS1972}, defined as
\begin{equation}
\label{legendre_poly}
\LegP{\llat} \left( \xx \right) \define \frac{1}{2^\llat \llat !} \DDn{}{\xx}{\llat} \left( \xx^2 - 1 \right)^\llat. 
\end{equation}
The normalisation of the spherical harmonics $\Ylm{\llat}{\mm}$ is given by the integral
\begin{equation}
\integ{\integ{\abs{\Ylm{\llat}{\mm} \left( \col, \lon \right) }^2 \sin \col \,}{\col}{0}{\pi} }{\lon}{0}{2 \pi} = 1.
\end{equation}
As specific value of the spherical harmonics, $\Ylm{\llat}{\mm} \left( \frac{\pi}{2} , 0 \right) $, as used in \eq{Utidelmkm_gal} is expressed as
\begin{equation}
\Ylm{\llat}{\mm} \left( \frac{\pi}{2}, 0 \right) = 
\left\{
\begin{array}{ll}
  \left( -1 \right)^{\left( \llat + \mm \right)/2} \frac{\left( \llat + \mm - 1 \right) !!}{\left( \llat - \mm \right) !!} \sqrt{\frac{\left( 2 \llat + 1 \right) \left( \llat - \mm \right)!}{4 \pi \left( \llat + \mm \right)!}} & \mbox{for} \ \llat + \mm \ \mbox{even},  \\
0 &\mbox{for} \ \llat + \mm \ \mbox{odd}, \\
\end{array}
\right.
\end{equation}
where the double factorial of a positive integer $\nn$, denoted by $\nn !!$, is defined as 
\begin{equation}
\nn !! = \prod_{\kk = 0}^{\ceil{\frac{\nn}{2}} - 1} \left( \nn - 2 \kk \right) = \nn \left( \nn-2 \right) \left( \nn - 4 \right) \ldots,
\end{equation}
with $\ceil{\nn/2}$ representing the ceiling of $\nn/2$.

\section{Wigner D-functions}
\label{app:wigner_dfunctions}

The Wigner D-functions introduced in \eqs{Ylmpla_Ylmfix}{Ylmfix_Ylmpla} represent the matrix elements of the rotation operator applied to spherical harmonics \citep[][Sect.~4.1]{Varshalovich1988}. They are structured as follows:
\begin{equation}
\label{wignermat}
\wignermat \define
\begin{bmatrix}
\wignerDmati{0}{0}{0} & 0 & \ldots & \ldots & \ldots & 0  \\
0 & \wignerDmati{\qq}{\mm}{1} &  \ddots & & & \vdots   \\
 \vdots  &  \ddots & \ddots & \ddots & &\vdots  \\
 \vdots &  & \ddots & \wignerDmati{\qq}{\mm}{\llat} & \ddots & \vdots \\
 \vdots &  & & \ddots & \ddots & 0  \\
 0 & \ldots & \ldots & \ldots & 0 & \wignerDmati{\qq}{\mm}{\lmax} 
\end{bmatrix},
\end{equation}
where $\lmax$ indicates the chosen truncation degree, and $\wignerDmati{\qq}{\mm}{\llat} $ is the matrix
\begin{equation}
\wignerDmati{\qq}{\mm}{\llat} \define 
\begin{bmatrix}
\wignerD{-\llat}{-\llat}{\llat} & \ldots & \wignerD{-\llat}{\mm}{\llat} & \ldots &  \wignerD{-\llat}{\llat}{\llat} \\
\vdots & & \vdots & & \vdots \\
\wignerD{\qq}{-\llat}{\llat} & \ldots & \wignerD{\qq}{\mm}{\llat} & \ldots & \wignerD{\qq}{\llat}{\llat} \\
\vdots & & \vdots & & \vdots \\
\wignerD{\llat}{-\llat}{\llat} & \ldots & \wignerD{\llat}{\mm}{\llat} & \ldots & \wignerD{\llat}{\llat}{\llat} 
\end{bmatrix} .
\end{equation}
The Wigner D-functions, $\wignerD{\qq}{\mm}{\llat}$, are defined as \citep[][Sect.~4.3, Eq. (1)]{Varshalovich1988} 
\begin{equation}
\label{wignerD}
\wignerD{\qq}{\mm}{\llat} \left( \angalp , \angbet , \anggam \right) \define \expo{- \inumber \qq \angalp} \wignerd{\qq}{\mm}{\llat} \left( \angbet \right) \expo{- \inumber \mm \anggam},
\end{equation}
where $\wignerd{\qq}{\mm}{\llat}$ is a real function given by \citep[][Sect.~4.3.1, Eq.~(2)]{Varshalovich1988}
\begin{align}
\wignerd{\qq}{\mm}{\llat} \left( \angbet \right) = & \left( -1 \right)^{\llat - \mm} \sqrt{\left( \llat + \qq \right)! \left( \llat - \qq \right)! \left( \llat + \mm \right) ! \left( \llat - \mm \right)! } \\
& \times \sum_{\kk} \left( -1 \right)^\kk \frac{ \left( \cos \frac{\angbet}{2} \right)^{\qq + \mm + 2 \kk} \left( \sin \frac{\angbet}{2} \right)^{2 \llat - \qq - \mm - 2 \kk} }{\kk ! \left( \llat - \qq - \kk \right) ! \left( \llat - \mm - \kk \right) ! \left( \qq + \mm + \kk \right) ! }. \nonumber
\end{align}
The summation index $\kk$ runs over all integer values for which the factorial arguments are non-negative, namely $\max \left( 0 , - \qq - \mm \right) \leq \kk \leq \min \left( \llat - \qq , \llat - \mm \right)$. 

In practice, Wigner D-functions are computed recursively using the method described by \cite{GG2009}, starting with
\begin{equation}
\begin{array}{ll}
\wignerDmati{0}{0}{0} = 1, & \displaystyle
\wignerDmati{\qq}{\mm}{1} = 
\begin{bmatrix}
\conj{a}^2 & - \sqrt{2} \conj{a} b & b^2 \\
-\sqrt{2} \conj{a} \conj{b} & a \conj{a} - b \conj{b} & -\sqrt{2} a b \\
\conj{b}^2 & - \sqrt{2} a \conj{b} & a^2
\end{bmatrix},
\end{array}
\end{equation}
where the Cayley-Klein parameters, $a$ and $b$, are defined in terms of the 3-2-3 Euler angles from \eq{eulermat} as 
\begin{align}
& a = \cos \left( \frac{\angbet}{2} \right) \expo{- \inumber \frac{1}{2} \left( \angalp + \anggam \right) }, 
& b = \sin \left( \frac{\angbet}{2} \right) \expo{ \inumber \frac{1}{2} \left( \angalp - \anggam \right) }.
\end{align}
For matrices of degrees $\llat \geq 2$, we use the recursion relation given by \citep[e.g.,][Sect.~2.5]{Boue2017}\footnote{Equation~(\ref{wigner_recursion}) is obtained by combining the recursion relations given by Eqs.~(14) and~(15) of \cite{Varshalovich1988}, Sect.~4.8.2.}
\begin{equation}
\label{wigner_recursion}
\wignerD{\qq}{\mm}{\llat} = \cwminusi{\qq}{\mm}{\llat} \wignerD{1}{1}{1} \wignerD{\qq-1}{\mm-1}{\llat-1} + \cwzeroi{\qq}{\mm}{\llat} \wignerD{1}{0}{1}  \wignerD{\qq-1}{\mm}{\llat-1} + \cwplusi{\qq}{\mm}{\llat} \wignerD{1}{-1}{1} \wignerD{\qq-1}{\mm+1}{\llat-1},
\end{equation}
with the coefficients $\cwminusi{\qq}{\mm}{\llat}$, $\cwzeroi{\qq}{\mm}{\llat}$ and $\cwplusi{\qq}{\mm}{\llat}$ defined as
\begin{equation}
\begin{array}{l}
\displaystyle \cwminusi{\qq}{\mm}{\llat}  = \sqrt{\frac{\left( \llat + \mm \right) \left( \llat + \mm - 1 \right)}{\left( \llat + \qq \right) \left( \llat + \qq - 1 \right) }},\\
\displaystyle \cwzeroi{\qq}{\mm}{\llat}  = \sqrt{ \frac{2 \left( \llat + \mm \right) \left( \llat - \mm \right) }{\left( \llat + \qq \right) \left( \llat + \qq - 1 \right)} }, \\
\displaystyle \cwplusi{\qq}{\mm}{\llat}   = \sqrt{\frac{\left( \llat - \mm \right) \left( \llat - \mm - 1 \right)}{\left( \llat + \qq \right) \left( \llat + \qq - 1 \right)} }. 
\end{array}
\end{equation}
Terms $\wignerD{\qq-1}{\mm+\nu}{\llat-1}$ where $\abs{\mm + \nu} > \llat-1$, with $\nu \in \left\{ -1, 0 , 1 \right\}$, are discarded and replaced by zero. Wigner D-matrix elements with negative indices $\mm$ are derived from those with positive $\mm$ using the symmetry
\begin{equation}
\wignerD{\qq}{\mm}{\llat} = \left( -1 \right)^{\qq- \mm} \conj{\wignerD{-\qq}{-\mm}{\llat}}. 
\end{equation}

\section{Tidal torque and power}
\label{app:torque_power}

In this appendix, we detail the derivations leading to \eqsto{torquex_sum}{powertide_sum}. The time-averaged tidal torque about the $\xx$-axis of the Galilean reference frame, given by \eq{torquex}, is expressed as 
\begin{equation}
\torquex  = - \frac{1}{4 \pi \Ggrav} \timeav{ \ointeg{ \! \! \left[ \left( \amopi{\xx} \Utide \right) \grad  \Uresp - \Uresp \grad \left( \amopi{\xx} \Utide \right) \right] \cdot }{\surfacev}{\domainstarbd}{}}.
\end{equation}
Assuming that $\domainstar$ is a sphere of radius $\rdstar$ centred on the planet's centre of gravity, we rewrite the two components of the integral in the frequency domain as 
\begin{equation}
\label{torquex_eq1}
 \ointeg{ \! \!  \conj{\left( \amopi{\xx} \Utide \right)} \grad \Uresp  \cdot  }{\surfacev}{\domainstarbd}{} = \integ{\integ{ \conj{\left( \amopi{\xx} \Utide \right)} \dd{\Uresp}{\rr} \rdstar^2 \sin \col}{\col}{0}{\pi}}{\lon}{0}{2\pi},
\end{equation}
and 
\begin{equation}
\label{torquex_eq2}
 \ointeg{ \! \!   \Uresp  \grad \conj{\left( \amopi{\xx} \Utide \right)}  \cdot  }{\surfacev}{\domainstarbd}{} = \integ{\integ{\Uresp \dd{ \conj{\left( \amopi{\xx} \Utide \right)}}{\rr}  \rdstar^2 \sin \col }{\col}{0}{\pi}}{\lon}{0}{2\pi}.
\end{equation}
The perturbed potential and its radial gradient are given by 
\begin{align}
\Urespn = &   \sum_{\kk} \sum_{\llat = 0}^{+ \infty} \sum_{\mm= - \llat}^\llat \left(\sum_{\qq= - \llat}^\llat  \Uplmsigj{\llat}{\kk, \qq, \qq}{\mm}  \right) \left( \frac{\rr}{\Rpla} \right)^{-\left( \llat + 1\right)} \! \! \! \! \! \!  \Ylm{\llat}{\mm} \left( \col , \lon \right) \expo{ \inumber \ftidefixi{\kk} \time}, \\
\dd{\Urespn}{\rr} = &  - \sum_{\kk} \sum_{\llat = 0}^{+ \infty} \sum_{\mm= - \llat}^\llat \frac{ \llat + 1 }{\rr} \left(\sum_{\qq= - \llat}^\llat  \Uplmsigj{\llat}{\kk, \qq, \qq}{\mm}  \right) \left( \frac{\rr}{\Rpla} \right)^{-\left( \llat + 1\right)} \! \! \! \! \! \!  \Ylm{\llat}{\mm} \left( \col , \lon \right) \expo{ \inumber \ftidefixi{\kk} \time},
\end{align}
and $\conj{\amopi{\xx} \Utiden}$ and its radial gradient, by
\begin{align}
\conj{\amopi{\xx} \Utiden}  = & \frac{\inumber}{\sqrt{2}} \sum_{\kk,\nuco} \sum_{\llat=2}^{+ \infty} \sum_{\mm = - \llat}^\llat \nuco \amopci{\nuco}{\llat}{\mm} \conj{\Ulmsigj{\llat}{\kk}{\mm}} \left( \frac{\rr}{\Rpla} \right)^\llat \conj{\Ylm{\llat}{\mm+\nuco}} \left( \col , \lon \right) \expo{-\inumber \ftidefixi{\kk} \time}, \\
\dd{\conj{\amopi{\xx} \Utiden}}{\rr}  = & \frac{\inumber}{\sqrt{2}} \sum_{\kk,\nuco} \sum_{\llat=2}^{+ \infty} \sum_{\mm = - \llat}^\llat \nuco \amopci{\nuco}{\llat}{\mm} \conj{\Ulmsigj{\llat}{\kk}{\mm}} \frac{\llat}{\rr} \left( \frac{\rr}{\Rpla} \right)^\llat \conj{\Ylm{\llat}{\mm+\nuco}} \left( \col , \lon \right) \expo{-\inumber \ftidefixi{\kk} \time}.
\end{align}
with $\kk$ running from $0$ to $+ \infty$ and $\nuco = \pm 1$. Substituting the above spherical harmonic expansions in \eqs{torquex_eq1}{torquex_eq2}, we obtain
\begin{align}
 \ointeg{ \! \!   \Uresp  \grad \conj{\left( \amopi{\xx} \Utide \right)}  \cdot  }{\surfacev}{\domainstarbd}{} = &  
 - \sum_{\kk, \nuco} \sum_{\llat=2}^{+ \infty} \sum_{\mm = - \llat}^\llat  \left( \frac{\rdstar}{\Rpla} \right)^\llat  \left( \frac{\llat+1 }{\rdstar} \right) \left(\frac{\Rpla}{\rdstar} \right)^{\llat+1} \! \!  \rdstar^2 \nonumber \\
 & \times \frac{\inumber}{\sqrt{2}}  \nuco \amopci{\nuco}{\llat}{\mm} \conj{\Ulmsigj{\llat}{\kk}{\mm}} \left(\sum_{\qq= - \llat}^\llat  \Uplmsigj{\llat}{\kk, \qq, \qq}{\mm+\nuco}  \right) ,
\end{align}
and
\begin{align}
 \ointeg{ \! \!  \conj{\left( \amopi{\xx} \Utide \right)} \grad \Uresp  \cdot  }{\surfacev}{\domainstarbd}{} = &  
  \sum_{\kk, \nuco} \sum_{\llat=2}^{+ \infty} \sum_{\mm = - \llat}^\llat  \left( \frac{\rdstar}{\Rpla} \right)^\llat  \frac{\llat }{\rdstar}  \left(\frac{\Rpla}{\rdstar} \right)^{\llat+1} \! \!  \rdstar^2 \nonumber \\
 & \times \frac{\inumber}{\sqrt{2}}  \nuco \amopci{\nuco}{\llat}{\mm} \conj{\Ulmsigj{\llat}{\kk}{\mm}} \left(\sum_{\qq= - \llat}^\llat  \Uplmsigj{\llat}{\kk, \qq, \qq}{\mm+\nuco}  \right) .
\end{align}
where the dependence on $\rdstar$ vanishes, as discussed by \cite{Ogilvie2013}. Finally, using the formula given by \eq{timeav_formula}, established in \append{app:time_averaged_power}, we end up with
\begin{equation}
\torquex  = - \frac{\Ktorque}{\sqrt{2}} \Im  \left\{ \sum_{\kk,\nuco} \sum_{\llat=2}^{+\infty} \sum_{\mm = -\llat}^\llat  \nuco \left( 2 \llat + 1 \right) \amopci{\nuco}{\llat}{\mm} \conj{\Ulmsigj{\llat}{\kk}{\mm} }  \left( \sum_{\qq=-\llat}^{\llat} \Uplmsigj{\llat}{\kk, \qq, \qq}{\mm+\nuco}   \right)   \right\}  , 
\end{equation}
which corresponds to \eq{torquex_sum}. The expressions provided by \eqsto{torquey_sum}{powertide_sum} for the other components of the torque and the tidal power are derived following similar steps.

\section{Tide-raising gravitational potential}
\label{app:forcing_potential}

In this appendix, we derive the expression for the components of the tide-raising gravitational potential given \eq{Utidelmkm_gal}. First, the tidal potential, defined by \eq{tidalpot1}, is expanded as a series of Legendre polynomials \citep[e.g.,][]{EW2009},
\begin{equation}
\Utide = \frac{\Ggrav \Mpert}{\rpert} \sum_{\llat = 2}^{+ \infty} \left( \frac{\rr}{\rpert} \right)^\llat \LegP{\llat} \left( \cos \zenithphi \right),
\end{equation}
where $\zenithphi$ designates the stellar zenithal angle (with $\zenithphi = 0$ at the sub-stellar point), and $\LegP{\llat}$ denotes the degree-$\llat$ Legendre polynomial, defined by \eq{legendre_poly}. 

Next, by applying the addition theorem for spherical harmonics \citep[e.g.,][Sect.~12.8]{Arfken2005}, we express $\LegP{\llat} \left( \cos \zenithphi \right)$ as
\begin{equation}
 \LegP{\llat} \left( \cos \zenithphi \right) = \frac{4 \pi}{2 \llat + 1} \sum_{\mm=-\llat}^{\llat} \Ylm{\llat}{\mm} \left( \col , \lon \right) \conj{ \Ylm{\llat}{\mm}} \left( \colpert , \lonpert \right) ,
\end{equation}
where $\colpert $ and $\lonpert$ represent the colatitude and longitude of the perturber in the fixed reference frame. We proceed by rewriting $\conj{ \Ylm{\llat}{\mm}} $ as a function of the perturber's orbital elements. This is achieved through the use of Wigner D-functions, as introduced in \eq{Ylmfix_Ylmpla},
\begin{equation}
\conj{ \Ylm{\llat}{\mm}} \left( \col , \lon \right)  = \sum_{\qq = - \llat}^{\llat} \wignerD{\mm}{\qq}{\llat}  \left( \angapert , \angbpert , \anggpert \right) \conj{\Ylm{\llat}{\qq}} \left( \frac{\pi}{2} , \trueapert \right), \nonumber 
\end{equation}
which can be simplified further to
\begin{equation}
\conj{ \Ylm{\llat}{\mm}} \left( \col , \lon \right)  = \sum_{\qq = - \llat}^{\llat} \wignerD{\mm}{\qq}{\llat}  \left( \angapert , \angbpert , \anggpert \right) \Ylm{\llat}{\qq} \left( \frac{\pi}{2} , 0 \right) \expo{-\inumber \qq \trueapert}, 
\end{equation}
where $\angapert$, $\angbpert$, and $\anggpert$ are the Euler angles defining the reference frame associated with the perturber's orbit, as provided in \eq{euler_angles_orbit}. Thus, the tide-raising gravitational potential is given by
\begin{align}
\Utide =&  \frac{\Ggrav \Mpert}{\Rpla} \sum_{\llat =2}^{+\infty} \sum_{\mm=-\llat}^{\llat} \sum_{\qq = -\llat}^{\llat} \frac{4 \pi}{2 \llat + 1} \left( \frac{\Rpla}{\smaxispert} \right)^{\llat + 1} \left( \frac{\rr}{\Rpla} \right)^\llat \Ylm{\llat}{\mm} \left( \col, \lon \right) \nonumber \\
& \times \wignerD{\mm}{\qq}{\llat}  \left( \angapert , \angbpert , \anggpert \right) \Ylm{\llat}{\qq} \left( \frac{\pi}{2} , 0 \right) \left( \frac{\smaxispert}{\rpert} \right)^{\llat + 1} \expo{-\inumber \qq \trueapert},
\end{align}
expressed as a function of the planet-perturber distance, $\rpert$, and true anomaly, $\trueapert$. 

By introducing time dependence through Hansen coefficients, as in \eq{Hansen_series}, we write
\begin{equation}
\left( \frac{\smaxispert}{\rpert} \right)^{\llat+1} \expo{-\inumber \qq \trueapert}  = \sum_{\kk=-\infty}^{+ \infty} \Hansen{\kk}{- \left( \llat+1\right)}{- \qq}  \expo{\inumber \kk \meanapert}, \nonumber 
\end{equation}
which is further expressed as
\begin{equation}
\left( \frac{\smaxispert}{\rpert} \right)^{\llat+1} \expo{-\inumber \qq \trueapert}   =  \sum_{\kk=-\infty}^{+ \infty} \Hansen{\kk}{- \left( \llat+1\right)}{- \qq}  \expo{\inumber \kk \left( \npert \time + \meanapertc \right)}.
\end{equation}
Substituting this into the expression for the tidal potential, we obtain
\begin{equation}
\Utide =  \frac{\Ggrav \Mpert}{\Rpla} \sum_{\llat =2}^{+\infty} \sum_{\mm=-\llat}^{\llat} \sum_{\qq = -\llat}^{\llat} \sum_{\kk = - \infty}^{+\infty} \hklmq{\kk}{\llat}{\mm}{\qq} \left( \rr, \col, \lon , \time \right),
\end{equation}
where the dimensionless function $\hklmq{\kk}{\llat}{\mm}{\qq}$ is given by
\begin{align}
\label{hklmq}
\hklmq{\kk}{\llat}{\mm}{\qq} \left( \rr , \col , \lon , \time \right) = & \frac{4 \pi}{2 \llat + 1}  \left( \frac{\Rpla}{\smaxispert} \right)^{\llat + 1} \left( \frac{\rr}{\Rpla} \right)^\llat \Ylm{\llat}{\mm} \left( \col, \lon \right)  \wignerD{\mm}{\qq}{\llat}  \left( \angapert , \angbpert , \anggpert \right), \nonumber \\
& \times \Ylm{\llat}{\qq} \left( \frac{\pi}{2} , 0 \right) \Hansen{\kk}{- \left( \llat+1\right)}{- \qq} \left( \eccpert \right)  \expo{\inumber \kk \left( \npert \time + \meanapertc \right)}.
\end{align}

By using the symmetry properties of spherical harmonics, Wigner D-functions \citep[e.g.,][Chapter~4]{Varshalovich1988}, and Hansen coefficients \citep[e.g.,][]{Hughes1981,Laskar2005}, we find
\begin{align}
\conj{\Ylm{\llat}{\mm} }\left( \col, \lon \right) & = \left( - 1 \right)^\mm \Ylm{\llat}{-\mm} \left( \col , \lon \right), \\
\conj{\wignerD{\mm}{\qq}{\llat} } & = \left( - 1 \right)^{\mm+\qq} \wignerD{-\mm}{-\qq}{\llat} \\
 \Hansen{-\kk}{- \left( \llat+1\right)}{- \qq} & =  \Hansen{\kk}{- \left( \llat+1\right)}{\qq}.
\end{align}
These relations imply that $\hklmq{\kk}{\llat}{\mm}{\qq}$ satisfies
\begin{equation}
\hklmq{-\kk}{\llat}{-\mm}{-\qq} = \conj{ \hklmq{\kk}{\llat}{\mm}{\qq}}, 
\end{equation}
allowing us to start the summation over $\kk$ from $\kk=0$ and group terms in pairs of complexes conjugates. Consequently, the forcing tidal potential is expressed as
\begin{equation}
\Utide \left(  \rr, \col, \lon , \time \right)   =  \Re \left\{ \sum_{\kk= 0}^{+ \infty}  \Utidensig \left( \rr, \col, \lon  \right) \expo{ \inumber \ftidefixi{\kk} \time} \right\} ,
\end{equation}
with the spatial functions $\Utidensig$ defined as
\begin{equation}
\Utidensig \left( \rr, \col, \lon \right) =  \sum_{\llat=2}^{+ \infty} \sum_{\mm = -\llat}^{\llat}  \Ulmsigj{\llat}{\kk}{\mm} \left( \frac{\rr}{\Rpla} \right)^{\llat} \Ylm{\llat}{\mm} \left( \col , \lon \right) ,
\end{equation}
and their coefficients $ \Ulmsigj{\llat}{\kk}{\mm}$ given by
\begin{align}
\label{Utidelmkm_gal_app}
\Ulmsigj{\llat}{\kk}{\mm} = &  \left( 2 - \kron{\kk}{0} \right) \frac{4 \pi}{2 \llat + 1} \frac{\Ggrav \Mpert}{\Rpla} \left( \frac{\Rpla}{\smaxispert} \right)^{\llat +1} \nonumber \\ 
& \times \sum_{\qq=-\llat}^{\llat} \wignerD{\mm}{\qq}{\llat} \left( \angapert , \angbpert, \anggpert \right) \Ylm{\llat}{\qq} \left( \frac{\pi}{2} , 0 \right) \Hansen{\kk}{- \left( \llat + 1 \right)}{- \qq} \left( \eccpert \right).
\end{align}

Finally, the phase factor $\expo{\inumber \kk \meanapert}$ from \eq{hklmq} is ignored in \eq{Utidelmkm_gal_app}, as it does not affect the tidal torque or dissipated power. The mean anomaly at $t = 0$, $\meanapertc$, can be set to zero. In the circular orbit case ($\eccpert=0$), the true anomaly equals the mean anomaly, and the Hansen coefficients simplify to 
\begin{equation}
\Hansen{\kk}{\llat}{\mm} \left( 0 \right) = \kron{\kk}{\mm}.
\end{equation}
Thus, the coefficients $ \Ulmsigj{\llat}{\kk}{\mm}$ from \eq{Utidelmkm_gal_app} become
\begin{equation}
\Ulmsigj{\llat}{\kk}{\mm} =   \left( 2 - \kron{\kk}{0} \right) \frac{4 \pi}{2 \llat + 1} \frac{\Ggrav \Mpert}{\Rpla} \left( \frac{\Rpla}{\smaxispert} \right)^{\llat +1} \! \!  \wignerD{\mm}{-\kk}{\llat} \left( \angapert , \angbpert, \anggpert \right) \Ylm{\llat}{-\kk} \left( \frac{\pi}{2} , 0 \right).
\end{equation}

\section{Convergence tests for tidal solutions}
\label{app:convergence_tests}

The LTEs given by \eqs{momentum}{continuity} are solved in the frequency domain using the spectral method described in detail in \cite{Auclair2023}. These solutions are expressed as series of spherical harmonics, which are theoretically infinite. However, to solve the problem numerically, we must truncate these series. The truncation degree, $\lmax$, is chosen to ensure that the error induced by this approximation remains negligible. To determine an appropriate value for $\lmax$, we performed convergence tests. In all cases, we used the simplified Earth-Moon system described in \sect{sec:application_earth_moon}, considering the coplanar-circular configuration and the parameter values given by Table~\ref{tab:param_refcase}. Under this configuration, the planet is distorted solely by the semidiurnal tide, and the tidal torque is applied around the planet's spin axis, which aligns with the total angular momentum vector. This means that $\torquex=0$ and $\torquey=0$. We calculated the evolution of the component $\torquez$ across a uniformly sampled range of semidiurnal tidal frequencies by varying the planet's rotation rate. 

\begin{figure}[t]
   \centering
   \raisebox{2cm}{\rotatebox{90}{$ \log_{10} \left( \torquez \right) \  {\rm [J]} $}}
   \includegraphics[width=0.45\textwidth,trim = 0.8cm 0.6cm 0.5cm 0.cm,clip]{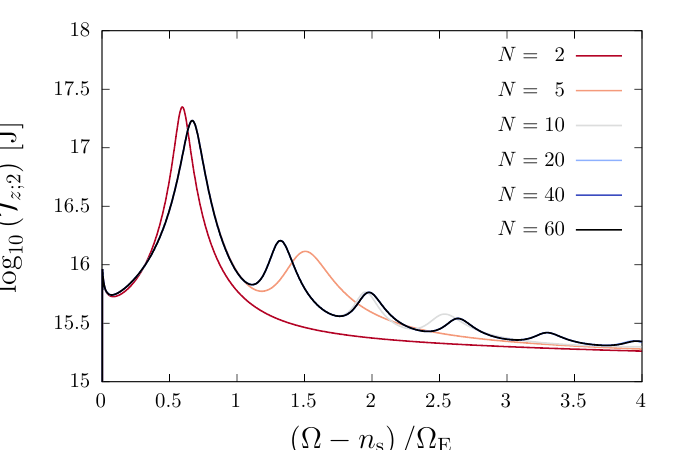} \\
   $\left( \spinrate - \npert \right) / \spinrate_{\rm E}$
      \caption{Semidiurnal tidal torque about the spin axis of a rocky planet with a global, uniform-depth ocean in the coplanar-circular configuration for various truncation degrees of the tidal solution ranging between $\lmax = 2$ and $\lmax = 60$. The torque is computed using \eq{torquez_sum}. It is plotted as a function of the normalised tidal frequency $\omega = \left( \spinrate - \npert \right)/\spinrate_{\rm E}  $, with $\spinrate_{\rm E} = 7.2921 \times 10^{-5} \ {\rm rad \ s^{-1}}$ being actual Earth's angular velocity \citep[][]{Fienga2021}. }
       \label{fig:convergence_tests}%
\end{figure}

Figure~\ref{fig:convergence_tests} presents the results for truncation degrees between $\lmax = 2$ (very low resolution) and $\lmax= 60$ (high resolution).The spectra display a series of peaks, as seen in \figs{fig:pdiss_solid_globoc}{fig:spinvel_solid_globoc}, corresponding to the typical pattern of oceanic tidal solutions in Laplace's theory \citep[e.g.,][]{ADLML2018,Motoyama2020,Tyler2021}. Each peak arises from the resonant excitation of an oceanic surface gravity mode by the tide-raising potential. In the global, uniform-depth ocean we studied, Laplace's tidal problem admits semi-analytical solutions, expressed as series of the so-called Hough functions. These are spherical harmonics distorted by the planet's rotation \citep[e.g.,][]{LH1968}. Since these functions differ from $\Ylm{2}{2}$, the semidiurnal tidal potential couples with several modes. The lowest frequency peak in \fig{fig:convergence_tests} corresponds to the ocean mode with the largest horizontal structure, which is also closest to the~$\Ylm{2}{2}$ harmonic. As the latitudinal wavelength of the resonantly excited modes decreases, the frequency of the peaks increases. \cite{GM1971} provide a closed-form solution for the LTEs, offering a discussion about the dependence of oceanic tidal elevation on tidal frequency \citep[see also][Eq.~(55), for the formulation of the oceanic tidal torque as a function of the tidal frequency]{Auclair2019}. The peak near synchronisation ($\spinrate=\npert$) corresponds to the maximum predicted by the Andrade model, which accounts for the tidal response of solid regions \citep[see, e.g.,][Fig.~2]{Efroimsky2012}. 

As the truncation degree increases, the tidal solutions capture more resonant peaks. A truncation degree of $\lmax=5$ appears sufficient to define the main peak, while setting $\lmax= 10$ captures the second peak, and so forth. It is worth noting the dynamical ocean tide diminishes at higher tidal frequencies because the overlap between the ocean modes and the $\Ylm{2}{2}$ spherical harmonic decreases. In the high-frequency range, the solution is dominated by the contribution of solid tides, which leads to a smoother evolution of the torque with the tidal frequency. Consequently, very high spectral resolutions are not required to accurately capture the planet's tidal response. Beyond $\lmax = 30$, the difference between numerical and exact solutions becomes negligible. Therefore, for practical purposes, we set $\lmax=30$, as used in the calculations in \sect{sec:application_earth_moon}.

\section{Tidal gravitational potentials in the rotating frame of reference of the perturber}
\label{app:gravpot_perturber_frame}

Since the tidal perturbation follows the perturber's motion, the coordinate system in which the perturber remains fixed is the most appropriate for representing both the tidal forcing gravitational potential, $\Utide$, and the gravitational potential induced by the tidal response, $\Uresp$. In this section, we derive the expressions for these two potentials in the rotated coordinate system, $\left( \rrrot,  \colrot, \lonrot  \right)$, where $\colrot=90^\circ$ and $\lonrot=0$ correspond to the sub-satellite point, while $\rrrot=\rr$. It is important to note that the notations $\colrot$ and $\lonrot$ are also used to refer to the coordinate system associated with the planet's rotating reference frame, which should not be confused with the frame associated with the perturber.  

The gravitational tidal potential is expressed by \eq{Utide} in the Galilean reference frame, $\rframe{\ifix}{O}{\ex}{\ey}{\ez}$, corresponding to the coordinates $\left( \rr , \col , \lon \right)$. Each Fourier component of this potential is expanded as a series of spherical harmonics (see \eq{Utidensig_sph}). Therefore, we transition from the fixed coordinate system, $\left( \rr, \col , \lon \right)$, to the rotated coordinate system, $\left( \rrrot, \colrot, \lonrot \right)$, by employing the Wigner-D elements introduced in \eq{Ylmfix_Ylmpla}, as follows:
\begin{align}
\Ylm{\llat}{\mm} \left( \col , \lon \right) &= \sum_{\qq = - \llat}^{\llat} \conj{\wignerD{\mm}{\qq}{\llat}} \left( \angapert , \angbpert , \anggpert + \trueapert \right) \Ylm{\llat}{\qq} \left( \colrot , \lonrot \right), \nonumber \\ 
& = \sum_{\qq = - \llat}^{\llat} \conj{\wignerD{\mm}{\qq}{\llat}} \left( \angapert , \angbpert , \anggpert  \right) \expo{\inumber \qq \trueapert } \Ylm{\llat}{\qq} \left( \colrot , \lonrot \right).
\end{align}
The Euler rotation angles $\left( \angapert, \angbpert, \anggpert \right)$ are those defined in \eq{euler_angles_orbit}. These angles specify the rotation that transitions from the Galilean reference frame to the reference frame associated with the perturber's orbital motion.

By replacing the index $\mm$ with $\jj$ and the index $\qq$ with $\mm$, we can rewrite \eq{Utidensig_sph} as 
\begin{equation}
\Utidensig \left( \rr, \col, \lon \right) =  \sum_{\llat=2}^{+ \infty} \sum_{\mm = -\llat}^{\llat} \sum_{\jj=-\llat}^{\llat}  \Ulmsigj{\llat}{\kk}{\jj} \conj{\wignerD{\jj}{\mm}{\llat}} \left( \angapert , \angbpert , \anggpert  \right) \left( \frac{\rr}{\Rpla} \right)^{\llat} \expo{\inumber  \mm \trueapert} \Ylm{\llat}{\mm} \left( \colrot , \lonrot \right),
\end{equation}
where $\Utidensig$ now depends implicitly on time via the perturber's true anomaly, $\trueapert$. To account for this, we express the factor $\expo{\inumber  \mm \trueapert} $ as a Fourier series using \eq{Hansen_series}, yielding
\begin{align}
\expo{\inumber \mm \trueapert} & = \sum_{\pp = - \infty}^{+\infty} \Hansen{\pp}{0}{\mm} \left( \eccpert \right) \expo{\inumber \pp \meanapert}, \nonumber \\
& = \sum_{\pp = - \infty}^{+\infty} \Hansen{\pp}{0}{\mm} \left( \eccpert \right) \expo{\inumber \pp \left(  \npert \time + \meanapertc \right)}. 
\end{align}
By grouping the time-dependent factors, we can identify the tidal frequencies associated with each Fourier component of the tidal potential in the perturber's frame, $\ftideroti{\kk,\pp} = \ftidefixi{\kk} + \pp \npert $. Given that $\ftidefixi{\kk} = \kk \npert$, only a single index is needed to define these tidal frequencies. Thus, we introduce the index $\qq = \kk + \pp$, allowing us to rewrite the forcing tidal potential as 
\begin{equation}
\Utide \left( \rrrot, \colrot, \lonrot, \time \right) = \Re \left\{ \sum_{\qq = - \infty}^{+ \infty} \Utidenrotsigi{\qq} \left( \rrrot, \colrot, \lonrot \right) \expo{\inumber \qq \npert \time}   \right\}.
\end{equation}
In this expression, the spatial functions $\Utidenrotsigi{\qq} $ are given by
\begin{equation}
\Utidenrotsigi{\qq}  \left( \rrrot , \colrot, \lonrot \right) = \sum_{\llat=2}^{+ \infty} \sum_{\mm = -\llat}^{\llat} \Ulmrotsigj{\llat}{\qq }{\mm} \left( \frac{\rrrot}{\Rpla} \right)^\llat \Ylm{\llat}{\mm} \left( \colrot , \lonrot \right) ,
\end{equation}
where the complex coefficients $\Ulmrotsigj{\llat}{\qq }{\mm} $ are defined as
\begin{equation}
\label{Ulmrotsigi_app}
\Ulmrotsigj{\llat}{\qq }{\mm} = \sum_{\kk=0}^{+\infty} \sum_{\jj = -\llat}^{+\llat}   \Hansen{\qq - \kk}{0}{\mm} \left( \eccpert \right) \expo{\inumber \left( \qq - \kk \right) \meanapertc} \Ulmsigj{\llat}{\kk}{\jj} \conj{\wignerD{\jj}{\mm}{\llat}} \left( \angapert , \angbpert , \anggpert  \right).
\end{equation}
As discussed in \append{app:forcing_potential}, the mean anomaly at $\time=0$ can be set to zero, snce it does not affect the tidal torque and dissipated power. Thus, the phase factor in \eq{Ulmrotsigi_app} can be ignored. 

For the components of the tidal potential $\Uresp$ that correspond to the same frequencies as the tidal forcing (i.e. components where $\pp = \qq$ in \eq{Uresp}), similar expressions can be obtained. These are the components that contribute to the tidal power and torque, while others cannot generally couple with the forcing. Neglecting these other components, Equation~(\ref{Uresp}) becomes 
\begin{equation}
\Uresp \left(  \rr, \col, \lon , \time \right)   =  \Re \left\{\sum_{\kk=0}^{+\infty}  \Urespnk{\kk} \left( \rr, \col, \lon  \right) \expo{ \inumber \ftidefixi{\kk} \time} \right\} ,
\end{equation}
where the spatial functions $\Urespnk{\kk} $ are expressed as
\begin{equation}
\Urespnk{\kk}  \left( \rr, \col, \lon  \right) =   \sum_{\llat=0}^{+ \infty} \sum_{\mm = - \llat}^{\llat} \Uplmsigj{\llat}{\kk}{\mm}  \left( \frac{\rr}{\Rpla} \right)^{-\left( \llat + 1 \right)} \Ylm{\llat}{\mm} \left( \col , \lon \right).
\end{equation}
The complex coefficients $\Uplmsigj{\llat}{\kk}{\mm} $ in the above equations are given by
\begin{equation}
\Uplmsigj{\llat}{\kk}{\mm} = \sum_{\qq=-\llat}^{+\llat}  \Uplmsigj{\llat}{\kk,\qq,\qq}{\mm}.
\end{equation}
The same coordinate transformation steps as those applied to the forcing tidal potential are used to rewrite the tidal response potential as
\begin{equation}
\Uresp \left( \rrrot, \colrot, \lonrot, \time \right) = \Re \left\{ \sum_{\qq = - \infty}^{+ \infty} \Urespnrotsigi{\qq} \left( \rrrot, \colrot, \lonrot \right) \expo{\inumber \qq \npert \time}   \right\},
\end{equation}
where $\Urespnrotsigi{\qq}$ is defined as
\begin{equation}
\Urespnrotsigi{\qq}  \left( \rrrot , \colrot, \lonrot \right) = \sum_{\llat=0}^{+ \infty} \sum_{\mm = -\llat}^{\llat} \Uplmrotsigj{\llat}{\qq }{\mm} \left( \frac{\rrrot}{\Rpla} \right)^{-\left(\llat+ 1 \right)} \Ylm{\llat}{\mm} \left( \colrot , \lonrot \right).
\end{equation}
The complex coefficients $\Uplmrotsigj{\llat}{\qq }{\mm}$ are expressed as
\begin{equation}
\label{Uplmrotsigi_app}
\Uplmrotsigj{\llat}{\qq }{\mm} = \sum_{\kk=0}^{+\infty} \sum_{\jj = -\llat}^{+\llat}   \Hansen{\qq - \kk}{0}{\mm} \left( \eccpert \right) \expo{\inumber \left( \qq - \kk \right) \meanapertc} \Uplmsigj{\llat}{\kk}{\jj} \conj{\wignerD{\jj}{\mm}{\llat}} \left( \angapert , \angbpert , \anggpert  \right).
\end{equation}

\section{Sensitivities of the tidal potential and torque to ocean tides}
\label{app:sensitivities_bulge_torque}

\begin{figure*}[t]
   \RaggedRight \twolabelsat{5.5cm}{\textsc{Full}}%
                           {13cm}{\textsc{Approx}}\\[0.1cm]
  \raisebox{\hraisebox}[1cm][0pt]{%
   \begin{minipage}{1.4cm}%
  $\obli = 90^\circ$
\end{minipage}}
\centering
   \includegraphics[width=\wpanel,trim = 0.cm 0.cm 9.cm 2.8cm,clip]{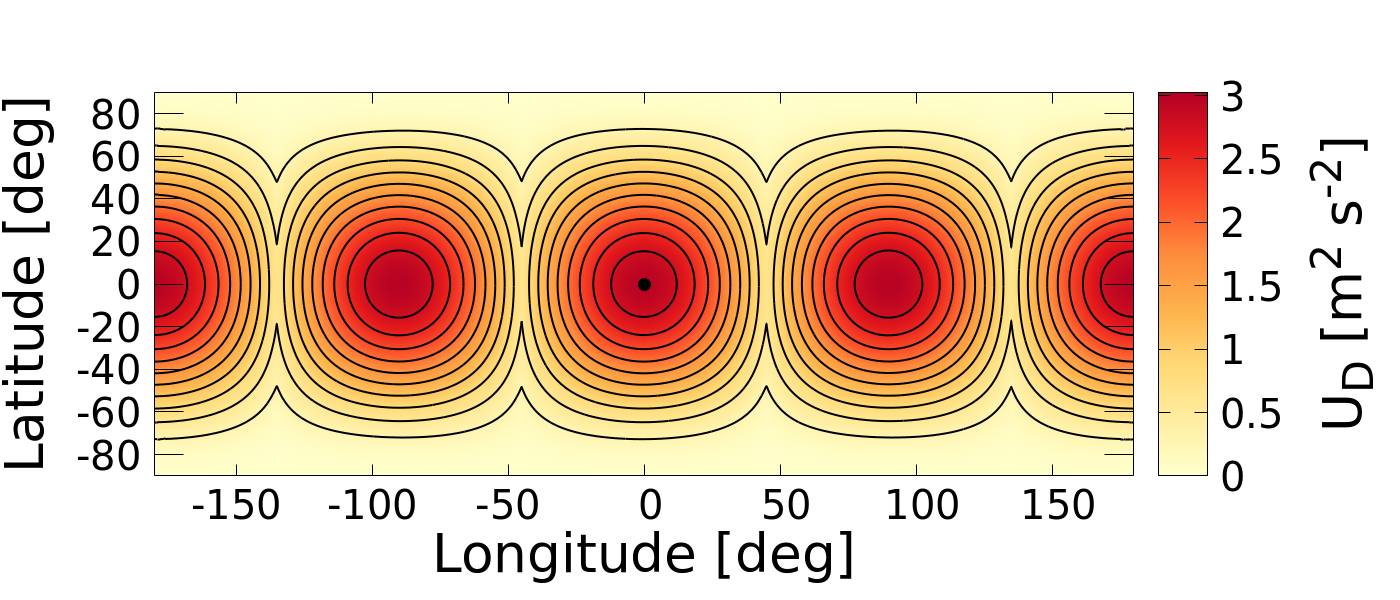} \hspace{\hsval}
   \includegraphics[width=\wpanel,trim = 0.cm 0.cm 9.cm 2.8cm,clip]{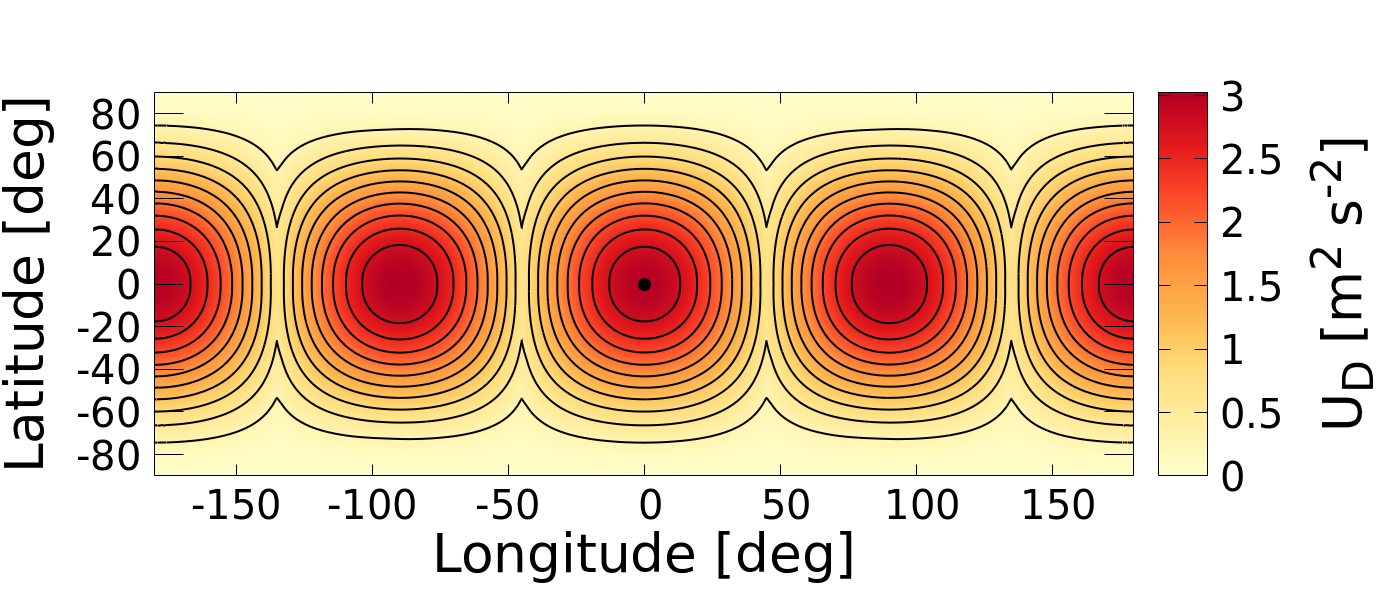} \hspace{0.36cm}  \hspace{\wkey}~ \\
   \raisebox{\hraisebox}[1cm][0pt]{%
   \begin{minipage}{1.4cm}%
   $\obli = 60^\circ$
\end{minipage}}
   \includegraphics[width=\wpanel,trim = 0.cm 0.cm 9.cm 2.8cm,clip]{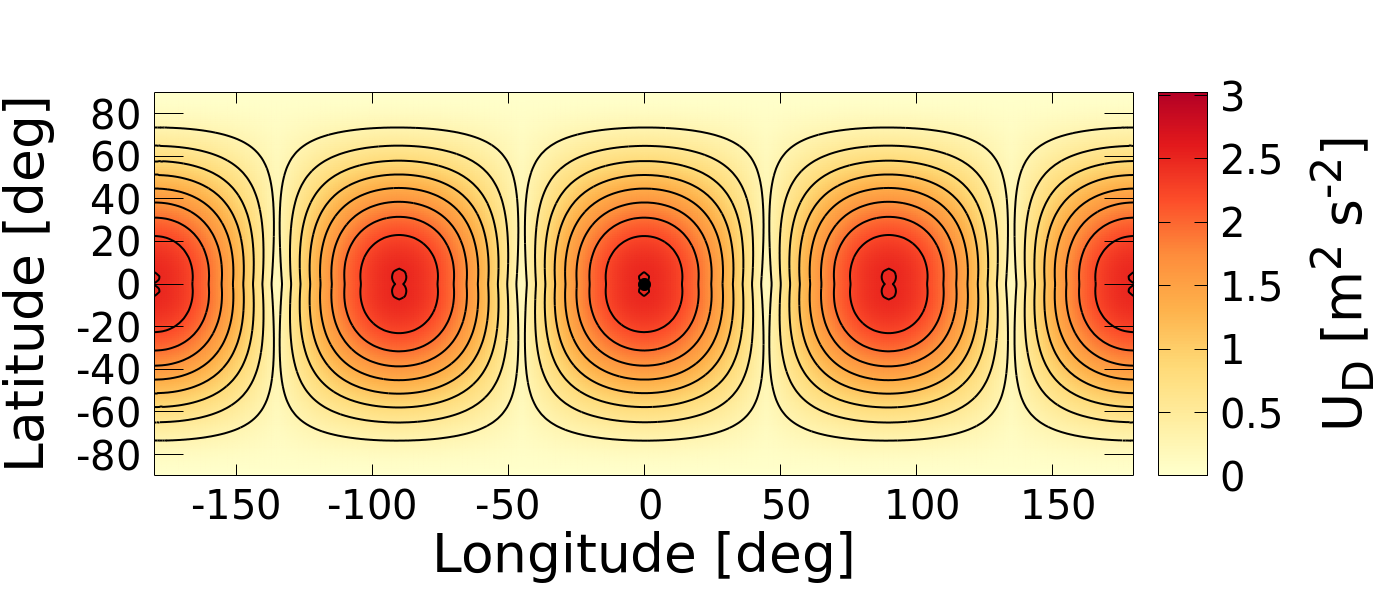} \hspace{\hsval}
   \includegraphics[width=\wpanel,trim = 0.cm 0.cm 9.cm 2.8cm,clip]{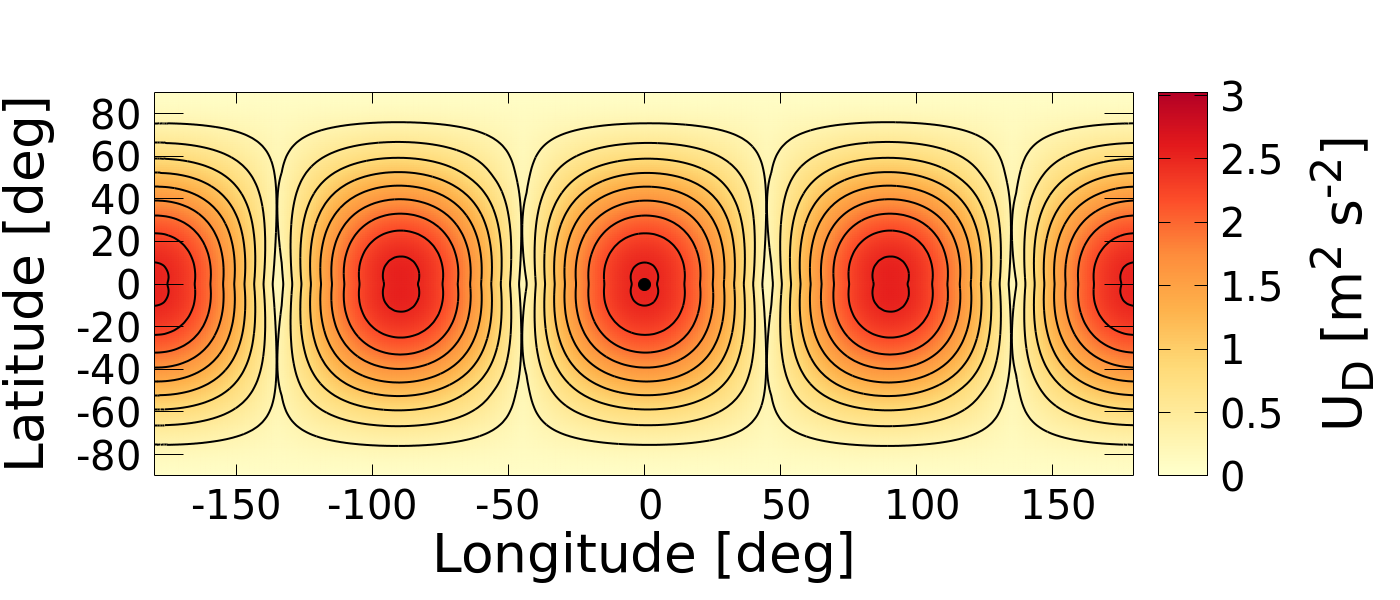} \hspace{0.36cm}  \hspace{\wkey}~ \\
   \raisebox{\hraisebox}[1cm][0pt]{%
   \begin{minipage}{1.4cm}%
   $\obli = 30^\circ$
\end{minipage}}
   \includegraphics[width=\wpanel,trim = 0.cm 0.cm 9.cm 2.8cm,clip]{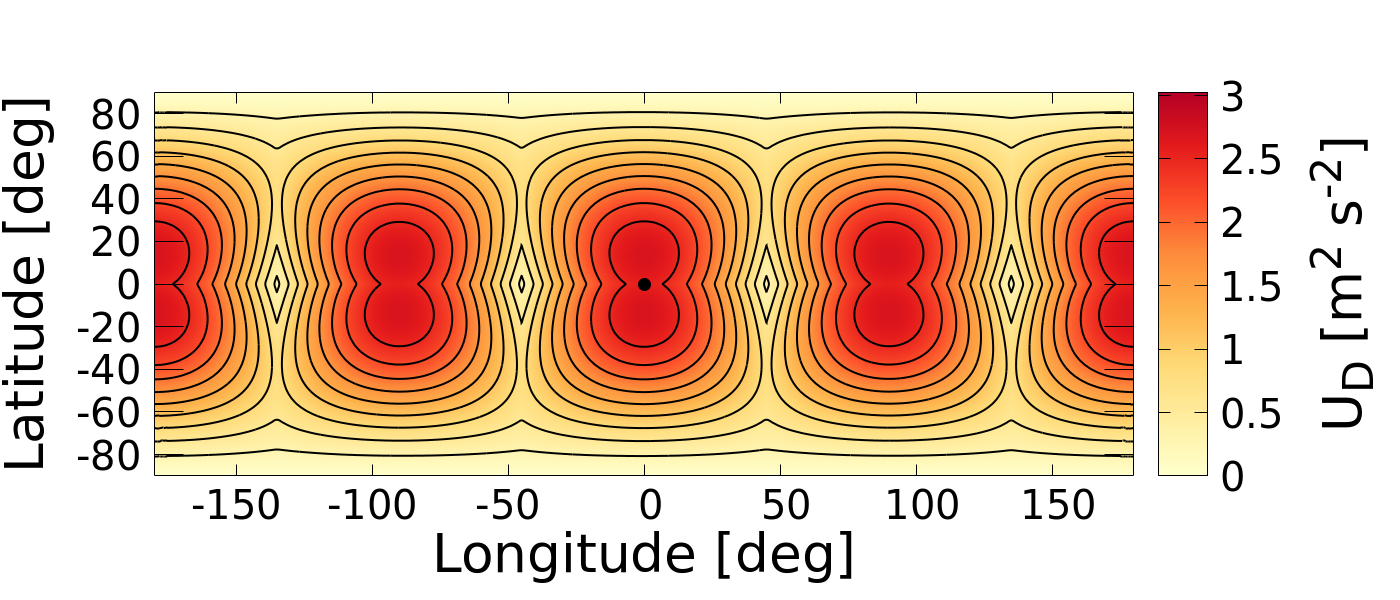} \hspace{\hsval}
   \includegraphics[width=\wpanel,trim = 0.cm 0.cm 9.cm 2.8cm,clip]{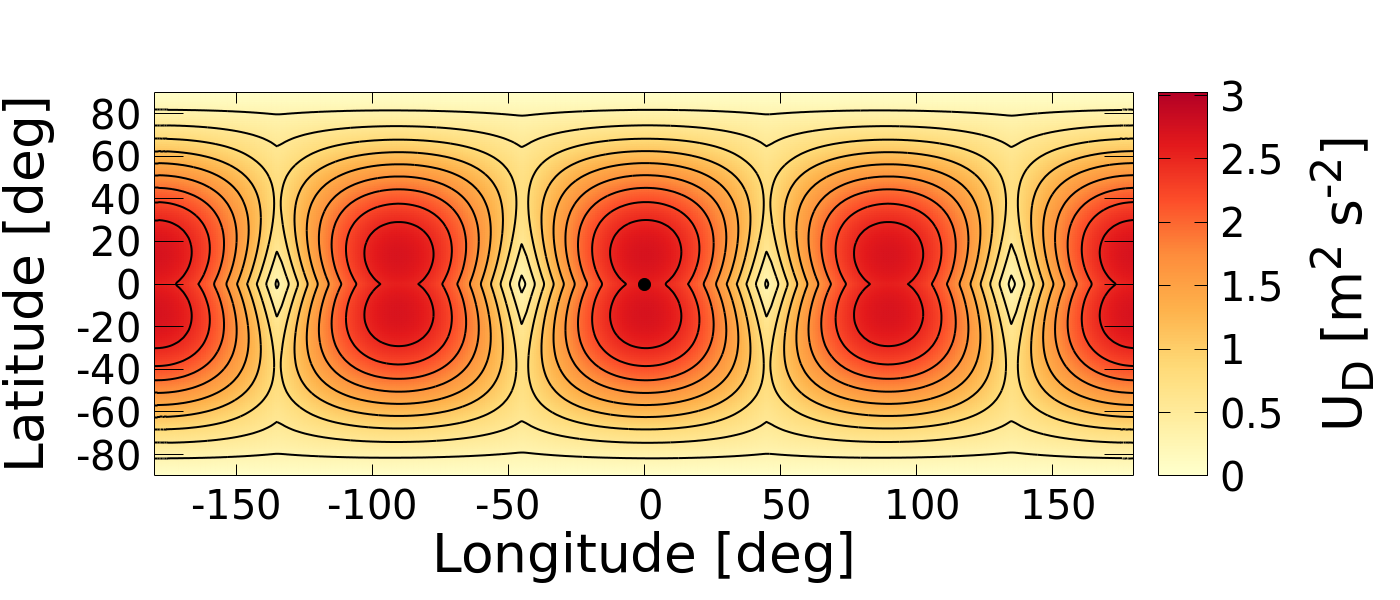} \hspace{0.36cm}  \hspace{\wkey}~ \\
   \raisebox{\hraisebox}[1cm][0pt]{%
   \begin{minipage}{1.4cm}%
   $\obli = 0^\circ$
\end{minipage}}
   \includegraphics[width=\wpanel,trim = 0.cm 0.cm 9.cm 2.8cm,clip]{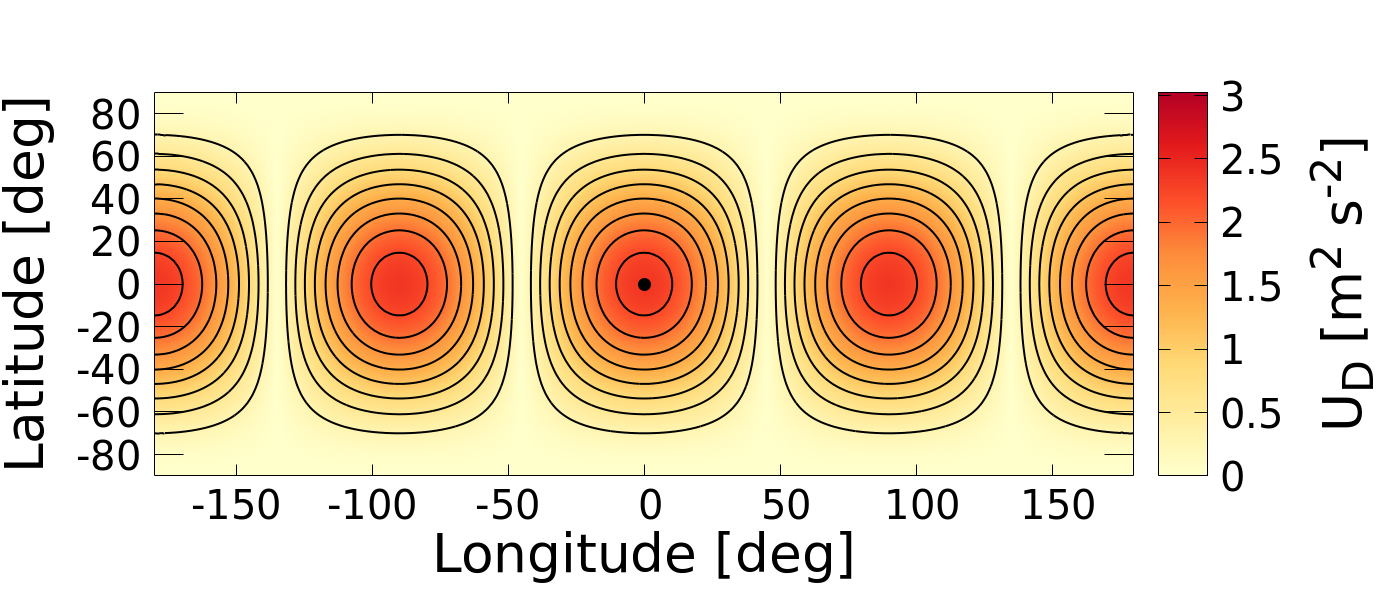} \hspace{\hsval}
   \includegraphics[width=\wpanel,trim = 0.cm 0.cm 9.cm 2.8cm,clip]{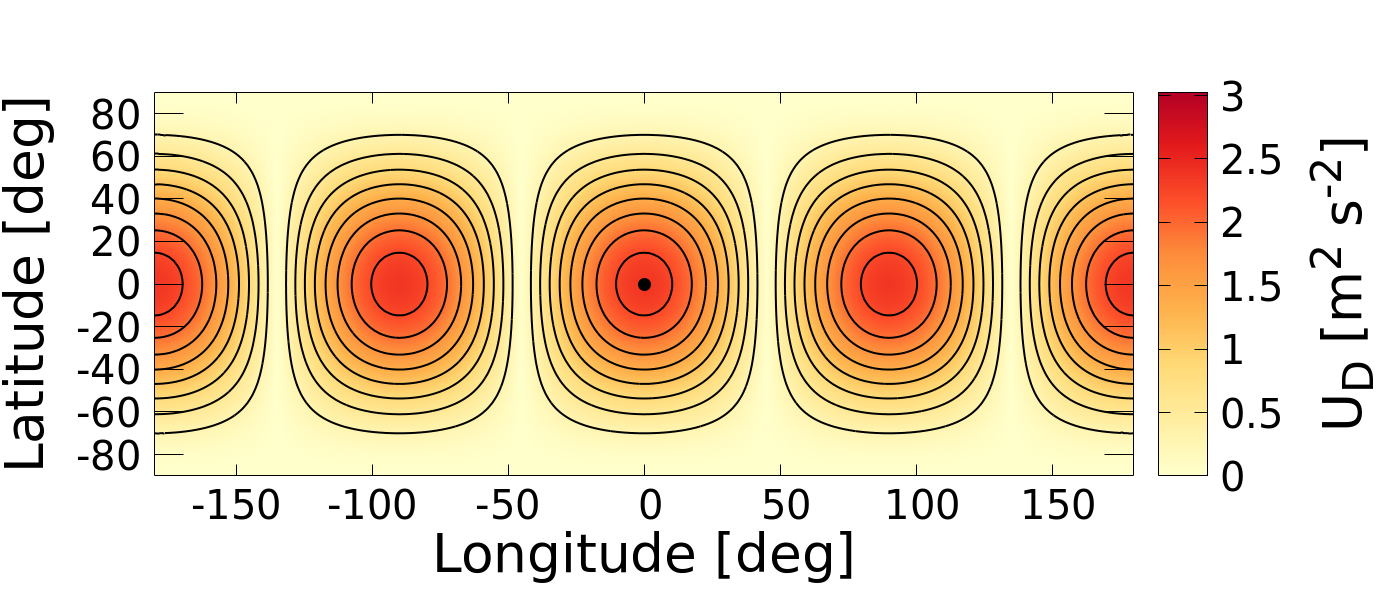} 
   \includegraphics[width=\wkey,trim = 40.5cm 0.cm 3.9cm 2.8cm,clip]{auclair-desrotour_figJ1h.png} 
   \raisebox{0.9cm}{\rotatebox{90}{$ \Urespub \  {\rm [m^2 \ s^{-2}]} $}}
      \caption{Gravitational potential induced by the tidal response of an ocean planet with deformable solid regions in the system of coordinates rotating with the perturber for $\Prot = 10$ hr and obliquity values ranging between $0^\circ$ and $90^\circ$. {\it Left:} Maximum amplitude of the tidal potential obtained from the full calculation (\textsc{Full}). {\it Right:}  Maximum amplitude of the tidal potential obtained with the standard approximation based on the equatorial degree-2 Love number (\textsc{Approx}). Amplitudes are plotted as functions of longitude, $\lonrot$ (horizontal axis), and latitude, $\colrot^\prime = 90^\circ - \colrot$ (vertical axis). Red areas indicate large amplitudes, and yellow areas small amplitudes. The black dot at $\left( \colrot^\prime, \lonrot \right) = \left( 0^\circ , 0^\circ \right)$ designates the sub-satellite point. }
       \label{fig:map_prot_10h_soloc}%
\end{figure*}

For comparison, we computed the maps in \fig{fig:map_prot_10h} for a global ocean with rigid solid regions and repeated this analysis for deformable solid regions. The results are shown in \fig{fig:map_prot_10h_soloc}. When considering the coupled tidal response of both the solid part and ocean components, we find that the tidal potential and torque exhibit markedly different sensitivities to the oceanic tidal response. Variations in self-attraction are largely unaffected by oceanic tides. The predominant patterns in \fig{fig:map_prot_10h_soloc} arise from the distortion of solid regions. In contrast, the tidal torque and dissipated power are heavily influenced by oceanic tides, as highlighted in \figsto{fig:pdiss_solid_globoc}{fig:obli_solid_globoc}. Notably, the tidal torque can be significantly amplified by the resonances of the tidally forced gravity modes, unlike the tidal potential. We elucidate this seemingly counterintuitive behaviour using a simple toy model. 

We consider a coplanar-circular configuration, where the perturber orbits the planet in a circular path within its equatorial plane. In this scenario, the only tidal component contributing to the orbital and rotational evolution of the planet-perturber system is the semidiurnal tide. Additionally, we adopt the perturber's reference frame introduced in \sect{ssec:mapping_self_attraction}, where the perturber is static, and the associated system of coordinates, $\left( \colrot , \lonrot \right)$. The sub-perturber point is defined as $\left( \colrot , \lonrot \right) = \left( 90^\circ , 0^\circ \right)$. Under these assumptions, the gravitational potential induced by the tidal deformation of the body is expressed as
\begin{equation}
\Uresp \left( \colrot, \lonrot \right) = \Re \left\{ \compcoef \Ylm{2}{2} \left( \colrot , \lonrot \right) \right\},
\end{equation}
where $\compcoef$ is a complex coefficient. In linear theory, the tidal torque scales as $\torquez \scale \Im \left\{ \compcoef \right\} $ while the spatial distribution of the tidal potential scales as $\Uresp \scale \abs{\compcoef} \cos \left( \mm \lonrot + \arg \compcoef \right)$, with $\arg \compcoef$ denoting the argument of $\compcoef$. Therefore, it is essential to examine how $\Im \left\{\compcoef \right\}$, $\abs{\compcoef}$, and $\arg \compcoef$ depend on the solid and oceanic tidal responses. 

The complex coefficient $\compcoef$ can be decomposed as 
\begin{equation}
\compcoef= \compcoef_\isol + \compcoef_\iocean,
\end{equation}
where $\compcoef_\isol$ and $\compcoef_\iocean$ represent the contributions from solid and oceanic tides, respectively. These complex coefficients can be written as
\begin{align}
& \compcoef_\isol = \abs{\compcoef_\isol} \expo{\inumber \alpha_\isol} , & \compcoef_\iocean = \abs{\compcoef_\iocean} \expo{\inumber \alpha_\iocean},
\end{align}
where $\alpha_\isol$ and $\alpha_\iocean$ are the phase angles associated with each response. For simplicity, we assume that both the solid and oceanic responses experience slight delays relative to the tide-raising potential, such that $\alpha_\isol \ll1 $ and $\alpha_\iocean \ll1$. We denote the ratio of the moduli as $\eta = \abs{\compcoef_\iocean}/\abs{\compcoef_\isol}$. The real and imaginary parts of $\compcoef$ are given by 
\begin{align}
\Re \left\{ \compcoef \right\} & = \abs{\compcoef_\isol} \cos \alpha_\isol + \abs{\compcoef_\iocean} \cos \alpha_\iocean, \\
\Im \left\{ \compcoef \right\}  & = \abs{\compcoef_\isol}  \sin \alpha_\isol +\abs{\compcoef_\iocean} \sin \alpha_\iocean .
\end{align}
Using first-order Taylor expansions in $\alpha_\isol$ and $\alpha_\iocean$, these simplify to 
\begin{align}
\Re \left\{ \compcoef \right\} & \approx \abs{\compcoef_\isol} \left( 1 + \eta \right), \\
\Im \left\{ \compcoef \right\}  & \approx \abs{\compcoef_\isol} \left(  \alpha_\isol + \eta  \alpha_\iocean \right).
\end{align}

From these equations, we can deduce the modulus and argument of $\compcoef$,
\begin{align}
\abs{\compcoef} & \approx \abs{\compcoef_\isol} \left( 1 + \eta \right) , \\
\arg \compcoef & \approx  \frac{\alpha_\isol + \eta \alpha_\iocean}{1 + \eta} . 
\end{align}
The solid tide is characterised as an equilibrium tide, meaning that $\alpha_\isol$ is very small and $\abs{\compcoef_\isol}$ is relatively independent on tidal frequency. Conversely, the oceanic tide is highly sensitive to tidal frequency within the resonant regime, leading to significant variations in $\alpha_\iocean$. It follows that $\alpha_\isol \ll \alpha_\iocean$. However, due to the density difference between solid rock and liquid water, $\eta \ll 1$ generally holds true. Consequently, $\abs{\compcoef} $ remains nearly constant, and $\arg \compcoef \ll 1$, while $\Im \left\{ \compcoef \right\} \scale \alpha_\iocean $ under the condition that $\alpha_\isol \ll \eta \alpha_\iocean$. This illustrates how the sensitivities of the tidal torque and potential to the oceanic tidal response can differ significantly. 

\end{appendix}

\end{document}